\documentstyle[preprint,aps,epsf]{revtex}
\begin{document}
\tightenlines
\preprint{\begin{minipage}[t]{1.5in}
          ITP-SB-98-28\\
          BNL-HET-98/17
          \end{minipage}}
\vspace{0.3in}

\title{Single Transverse-Spin Asymmetries in Hadronic
       Pion Production}
\author{Jianwei Qiu$^1$\footnote{On-leave from Department of 
                        Physics and Astronomy, Iowa State University,
                        Ames, Iowa 50011, USA.} 
    and George Sterman$^2$}
\address{$^1$Physics Department, Brookhaven National Laboratory\\
             Upton, New York 11973-5000, USA \\
         $^2$Institute for Theoretical Physics, 
             State University of New York \\
             Stony Brook, New York 11794-3840, USA}
\date{June 4, 1998}
\maketitle

\begin{abstract} 
We analyze single transverse-spin 
asymmetries for hadronic pion production at 
large transverse momenta using QCD factorization.  In the large $x_F$ region, 
leading 
contributions to the asymmetries are naturally produced by twist-3 parton 
correlation functions that couple quark fields and gluon 
field strengths.  With a simple model for these matrix elements, 
leading-order asymmetries 
calculated from QCD are consistent with 
data on pion production from Fermilab,
and can be used to predict single-spin asymmetries at RHIC.
We argue that our perturbative calculation
for the asymmetries is relevant to pion transverse momenta as 
low as a few GeV.

\vspace{0.2in}

\end{abstract} 

\newpage

\section{Introduction}
\label{sec:1}

Perturbative Quantum Chromodynamics (QCD) has been successful in
interpreting and predicting spin-averaged scattering
cross sections at large momentum transfer.  
Since quarks and gluons carry
spin, we expect QCD to
apply to hard spin-dependent scattering as well.  However, high
energy experiments with polarized beam and/or target have provided
many theoretical challenges.  For example, data on the
spin asymmetries in deep-inelastic scattering (DIS) of polarized
leptons on polarized hadrons \cite{EXP:EMC} sparked a wave of
theoretical effort in understanding the nature of the nucleon's spin
\cite{THY:EMC}.   

A spin
asymmetry is the difference of two polarized cross
sections, with opposite directions of polarization, divided by their
sum.  Asymmetries can be obtained with both beams (or beam and
target) polarized or only one beam (or target) polarized.  The former
is a double spin asymmetry, and the latter 
a single spin asymmetry.  Depending on the direction of
the polarization, we can have longitudinal-spin asymmetries, if the
polarization is along the beam direction, and/or transverse-spin
asymmetries, when the spin is polarized perpendicular to the beam direction. 

Because of parity and time-reversal invariance, 
single longitudinal-spin asymmetries for single-particle
inclusive production vanish for the strong interactions.  However,
experimentally-significant 
single transverse-spin asymmetries have been observed in $\Lambda$
production, as well as pion production, for almost twenty years 
\cite{EXP:Lambda,EXP:Pion}.  These single transverse-spin
asymmetries are of the order of ten or more percent of the unpolarized
cross section.  Experimental results on pion
production have been very consistent, and the effects persist to pion
transverse momenta of several GeV, into the hard-scattering
region, where perturbative QCD (pQCD) has had
success in describing spin-averaged cross sections \cite{JFO:Rev}. The
extension of the pQCD formalism to spin-dependent cross 
sections, however, has not been completely straightforward.
It was pointed out long ago \cite{AN:KPR} that QCD perturbation 
theory predicts vanishing single transverse-spin asymmetries 
at high $p_T$.  Efremov and Teryaev later pointed out that a nonvanishing
single transverse-spin asymmetry can be obtained in pQCD if one goes
beyond the leading power \cite{AN:ET1,AN:ET2}.  However,
the relatively large size and peaking in the forward direction of
observed effects remained a difficulty \cite{AN:ET3}.  

Some time ago, using the example of hadronic
direct photon production \cite{AN:QS}, we demonstrated  that
single transverse-spin asymmetries can be consistently evaluated 
in terms of generalized
factorization theorems in perturbative QCD \cite{QS:FACs}.  The
asymmetries are presented as a sum of terms, each of which
consists of a convolution of a twist-2 parton distribution
from the unpolarized hadron, a twist-3 quark-gluon correlation
function from the polarized hadron, and a short-distance partonic hard
part calculable in perturbative QCD.  The twist-3 quark-gluon
correlation functions
reflect the interaction of quarks with the color
field of the hadron \cite{AN:QS,AN:Ryskin,TF:Schafer}.  
In order to test this formalism, we
need to have more than one physical process to extract information
on these new and fundamental correlation functions, and to 
test their universality.  Recent work has explored their
role in the Drell-Yan process \cite{TFDY:Boer}.
In this paper, we will not explore the physical interpretation
of the correlation functions beyond what is currently in the
literature.  Rather, we concentrate on the extension
of the formalism to pion production.

In the forward region for pion production, 
where $x_F$ is large, we shall argue that
leading contributions to the asymmetry depend
on only one twist-3 matrix element
 (given in Eq.\ (\ref{s3e18}) below), which couples two quark fields and one
gluon field strength.  This is the same matrix element that gives the 
leading contribution to single transverse-spin asymmetries in direct
photon production \cite{AN:QS}.  
With a simple model for this twist-3 matrix element, we show that significant 
asymmetries can
be generated, and that the asymmetries increase naturally as a
function of $x_F$.  Our simple model has two parameters: one for the 
normalization, and the other for the relative sign between the up and down 
quark correlation functions.  Extrapolating from measured single
transverse-spin asymmetries in $\pi^+$ and $\pi^-$ production in 
proton($\uparrow$)-proton collisions \cite{EXP:Pion}, we 
fix these two parameters in our model.
We can then derive both the sign and shape of the asymmetries
for $\pi^0$ production, as well as pion production in collisions 
with a polarized antiproton beam.  Our results are consistent 
with data from Fermilab experiments.  The model then
predicts the normalization, $x_F$ and
transverse-momentum dependence of the
asymmetries at higher energies.  These predictions can be
tested at RHIC.

The naive expectation for the twist-3
asymmetry, $A_N$, is $\lambda/\ell_T$, with $\lambda$ 
a nonperturbative scale from the twist-3 matrix element and $\ell_T$ 
the transverse momentum of the observed particle.  A pure $1/\ell_T$ 
dependence, however, decreases quickly as 
$\ell_T$ increases, and becomes 
ill-defined when $\ell_T$ is small.  
Consequently, one might worry that the range of $\ell_T$ where the asymmetry 
is not too small, while $\ell_T$ is large enough to use pQCD, 
is very limited, and that the region to study twist-3 physics 
might be too limited to be interesting.  In fact, we 
shall see below that single 
transverse-spin asymmetries are a very good observable to study
twist-3 physics perturbatively.  

In contrast to the naive expectation, for the kinematics of the Fermilab data, 
the
$\lambda/\ell_T$ contribution to $A_N$ is not the dominant source of 
the asymmetry.  From dimensional analysis alone, the asymmetry $A_N$ admits two 
types of contributions, which are proportional to 
$\lambda \ell_T/(-U)$ as well as $\lambda \ell_T/(-T) \sim \lambda/\ell_T$,
with $U$ and $T$ Mandelstam variables.  
Their relative contributions can be determined by perturbative calculation.
For large $x_F$, where the asymmetry is large
experimentally, $U$ is larger than
$T$, but we shall show in this paper that
the coefficient for the $\lambda\ell_T/(-U)$ term is much
larger than that of the $\lambda/\ell_T$ term in this
region (see Eq.~(\ref{s4e8})).  As we will see in Sec.~\ref{sec:5},
the transverse momentum dependence of the asymmetry is 
actually quite mild for 
$\ell_T$ from less than 2 up to 6 GeV at $x_F=0.4$, where
much of the Fermilab data were collected.  This conclusion is very encouraging
for future applications of perturbative QCD beyond the level of leading twist.

Our method and results can be generalized to single
transverse-spin asymmetries in other single particle production.
With the extracted information on twist-3 matrix elements, we can
predict both the sign and magnitude of single transverse-spin asymmetries 
for any inclusive single-particle production, such as for 
direct photons, kaons, or other hadrons. 

Related work on single-spin asymmetries involves the incorporation
of parton transverse momenta, either in parton distributions
\cite{AN:Sivers,AN:Anselmino}    or fragmentation functions 
\cite{FRAG:Collins,FRAG:JJ,AN:Artru}.
There is considerable evidence that at transverse momenta in
the range of a few GeV, ``$k_T$-smearing" effects can be important
\cite{JFO:Rev,ktsmear} in spin-averaged cross sections.  It would seem natural
to include them in the same range for single-spin  asymmetries as well.
Whether they should be thought of as the dynamical  source of the
asymmetry remains to be seen.  
The fragmentation analysis requires the introduction of ``chiral-odd" 
distribution
functions \cite{FRAG:Collins,AN:Artru}, which combine with the leading-twist
transversity function \cite{Ralston:Soper,Jaffe:Ji} to produce nonvanishing 
asymmetries.  
We shall discuss how these effects can arise in the context of twist-3
factorization theorems, but our explicit models will be based for
simplicity on chiral-even parton distributions only.

The twist-three analysis described
here is in some sense a minimalist approach, depending
on only light-cone variables, which we hope can
serve as  a benchmark for models which include models of 
both light-cone and transverse degrees of freedom.
Other descriptions of single-spin asymmetries are
based on multiquark interactions \cite{TT1} and orbital motion \cite{Mengetal1}.
Interesting comparisons 
of different approaches may be found in \cite{TT2} and \cite{Mengetal2}.

Our paper is organized as follows.
In Sec.~\ref{sec:2}, we 
define single transverse-spin asymmetries in single particle 
production in hadronic collisions.  We introduce generalized 
factorization formulas for the asymmetries, identify terms that we expect to 
dominate in the large-$x_F$ region, review the
factorization procedure  at twist-three and leading order,
and recall the leading-order spin-averaged cross sections
to which we compare.
In Sec.~\ref{sec:3}, we present our explicit calculations of 
single transverse-spin asymmetries in hadronic pion production.
We express these asymmetries in terms of short-distance
partonic cross sections (coefficient functions), calculated in perturbative QCD, 
and 
non-perturbative twist-3 matrix elements.  Using a simple
model for the twist-3 matrix elements, we compare our calculated asymmetries
with experimental data in Sec.~\ref{sec:4}.  Finally, in
Sec.~\ref{sec:5}, we give a summary of our results, and  
an outlook for the subject.  We have also included an appendix,
in which we review the application
of parity and time reversal symmetry, and
identify the list of chiral-even and chiral-odd twist-3 distributions and
fragmentation functions that can contribute to the single-spin asymmetry
for pion production.



\section{Single Transverse-Spin Asymmetries}
\label{sec:2}

\subsection{Definition and General Considerations}
\label{sec:2a}

Single spin asymmetries are introduced for reactions
in which only one particle is polarized.  For example, consider
single-particle inclusive production in a high energy collision,
\begin{equation}
A(P,\vec{s}\,) + B(P') \longrightarrow C(\ell) + X \ ,
\label{s2e1}
\end{equation}
where $A$ and $B$ are the initial particles, with $A$ polarized, 
where $C$ is the observed particle (say, a pion) of momentum $\ell$, 
and where $X$ represents all other particles in the final state.  
In order to fix the kinematics, we
choose the center of mass frame of the incoming hadrons, with the $z$-axis 
along the momentum of the polarized hadron.  We introduce two four-vectors, 
$\bar{n}^{\mu}$ and $n^{\mu}$, 
\begin{eqnarray}
\bar{n}^{\mu} &\equiv & (\bar{n}^+,\bar{n}^-,\bar{n}_T)
\equiv (1, 0, 0_T) \ ,
\nonumber \\
n^{\mu} &\equiv & (0, 1, 0_T) \ ,
\label{s3e1}
\end{eqnarray}
with $\bar{n}^2=0=n^2$, and $\bar{n}\cdot n=1$.
The incoming hadrons' momenta are 
$P^{\mu}\sim \bar{n}^{\mu}\, \sqrt{S/2}$, and 
$P'^{\mu}\sim n^{\mu}\, \sqrt{S/2}$, respectively.  
Invariants at the hadron level are defined as
\begin{eqnarray}
S &=& (P+P')^2 \approx 2P\cdot P' \nonumber \\
T &=& (P-\ell)^2 \approx -2P\cdot\ell 
\label{s3e2} \\
U &=& (P'-\ell)^2 \approx -2P'\cdot\ell ,\nonumber 
\end{eqnarray}
where hadron masses are neglected.  Given Eq.~(\ref{s3e2}), we next introduce
\begin{eqnarray}
x_F &=& \frac{2\ell_{z}}{\sqrt{S}} 
      = \frac{T-U}{S} \ , \nonumber \\
x_T &=& \frac{2\ell_T}{\sqrt{S}} \ .
\label{s3e3}
\end{eqnarray}

We now introduce $\sigma(\ell,\vec{s}\,)$ as the cross section of the process 
given in 
Eq.~(\ref{s2e1}).  The spin-averaged cross section
for single-particle inclusive production may be represented as
\begin{equation}
\sigma(\ell) \equiv \frac{1}{2}
    \left[\sigma(\ell,\vec{s}\,) + \sigma(\ell,-\vec{s}\,)\right]\, ,
\label{s2e2}
\end{equation}
and the corresponding spin-dependent cross section as
\begin{equation}
\Delta\sigma(\ell,\vec{s}\,) \equiv \frac{1}{2}
    \left[\sigma(\ell,\vec{s}\,) - \sigma(\ell,-\vec{s}\,)\right]\ .
\label{s2e3}
\end{equation}
The single spin asymmetry is often defined as a dimensionless ratio 
of spin-dependent and spin-averaged cross sections, 
\begin{equation}
A(\ell,\vec{s}\,) 
    \equiv \frac{\Delta\sigma(\ell,\vec{s}\,)}{\sigma(\ell)}
         = \frac{\sigma(\ell,\vec{s}\,) - \sigma(\ell,-\vec{s}\,)}
                {\sigma(\ell,\vec{s}\,) + \sigma(\ell,-\vec{s}\,)}
\ .
\label{s2e4}
\end{equation}
A single longitudinal-spin asymmetry is denoted as $A_L$, 
and a single transverse-spin asymmetry as $A_N$.  
We shall be concerned in this paper with $A_N$.  
For differential cross sections, the asymmetry
can be defined as
\begin{equation}
A_N(\ell,s_T) = {E_\ell\; {d^3\Delta\sigma(\ell,\vec{s}_T)}/{d^3\ell} \over
      E_\ell\; {d^3\sigma(\ell)}/{d^3\ell}} \ ,
\label{s2e5}
\end{equation}
where $E_{\ell}d^3\sigma/d^3\ell$ and $E_{\ell}d^3\Delta\sigma/d^3\ell$
are the Lorentz invariant spin-averaged and spin-dependent cross section, 
respectively.  
In this paper, we will concentrate on single transverse-spin 
asymmetries in the forward region (i.e., large $x_F$) where 
the asymmetries are largest \cite{EXP:Pion}.

Due to the symmetries of fundamental interactions, it is possible to have 
a vanishing single transverse-spin asymmetry, even though the 
corresponding total cross section 
$\sigma(\ell,\vec{s}\,)$ itself is finite.  
For example, it was pointed out by Christ and Lee over 30 years ago 
\cite{Christ:Lee} that time-reversal invariance forbids single 
transverse-spin asymmetries in inclusive deep-inelastic scattering (DIS)
to lowest order in $\alpha_{\rm EM}$.  Let us
review the reason.  

\begin{figure}
\epsfysize=1.8in
\epsfbox{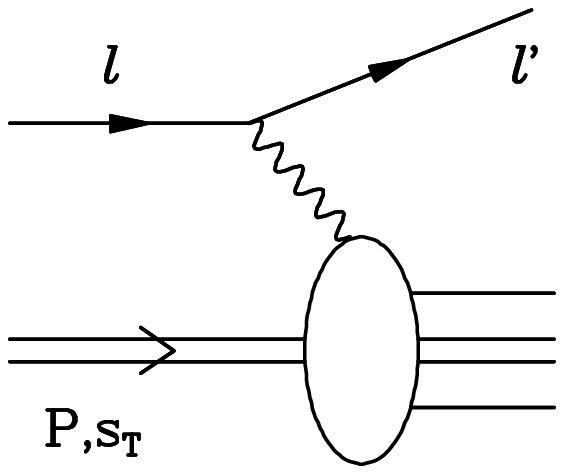}
\caption{Inclusive lepton-hadron deep-inelastic scattering, with the 
target hadron polarized transversely.}
\label{fig1}
\end{figure}

Consider a general inclusive
lepton-hadron deep-inelastic scattering, which is the  analog of Eq.\ 
(\ref{s2e1}),
\begin{equation}
L(\ell) + H(P,\vec{s}_T) \longrightarrow L(\ell') + X \ ,
\label{s2e6}
\end{equation}
where $L(\ell)$ and $L(\ell')$ are unpolarized incoming and outgoing 
leptons of 
momenta, $\ell$ and $\ell'$, respectively, and $H(P,\vec{s}_T)$ 
represents the polarized target hadron with its spin $\vec{s}_T$ 
perpendicular to the beam momentum.  In the approximation of
one-photon exchange, as shown in Fig.~\ref{fig1}, the inclusive
cross section $\sigma(\vec{s}_T)$ can be expressed as 
\begin{equation}
\sigma(\vec{s}_T) \propto L^{\mu\nu}\, W_{\mu\nu}(\vec{s}_T) \ ,
\label{s2e7}
\end{equation}
where the leptonic tensor, $L^{\mu\nu}$, is 
symmetric, and the hadronic tensor is given in terms of matrix elements
of electromagnetic currents,
\begin{equation}
W_{\mu\nu}(\vec{s}_T)  \propto
\langle P,\vec{s}_T|\,j_{\mu}^{\dagger}(0)\,j_{\nu}(y)\,
                   |P,\vec{s}_T\rangle  \ .
\label{s2e8}
\end{equation}
Applying parity and time-reversal (PT) and 
translation invariance to the matrix
element in Eq.~(\ref{s2e8}), we obtain following relation,
\begin{equation}
\langle P,\vec{s}_T|\,j_{\mu}^{\dagger}(0)\,j_{\nu}(y)\,
                   |P,\vec{s}_T\rangle  
=
\langle P,-\vec{s}_T|\,j_{\nu}^{\dagger}(0)\,j_{\mu}(y)\,
                    |P,-\vec{s}_T\rangle  \ .
\label{s2e9}
\end{equation}
Combining Eqs.~(\ref{s2e8}) and (\ref{s2e9}), we find
\begin{equation}
W_{\mu\nu}(\vec{s}_T)  =
W_{\nu\mu}(-\vec{s}_T)  \  .  
\label{s2e10}
\end{equation}
>From Eq.~(\ref{s2e3}), we obtain the spin-dependent cross section 
for inclusive deep-inelastic scattering,
\begin{eqnarray}
\Delta\sigma(\vec{s}_T) & \propto &
L^{\mu\nu}\, \left[ W_{\mu\nu}(\vec{s}_T) - W_{\mu\nu}(-\vec{s}_T)
             \right]  \nonumber \\
&=& L^{\mu\nu}\, 
    \left[ W_{\mu\nu}(\vec{s}_T) - W_{\nu\mu}(\vec{s}_T) \right] 
                      \nonumber \\
&=& 0 \, ,
\label{s2e11}
\end{eqnarray}
where in the second line we use Eq.~(\ref{s2e10}) and in the third 
the symmetry of $L^{\mu\nu}$
when  the lepton is unpolarized.   From Eqs.~(\ref{s2e4})
and (\ref{s2e11}), it is clear that the single transverse-spin asymmetry 
for inclusive deep-inelastic scattering, $A_N^{\rm DIS}$, vanishes
to lowest order in $\alpha_{\rm EM}$. 

In hadron-hadron scattering, in contrast, the presence of 
multiple (initial-state or final-state)
interactions prevents a simple decomposition like Eq.\ (\ref{s2e7}), and
allows single transverse-spin asymmetries for final-state 
photons as well as hadrons \cite{AN:ET2,AN:QS}.  Experimentally, 
data from Fermilab show large single transverse-spin asymmetries 
in single pion production \cite{EXP:Pion}, and at the same time, show 
no apparent single transverse-spin asymmetries in prompt photon 
production in the central (low $x_F$) region \cite{EXP:Photon}.  

Experiments at Fermilab for pion 
($\pi^{\pm}, \pi^0$) and prompt photon production 
were carried out with a 200 GeV polarized proton 
(or antiproton) beam on an unpolarized proton target.  
The observed single transverse-spin asymmetries of inclusive single 
pion production can be as large as 20 to 30\% in the forward region.  
In addition to the large values of the asymmetries, a number of other 
interesting 
features are evident in the data.  For example,
a strong rise of the asymmetries with $x_F$ was observed for
all pion charges.  
When the beam was switched from polarized proton to polarized 
antiproton, the same sign of the asymmetry was observed for $\pi^0$, 
while the sign of the asymmetry for $\pi^+$, as well as $\pi^-$, changed. 
Both beams had opposite signs of the asymmetries of 
$\pi^+$ and $\pi^-$.

\begin{figure}
\epsfysize=3.0in
\epsfbox{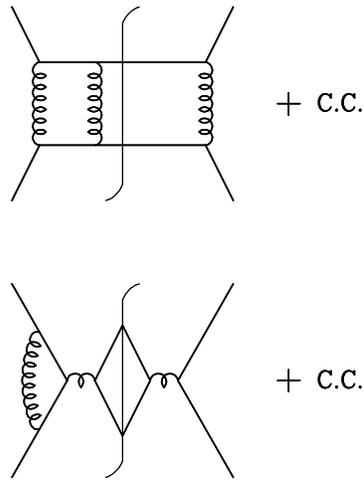}
\caption{Quark-quark scattering diagrams that give a nonvanishing single 
transverse-spin asymmetry in large-$p_T$ reactions \protect
\cite{AN:KPR}.}
\label{fig2}
\end{figure}

Perturbative QCD was first used to study the effects of single 
transverse-spin asymmetries by Kane, Pumplin, and Repko (KPR)
\cite{AN:KPR}.  KPR calculated the single transverse-spin asymmetry 
for single hadron (pion) production in terms of a QCD parton model.
By calculating the quark-quark scattering diagrams shown in 
Fig.~\ref{fig2}, KPR found that the nonvanishing single transverse-spin 
asymmetry for large-$p_T$ reactions is proportional to the quark mass:
$A_N \propto T_m \sim m_q \langle P,\vec{s}_T| \bar{\psi} \Gamma
\psi |P,\vec{s}_T\rangle$, where, for example, $\Gamma=
\gamma^+ \gamma_5 \gamma_T$.  
Consequently, the asymmetry 
vanishes in the scaling limit ($m_q\rightarrow 0$).  
Although this calculation does not explain the observed large single 
transverse-spin 
asymmetries \cite{AN:KPR,AN:ET1}, the fact that the result 
is proportional to the quark mass indicates that the
single transverse-spin asymmetry is a twist-3 effect in QCD perturbation
theory \cite{AN:ET2,AN:QS,Ratcliffe}.  

QCD dynamics, however, is much richer than the parton model.  In addition to the
parton mass effects just discussed, 
there are other twist-3 contributions.  Because quarks are not exactly 
parallel to the incoming hadron beam, twist-3 contributions also arise 
from ``intrinsic'' transverse momentum, which is proportional to 
$T_{k_T} \sim \langle P,\vec{s}_T| \bar{\psi} \Gamma \partial_T \psi 
|P,\vec{s}_T\rangle$.  In addition, there are twist-3 contributions from 
the interference between a quark state and a quark-gluon state, which is 
proportional to $T_{A_T} \sim \langle P,\vec{s}_T | \bar{\psi} 
\Gamma A_T \psi 
|P,\vec{s}_T\rangle$.  Due to gauge invariance, $T_{k_T}$ and 
$T_{A_T}$ are not independent, and can be combined to form
$T_{D_T} \sim \langle P,\vec{s}_T | \bar{\psi} \Gamma D_T \psi 
|P,\vec{s}_T\rangle$, and/or $T_{F} \sim \langle P,\vec{s}_T | 
\bar{\psi} \Gamma F^+{}_T \psi |P,\vec{s}_T\rangle$, with $F^+{}_T \propto
[D^+,D_T]$, where $D_{\mu}$ is the covariant derivative.  Therefore, 
in addition to parton mass effects, single transverse-spin asymmetries 
can be proportional to the twist-3 matrix elements $T_{D_T}$ and 
$T_{F}$ \cite{AN:ET2,AN:QS,Ratcliffe}.  These twist-3 matrix elements 
involve three field operators ($\bar{\psi}\Gamma D_T\psi,\
\bar{\psi}\Gamma F^+{}_T\psi$, or with the quark fields replaced by 
gluon field strengths \cite{AN:JI}).  
Also, different choices for
the Dirac matrices $\Gamma$ in the operators give different twist-3 matrix 
elements (see the Appendix) \cite{AN:QS}. 

Because of their odd numbers of field operators, 
three-field twist-3 matrix elements 
do not have the probability interpretation of parton distributions,
which are proportional to matrix elements of twist-2 operators, 
$\bar{\psi}\Gamma \psi$ or $F^+{}_TF^+{}_T$.  
In principle, however, they are as fundamental 
as the parton distributions.  Measurements of twist-3 
distributions, or three-field correlation 
functions, provide us new opportunities to study QCD dynamics.  

\subsection{Factorization and the Valence Quark Approximation} 
\label{sec:2b}

\begin{figure}
\epsfysize=2.2in
\epsfbox{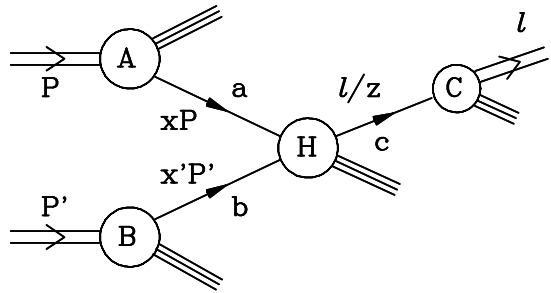}
\caption{Sketch of  single pion production in spin-averaged
hadron-hadron collisions.}
\label{fig3}
\end{figure}

As we have seen, 
spin-dependent asymmetries for hadronic pion production with one 
hadron transversely polarized vanish at large momentum transfer \cite{AN:KPR}.  
Nonvanishing values of the single transverse-spin asymmetry signal  
non-leading power contributions. 
According to 
the basic factorization theorems \cite{CSS:FAC}, the leading power
spin-averaged cross section for the production of a pion with large 
transverse momentum $\ell_T$ can be factorized into 
{\it four} separated functions, as sketched in Fig.~\ref{fig3},
\begin{equation}
\sigma_{A+B\rightarrow \pi}
= \sum_{abc} \phi_{a/A}(x) \otimes \phi_{b/B}(x')
             \otimes \hat{\sigma}_{a+b\rightarrow c} 
             \otimes D_{c\rightarrow \pi}(z)\ ,
\label{s3e4}
\end{equation}
where $\sum_{abc}$ represents the sum over parton flavors: quark, 
antiquark and gluon.  In Eq.~(\ref{s3e4}), 
$\phi_{a/A}(x)$ and $\phi_{b/B}(x')$ are  
probability densities to find parton $a$ of momentum $xP$ 
in hadron $A$ and parton $b$ of momentum $x'P'$ in hadron $B$, 
respectively.  
As noted above, they may be interpreted in terms of
expectation values in the hadronic state of two-field matrix elements,
for example $\bar{\psi}\Gamma \psi$ or $F^+{}_TF^+{}_T$.
$D_{c\rightarrow \pi}(z)$ is the 
fragmentation function for a parton $c$ of momentum $p_c=\ell/z$ 
to fragment into a pion of momentum $\ell$, and 
$\hat{\sigma}_{a+b\rightarrow c}$ is a short-distance partonic 
part (the Born cross section plus corrections), calculable perturbatively 
order-by-order in $\alpha_s$.
The symbol $\otimes$ in Eq.~(\ref{s3e4}) represents the 
convolution over the corresponding parton momentum fraction.  In terms of 
the Lorentz invariant differential cross section, Eq.~(\ref{s3e4}) can be
written as \cite{JFO:Rev}
\begin{equation}
E_{\ell}\frac{d^3\sigma_{A+B\rightarrow \pi}}{d^3\ell} 
= \sum_{abc}
\int dx\, \phi_{a/A}(x)\, 
\int dx'\, \phi_{b/B}(x')\,
\int \frac{dz}{z}\, 
\left(E_{c}\frac{d^3\hat{\sigma}_{a+b\rightarrow c}}{d^3p_c}\right) \,
\frac{D_{c\rightarrow\pi}(z)}{z} \ . 
\label{s3e5}
\end{equation}
The predictive power of Eq.~(\ref{s3e5}) depends on independent 
measurements of the non-perturbative functions, $\phi_{a/A}$,
$\phi_{b/B}$ and $D_{c\rightarrow \pi}$, and the calculation of
the partonic part
$E_{c}d^3\hat{\sigma}_{a+b\rightarrow c}/d^3p_c$.  

Just as for most other physical observables calculated in perturbative QCD,
the predictive power of the theory for twist three relies on factorization 
theorems
\cite{CSS:FAC}.  Physical observables that depend on the transverse 
polarization of a single hadron are typically power corrections to
the total cross section, in
comparison with spin-averaged 
or longitudinally polarized cross sections.  In Ref.~\cite{QS:DYQ},
for a physical observable with a large momentum transfer $Q$,
we extended the factorization program to $O(1/Q^2)$ corrections for 
unpolarized hadron-hadron cross sections, and in \cite{QS:FACs,QS:FAC}
to $O(1/Q)$ corrections in polarized cross sections.

\begin{figure}
\epsfysize=5.0in
\epsfbox{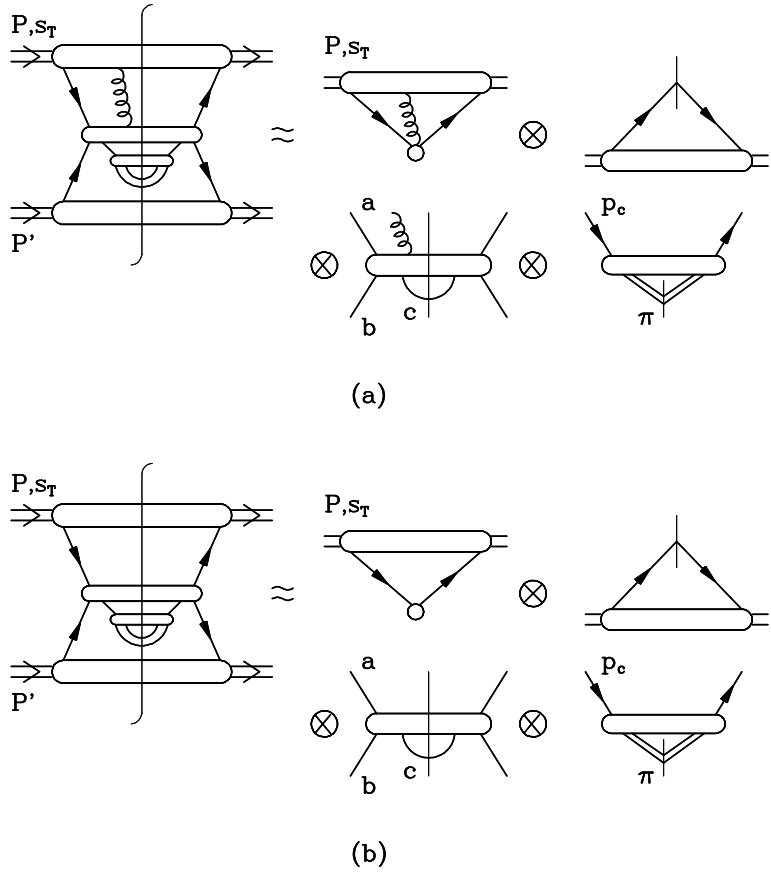}
\caption{Factorization of a typical forward scattering amplitude 
contributing to the spin-dependent cross section for
hadronic pion production: (a) with chiral-even three-parton matrix element,
(b) with chiral-odd transversity. }
\label{fig5}
\end{figure}

Following the generalized 
factorization theorem \cite{QS:FACs,QS:FAC}, the transverse spin-dependent
cross section for large $\ell_T$ pions, 
$\Delta\sigma(\vec{s}_T)$, can be written in much the same way
as the spin-averaged cross section, Eq.\ (\ref{s3e4}), as a sum of three generic 
higher-twist 
contributions, each of which can also be factorized 
into four functions, 
\begin{eqnarray}
\Delta\sigma_{A+B\rightarrow \pi}(\vec{s}_T)
&=& \sum_{abc} \phi^{(3)}_{a/A}(x_1,x_2,\vec{s}_T) 
             \otimes \phi_{b/B}(x')
             \otimes H_{a+b\rightarrow c}(\vec{s}_T) 
             \otimes D_{c\rightarrow \pi}(z)
\nonumber \\
&+& \sum_{abc} \delta q^{(2)}_{a/A}(x,\vec{s}_T) 
             \otimes \phi^{(3)}_{b/B}(x_1',x_2')
             \otimes H_{a+b\rightarrow c}''(\vec{s}_T) 
             \otimes D_{c\rightarrow \pi}(z)
\label{s3e11} \\
&+& \sum_{abc} \delta q^{(2)}_{a/A}(x,\vec{s}_T) 
             \otimes \phi_{b/B}(x')
             \otimes H_{a+b\rightarrow c}'(\vec{s}_T) 
             \otimes D^{(3)}_{c\rightarrow \pi}(z_1,z_2)  
\nonumber \\  
&+& \mbox{higher power corrections\, ,} 
\nonumber
\end{eqnarray}
where $\sum_{abc}$ represents sums over parton flavors: quark, 
antiquark and gluon, and where
$\phi_{b/B}(x')$ and $D_{c\rightarrow \pi}(z)$ are 
standard twist-two  parton distributions and fragmentation
functions, respectively.  In Eq.~(\ref{s3e11}), 
the first term corresponds to the process sketched 
in Fig.~\ref{fig5}a, and the second and third terms 
correspond to the ones sketched in Fig.~\ref{fig5}b.

For the first term in Eq.\ (\ref{s3e11}), nonvanishing contributions to 
$\Delta\sigma(\vec{s}_T)$ 
come from twist-3 parton distributions (correlation functions) 
$\phi^{(3)}_{a/A}(x_1,x_2,\vec{s}_T)$ 
in the polarized hadron.  For the second and third terms, 
the contributions to 
$\Delta\sigma(\vec{s}_T)$ involve the twist-2 transversity 
distributions $\delta q^{(2)}_{a/A}(x,\vec{s}_T)$ \cite{Ralston:Soper,Jaffe:Ji}. 
 Because
the operator in the transversity distribution requires an
even number of $\gamma$-matrices \cite{Ralston:Soper,Jaffe:Ji}, 
the second term and third terms
in Eq.~(\ref{s3e11}) also include a
twist-3, chiral-odd
parton distribution $\phi^{(3)}_{b/B}(x_1',x_2')$ from the unpolarized hadron 
$B$,
or a twist-3, chiral-odd fragmentation function, 
$D^{(3)}_{c\rightarrow \pi}(z_1,z_2)$.  
In the factorized form of Eq.\ (\ref{s3e11}), PT invariance may be
applied in a manner analogous to the treatment of the DIS
cross section given above.  In this case, however, PT invariance allows nonzero 
$A_N$
for a limited number of
functions, as discussed in the Appendix.

As in the spin-averaged cross section, Eq.~(\ref{s3e4}), 
the hard-scattering functions $H_{a+b\rightarrow c}(\vec{s}_T)$ are the only 
factors in 
Eq.~(\ref{s3e11}) that are calculable in QCD perturbation theory.  
The calculation of the $H$'s depends on the explicit definitions 
of the twist-3 distributions, for example $\phi^{(3)}_{a/A}(x_1,x_2,\vec{s}_T)$, 
and
the predictive power of Eq.~(\ref{s3e11}) relies
on the universality of the new twist-3 
distributions \cite{QS:FACs,QS:FAC}.

Eq.\ (\ref{s3e11}) illustrates the typical complexity of higher-twist
analysis: even at first nonleading twist, whole new  classes of 
functions begin to contribute.   This complexity is
particularly difficult to sort out for physical observables 
to which leading-twist terms contribute.   The combination of small effects and 
complex
parameterizations has made the extraction of higher twist distributions
from the data difficult, despite the considerable effort that has been
invested in the formalism \cite{HT:Ellis,HT:QIU}. \footnote{We may note recent
progress based on models of higher twist in deeply inelastic
scattering and fragmentation inspired by renormalon analysis 
{\protect \cite{IRRDIS1,IRRDIS2,IRRFrag}}.}  The vanishing of single-spin 
asymmetries
at leading power solves one of these problems, the masking of higher twist
by leading twist.  Beyond this, however, it is clear that to fully
disentangle all of the functions contributing to Eq.\ (\ref{s3e11})
would require a constellation of data and a level of analysis far beyond 
what is currently available.   
Turning specifically to the first term in Eq.\ (\ref{s3e11}), we observe that 
the index
$a$ refers to pairs of partons, and that the functions $\phi_{a/A}^{(3)}$ are
correspondingly functions of two momentum fractions.  In addition, even assuming 
that
we knew this set of functions, we would still be faced with the
chiral-odd distributions and fragmentation functions in the
second and  third sums in Eq.\ (\ref{s3e11}).  
We would like to suggest, however, that by restricting
ourselves to the limited kinematic range of large $x_F$ for the observed
particle, we may simplify the analysis greatly, 
and construct a simple model that explains the available data,
and that provides extrapolations to higher energies and momentum
transfers.

We are going to present a calculation of
the large-$x_F$  asymmetry at moderate or large $\ell_T$,
in terms of the chiral even
functions $\phi_{a/A}^{(3)}(x_1,x_2,\vec{s}_T)$ only (first line
of Eq.\ (\ref{s3e11})).  In these functions, we will consider only
combinations of valence quark flavors with gluons.
We will not find it necessary to specify these
functions for all
values of $x_1$ and $x_2$, but only for the line $x_1=x_2$, at which the gluon
carries vanishingly small momentum fraction.  We will refer
to this set of simplifications as the {\it valence quark-soft gluon}
 approximation below.  
In this model, we thus neglect potential contributions from the  
transversity, coupled with the chiral-odd twist-3 distributions and
fragmentation functions identified in the Appendix.  We hope to explore these
contributions elsewhere, but in the absence of 
independent information on the transversity, 
it seems natural to test the plausibility of
a model based on chiral-even distributions alone.

First, consider our restriction to  valence quarks. 
Given that single transverse-spin 
asymmetries were measured at Fermilab with a 200 GeV polarized beam 
\cite{EXP:Pion}, only partons ($a$ and $b$ in Eq.~(\ref{s3e5})) with 
large momentum fractions will be relevant for large $x_F$
or $\ell_T$.  Because parton-to-pion
fragmentation functions vanish as $z\rightarrow 1$, the effective 
momentum of the fragmenting parton, $p_c=\ell/z$, 
should be much larger than the pion momentum $\ell$.  
Therefore, the dominant contribution to the cross sections 
in the central region should come from $x\sim x'$ in Eq.\ (\ref{s3e11}), 
with, in addition, $x$ much larger than $x_T\approx 0.25$, which corresponds to 
$\ell_T 
\approx 4$ GeV at $E_{\rm beam}=200$ GeV.  In our calculation
we will concentrate on the forward region, where $x_F$ is large.
Similarly, in this region the dominant contributions to the cross section come 
from  
$x$ considerably larger than $x_T$ (i.e., $x > 0.25$) even for relatively
small $\ell_T$.  
For large $x$, there are few gluons or sea quarks from the beam
hadron.  Therefore, in our numerical calculations, we will keep only 
valence quarks from the polarized beam.  That is, $\sum_a$ in 
Eq.~(\ref{s3e11}) now runs over only up and down valence quarks, coupled
with a single gluon field.

In presenting this argument, we are well aware that in principle
the flavor content of the twist three distributions may
be totally different than those of twist two. Nevertheless, we consider 
it by far more natural to assume that three-field correlations
at large $x_i$ will be dominated by the same flavors as in
the two-field, parton distribution, case.  
We recognize that this remains, however, an assumption.  
In our case, it means that we shall keep only valence quarks 
from the polarized beam, accompanied in twist-3 by gluons.
In particular, we shall not consider three-gluon matrix elements \cite{AN:JI}.

We now turn to the question of ``soft gluons".
To anticipate, the twist-three asymmetry 
involves only two classes of contributions in $H_{abc}(x_1,x_2)$.
One of these is 
proportional to $\delta(x_1-x_2)$, and the other to $\delta(x_i),\ i=1,2$.
The first case sets the  momentum 
carried by the gluon field in the twist-3 matrix element into the hard
scattering
to zero, leaving the momenta
carried by the two quark fields in the combination $\bar{\psi}F^{+T}\psi$
diagonal\footnote{Note, there are no ``soft gluons" in the short-distance
functions $H$.}.
In  the second, one  of the quark fields ($\psi$ or $\bar{\psi})$
carries vanishing momenta.
We refer to these two possibilities as ``soft gluon" and ``soft fermions" 
poles, respectively \cite{AN:QS}.
Soft gluon terms are typically accompanied by derivatives of 
the parton functions $\phi_{a/A}^{(3)}$, while soft fermion terms are not.  
We have emphasized
in Ref.\ \cite{AN:QS} that terms that involve 
derivatives with respect to distributions tend to be 
strongly enhanced near the edges of phase space, relative to
those without derivatives.   We shall see this in our
explicit model below.  We shall  assume, in fact, that it is
this effect that is primarily responsible for the experimentally-observed rise
in single-spin asymmetries toward $x_F=1$.  
We  therefore suggest that only
terms in which such derivatives occur need be kept, 
in order to describe 
the large-$x_F$ single-spin asymmetry.  
In summary, only soft-gluon terms, from the first line of Eq.\ (\ref{s3e11})
produce the shape of the large asymmetries observed in the data
in the forward region, and for these terms $x_1=x_2$. 

To set the stage for the explicit calculations of the
next section, we first give
an example of leading-order factorization at twist three
for the spin-dependent cross section,
following the method of Ref.\ \cite{AN:QS}.
This will enable us to trace the origin of twist-three 
spin distributions, and of the poles that
underly the valence quark-soft gluon approximation that
we have just described.

\subsection{Twist-3 Factorization at Leading Order}
\label{sec:2c}

\begin{figure}
\epsfysize=1.3in
\epsfbox{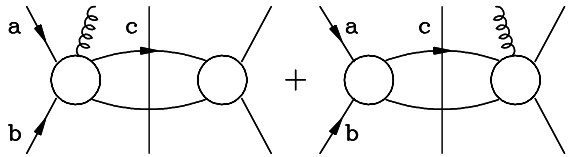}
\caption{General Feynman diagrams contributing to the partonic parts
$H$ in Eq.~(\protect\ref{s3e11}).}
\label{fig6}
\end{figure}

The twist-3 correlation functions, $\phi^{(3)}_{a/A}(x_1,x_2,\vec{s}_T)$,
depend on two parton momentum fractions, while twist-2 parton distributions, 
which are probability densities, depend on only one.  
Considering the effort and data needed to determine the 
parton distributions, it appears a difficult task 
to get a full description of these twist-3 distributions.  From 
the general Feynman graphs contributing to the $H$'s 
in Eq.~(\ref{s3e11}), as shown in Fig.~\ref{fig6}, it is also clear 
that there are many diagrams, even at lowest order.  
Their treatment is simplified, however, by taking advantage of the 
relation of the asymmetry to the pole 
structure of $H$ \cite{AN:QS}.
This will enable us to evaluate 
$\Delta\sigma(\vec{s}_T)$ in Eq.~(\ref{s3e11}) efficiently.  
Indeed, we will find that $\Delta\sigma(\vec{s}_T)$ 
depends on the
twist-3 distributions through only a single independent momentum fraction,
with the other fraction fixed 
by a pole.  To see how this comes about, we consider a specific set 
of contributions, associated with 
the three classes of diagrams shown in 
Fig.~\ref{fig7}.  We first discuss the analysis of these
diagrams according to the method of Ref.\ \cite{AN:QS}, and
then briefly discuss other possibilities, reviewing why we expect 
those of Fig.\ \ref{fig7} to dominate the asymmetry in the large-$x_F$
region.

In our valence quark-soft gluon approximation,
introduced in the last subsection, the fermion flavor $a$ 
from the polarized hadron
in Figs.\ \ref{fig6} and \ref{fig7} runs over valence 
quarks only, while parton $b$ from the unpolarized hadron can be 
a gluon, valence quark or sea quark. 
We start from these three classes 
of diagrams, and derive below the factorized form for
the spin-dependent cross section $\Delta\sigma(\vec{s}_T)$.
The hard-scattering diagrams of Fig.\ \ref{fig7} are all embedded in the overall
process shown in Fig.~\ref{fig8}a.  The top part of this 
general diagram is proportional to 
the expectation value  of an operator of the form
$\bar{\psi}A_{\sigma}\psi$ in the polarized incoming hadron
state $|P,\vec{s}_T\rangle$, while the bottom part includes the
hard subprocess, as well as the target hadron 
matrix element and the final-state pion fragmentation function.
In Fig.~\ref{fig8}, 
$k_1$ and $k_2$ are valence quark momenta, and $\sigma$ is the 
Lorentz index for the gluon field.  We work, as in Ref.\ \cite{AN:QS}, in 
Feynman gauge.
To derive a  factorized expression for these contributions,
we must separate spinor and color traces, as well as sums over
vector Lorentz indices between the functions  $T$ and $S$.

After separation of all traces by a Fierz projection (see Appendix), 
the two functions $T$ and $S$ are connected only by the
two momentum integrals that they share.
The leading contributions of the 
general diagram shown in Fig.~\ref{fig8}a can then be represented 
by the factorized diagrams shown in Fig.~\ref{fig8}b, and can be 
written as
\begin{equation}
d\Delta\sigma(\vec{s}_T) \equiv \frac{1}{2S} \sum_{a}
  \int \frac{d^4k_1}{(2\pi)^4}\, \frac{d^4k_2}{(2\pi)^4}\,
  \left[ T_a(k_1,k_2,\vec{s}_T)\, 
         S_a(k_1,k_2) \right]\ ,
\label{s3e12}
\end{equation}
where $1/2S$ is a flux factor, 
$\sum_a$ runs over only valence flavors, 
$T_a(k_1,k_2,\vec{s}_T)$ is proportional to the matrix 
element of the operator, $(2\pi)
[\bar{\psi}_{a}\gamma^+ A^+ \psi_{a}]/2P^+{}^2$, 
and $S_a(k_1,k_2)$ represents the bottom part of 
the general diagram shown in Fig.~\ref{fig8}b.   The 
function $S_a(k_1,k_2)$ is contracted with 
$[(1/2)\gamma\cdot P P_{\sigma}]\,C_a/(2\pi)$, where 
the factor $(2\pi)$ is due to the normalization of 
twist-3 matrix element $T$, which we will
specify below.  The color factor $C_a$ is left
from the factorization of color
traces between $T_a(k_1,k_2,\vec{s}_T)$ and
$S_a(k_1,k_2)$ \cite{AN:QS}.  With the function $T_a(k_1,k_2,\vec{s}_T)
\propto \bar{\psi}A\psi$, the corresponding $C_a$ is defined 
for all valence flavors $a$ as
\begin{equation}
(C_a^B)_{ij}=\left(\frac{2}{N^2-1}\right) (t^B)_{ij}\, , 
\label{s3e12n}
\end{equation}
with $N=3$ colors, $B$ the gluon color index, and with quark color 
indices $ij$.  The matrix $(t^B)_{ij}$ is the SU(3) generator
in the defining representation of the group. 

The next step in the factorization procedure is
the ``collinear" expansion \cite{QS:DYQ,QS:FAC}, which will enable us
to reduce the four-dimensional integrals in Eq.\ (\ref{s3e12})
to convolutions in the momentum fractions of partons,
as in Eq.\ (\ref{s3e11}). Expanding 
$S_a$ in the partonic momenta, $k_1$ and $k_2$,
around $k_1=x_1P$ and $k_2=x_2P$, respectively, we have
\begin{eqnarray}
S_a(k_1,k_2) = S_a(x_1,x_2) 
&+& \frac{\partial S_a}{\partial k_{1}^{\rho}}(x_1,x_2)
          \left(k_1-x_1P\right)^{\rho} \nonumber \\
&+& \frac{\partial S_a}{\partial k_{2}^{\rho}}(x_1,x_2)
          \left(k_2-x_2P\right)^{\rho}
 + \dots \ .
\label{s3e13}
\end{eqnarray}
This expansion, substituted in Eq.\ (\ref{s3e12}), 
allows us to integrate over three of the four 
components of each of the loop momenta $k_i$.
The top part of the diagram $T_a$ then becomes
a twist-three light cone matrix element, convoluted with the
terms of Eq.\ (\ref{s3e13}) in the
remaining fractional momentum variables $x_i$.

As stressed in Refs.\ \cite{AN:ET2,AN:QS}, some of the
matrix elements that result from the collinear 
expansion can have nontrivial spin-dependence.
It is at this stage that the pole structure of
the hard scattering begins to play an important
role.  In fact, as shown in Refs.\ \cite{AN:ET2,AN:QS} and below, 
nonzero spin dependence is found {\it only} from
pole terms in the hard scattering.  Without these
poles, the symmetries of the strong interaction 
force the asymmetry to vanish, in much the same fashion
as for DIS above.  Indeed, the poles provide exactly
the sort of multiple interactions that are absent
in DIS at lowest order in QED.
The first term in the expansion, Eq.\ (\ref{s3e13}), $S_a(x_1,x_2)$, 
 does not contribute to 
$\Delta\sigma(\vec{s}_T)$ when combined with 
$T_a(k_1,k_2,\vec{s}_T)$ in Eq.\ (\ref{s3e12}),
because it lacks true initial- or final-state
interactions.  
We will therefore drop it below.

Let us next look for poles in the diagrams of Fig.\ \ref{fig7}
from the remaining terms in Eq.\ (\ref{s3e13}), and identify
the relevant twist-three matrix element.
All of the diagrams in
Fig.~\ref{fig7} provide a pole at $x_1=x_2$ when $k_i=x_iP 
(i=1,2)$.  As we will show below, these poles have
the property that
\begin{equation}
\frac{\partial S_a}{\partial k_{2}^{\rho}}(x_1,x_2)
= - \frac{\partial S_a}{\partial k_{1}^{\rho}}(x_1,x_2)\, ,
\label{s3e14}
\end{equation}
for $x_1=x_2$.  This equality is to be interpreted
in the sense of distributions, since $S_a$ is singular at $x_1=x_2$.
Substituting Eq.~(\ref{s3e14}) into Eq.~(\ref{s3e13}) and neglecting 
higher order derivatives, we have
\begin{equation}
S_a(k_1,k_2) \approx 
\frac{\partial S_a}{\partial k_{2}^{\rho}}(x_1,x_2)
\left[ \omega^{\rho\sigma}(k_2-k_1)_{\sigma} \right]\ ,
\label{s3e15}
\end{equation}
where the projection operator $\omega^{\rho\sigma}$ is defined as 
$\omega^{\rho\sigma}\equiv g^{\rho\sigma}-\bar{n}^{\rho} n^{\sigma}$.
Substituting Eq.~(\ref{s3e15}) into Eq.~(\ref{s3e12}) and performing
the integration over the non-longitudinal components of the $k$'s, we derive
\begin{equation}
d\Delta\sigma(\vec{s}_T) = \frac{1}{2S}
\sum_{a} \int dx_1 dx_2 \left[
 i\epsilon^{\rho s_T n \bar{n}}\,
 \frac{\partial S_a}{\partial k_{2}^{\rho}}(x_1,x_2) \right]_{k_2^\rho=0}\,
 T_{F_a}^{(V)}(x_1,x_2) \ ,
\label{s3e16}
\end{equation}
where the integration over $x_1$ (or $x_2$) will be fixed by the 
corresponding
pole in $\partial S_a/\partial k_2$, and where  
$\epsilon^{\rho s_T n \bar{n}}$ is defined as
\begin{equation}
\epsilon^{\rho s_T n \bar{n}} 
= \epsilon^{\rho\sigma\mu\nu}\,
    \vec{s}_{T_{\sigma}}\, n_{\mu}\, \bar{n}_{\nu} \ .
\label{s3e17}
\end{equation}
The function $T_{F_a}^{(V)}(x_1,x_2)$ for flavor $a$ 
in Eq.~(\ref{s3e16}) is one of the twist-3 
distributions introduced in Ref.~\cite{AN:QS},
\begin{eqnarray}
T_{F_a}^{(V)}(x_1,x_2) &=&
\int \frac{dy_1^- dy_2^-}{4\pi} 
     \mbox{e}^{ix_1P^+y_1^- + i(x_2-x_1)P^+ y_2^-}
\nonumber \\
&& \quad \times
\langle P,\vec{s}_T |\bar{\psi}_a(0)\gamma^+
 \left[\epsilon^{s_T\sigma n \bar{n}}\, F_{\sigma}^{\ +}(y_2^-)\right]
 \psi_a(y_1^-) |P,\vec{s}_T\rangle \ .
\label{s3e18}
\end{eqnarray}
The ordered exponentials of the gauge field that make this matrix
element gauge invariant have been suppressed \cite{QS:DYQ,QS:FAC}.
It is easy to show that $T_{F_a}^{(V)}$ is real.
Parity ensures that $T_{F_a}^{(V)}\sim \epsilon^{\rho n\bar{n}s}$,
and time reversal invariance then implies that it is
an even function of $x_1$ and $x_2$,
\begin{equation}
T_{F_a}^{(V)}(x_1,x_2)=T_{F_a}^{(V)}(x_2,x_1)\, .
\label{TFiseven}
\end{equation}
These properties are valuable in isolating
nonvanishing asymmetries.  For instance, the
fact that $T_{F_a}^{(V)}$ is real ensures that
only the poles of $S$ in
Eq.\ (\ref{s3e16}) can contribute.  

Having factorized the twist-3 distribution $T_F^{(V)}$, 
we now factorize the remaining function $[i\epsilon^{\rho s_T n \bar{n}}
\partial S_a/\partial k_2]$ in Eq.~(\ref{s3e16}) 
into a perturbatively calculable partonic part
$H_{a+b\rightarrow c}$, a corresponding target parton 
distribution $\phi_{b/B}$ and a fragmentation function 
$D_{c\rightarrow\pi}$.  At the leading power, diagrams contributing
to $S_a(k_1,k_2)$ can be represented as in Fig.~\ref{fig9}a, and can be 
factorized as
\begin{eqnarray}
S_a(k_1,k_2) 
&\approx & \sum_{b} \int \frac{d^4k'}{(2\pi)^4} 
            [M_{a+b}(k_1,k_2,k') B_b(k',P')]
\nonumber \\
&\approx & \sum_{ b}\int \frac{dx'}{x'} 
            M_{a+b}(k_1,k_2,x') \phi_{b/B}(x')\ ,
\label{s3e19}
\end{eqnarray}
where $\sum_b$ runs over all parton flavors, and $\phi_{b/B}(x')$
is a twist-2 parton distribution for flavor $b$, for the 
unpolarized target hadron $B$.  We use the matrix 
element definitions of twist-2 
parton distributions given in Ref.~\cite{CS:PDF}.  Similarly, as
shown in Fig.~\ref{fig9}b, the factor $M_{a+b}(k_1,k_2,x')$ in 
Eq.~(\ref{s3e19}) can be further factorized into 
a convolution of Feynman diagrams,
calculable in perturbation theory, with standard twist-2 fragmentation 
functions,
\begin{equation}
M_{a+b}(k_1,k_2,x') 
\approx  \sum_{c} \int dz\, 
            H_{a+b\rightarrow c}(k_1,k_2,x',p_c=\ell/z)\, 
            D_{c\rightarrow \pi}(z=\ell/p_c)\ ,
\label{s3e20}
\end{equation}
where the $H_{a+b\rightarrow c}$ are given by the diagrams 
of Fig.~\ref{fig7}.  The fragmentation functions
$D_{c\rightarrow \pi}(z=\ell/p_c)$ are also defined 
as matrix elements in Ref.~\cite{CS:PDF}.

Finally, substituting Eq.~(\ref{s3e20}) into Eq.~(\ref{s3e19}), and 
Eq.~(\ref{s3e19}) into Eq.~(\ref{s3e16}), we derive a
factorized expression for $\Delta\sigma(\vec{s}_T)$ in the form of
Eq.~(\ref{s3e11}),
\begin{eqnarray}
d\Delta\sigma(\vec{s}_T) &=& \frac{1}{2S} \sum_{abc} 
\int dz\, D_{c\rightarrow\pi}(z)
\int \frac{dx'}{x'}\, \phi_{b/B}(x')
\int dx_1 dx_2\, T_{F_a}^{(V)}(x_1,x_2) 
\nonumber \\
&& \quad \times \left[
 i\epsilon^{\rho s_T n \bar{n}}\,
 \frac{\partial}{\partial k_{2}^{\rho}}
      H_{a+b\rightarrow c}(k_1=x_1P,k_2=x_2P,x',p_c=\ell/z) \right]_{k_2^\rho=0} 
\ ,
\label{s3e21}
\end{eqnarray}
where the integration over either $x_1$ or $x_2$ 
can be done by using the pole in $H_{a+b\rightarrow c}$.  
This results in a factorization with only
a single momentum fraction for each of the incoming hadrons, 
similar to that for the spin-averaged cross section 
in Eq.~(\ref{s3e5}), with $\phi_{a/A}(x)$ replaced by
$T_{F_a}^{(V)}(x,x)$.
In order to use this factorized formula for 
single transverse-spin asymmetries in pion production, 
in the following section we will evaluate the diagrams shown in 
Fig.~\ref{fig7} with off-shell momenta $k_1$ and $k_2$.
In each case, we will 
verify Eq.~(\ref{s3e14}), or equivalently, observe that
\begin{equation}
\frac{\partial H}{\partial k_{2_\rho}} (x_1,x_2=x_1)
=-\frac{\partial H}{\partial k_{1_\rho}} (x_1,x_2=x_1) \ ,
\label{s3e22}
\end{equation}
where, again, the equality is to be interpreted in terms
of distributions.

\subsection{Leading Contributions in the Forward Region}
\label{sec:2d}

Before entering into the detailed calculations of the hard-scattering
functions $H_{a+b\rightarrow c}$
in Eq.\ (\ref{s3e21}), we return to issue of 
why we believe that
the dominant contribution is given by the $T_{F_a}^{(V)}$ in Eq.\
(\ref{s3e21}).  We have already indicated that this
is due to the derivative structure of these contributions.
Let us see how these derivatives arise.

>From the diagrams shown in Fig.~\ref{fig7}, with the momenta $p_c$ and $x'P'$  
fixed, we get four typical sources of $k_i$ ($i=1,2$) 
dependence: (1) $k_i$-dependence in $\delta(L(k_i)^2)$ with $L$ the
momentum of the unobserved final-state parton, (2) $k_i$-dependence
in the propagators which go on-shell when $k_1=k_2$, (3) 
$k_i$-dependence in the off-shell propagators, and (4) 
$k_i$-dependence in the numerators.  The derivatives of 
$H_{a+b\rightarrow c}(k_1,k_2,x',p_c)$ with respective to 
$k_i$ have the following features: 
\begin{enumerate}
\item
$(\partial/\partial k_i) \delta(L(k_i)^2)$ gives 
$\delta'(L(x_iP)^2)$, and its contribution to $\Delta\sigma(\vec{s}_T)$ 
is proportional to $(\partial/\partial x) T_{F}^{(V)}(x,x)$ 
after integration by parts;
\item
$(\partial/\partial k_i)$ on a propagator
that is potentially on-shell changes a single pole
to a double pole, and the resulting integration over the double pole makes
the contribution to $\Delta\sigma(\vec{s}_T)$ 
proportional to $(\partial/\partial x) T_{F}^{(V)}(x,x)$;
\item
$(\partial/\partial k_i)$ on an off-shell propagator does not 
change the pole structure, and its contribution to 
$\Delta\sigma(\vec{s}_T)$ is proportional to $T_{F}^{(V)}(x,x)$ without 
a derivative;
\item
$(\partial/\partial k_i)$ on $k_i$-dependence in the numerator gives 
contributions to $\Delta\sigma(\vec{s}_T)$ proportional to 
$T_{F}^{(V)}(x,x)$ without derivatives.
\end{enumerate}

As we have pointed out earlier, we are interested in the asymmetries in
the forward region, where $x_F$ is large.  Asymmetries
in this region are dominated by large
net momentum fraction $x$ from the polarized beam parton,
coupled with relatively small momentum fraction $x'$ from the partons
of the unpolarized target hadron.  Since
all distributions vanish as a power
for large $x$, as $(1-x)^{\beta}$ with $\beta>0$, $(\partial/\partial 
x)T_F^{(V)}(x,x)
\gg T_F^{(V)}(x,x)$ when $x\rightarrow 1$.  Therefore, in the
forward region, terms proportional to derivative of the distributions 
$T_F^{(V)}$
dominate.  In order to simplify our calculations
of the largest effect, we keep only 
these terms. 
Thus, in Sec.\ \ref{sec:3} we will keep
only those contributions corresponding to 
items (1) and (2) listed above.

Turning, finally, to other possible contributions in Eq.\ (\ref{s3e11}), we 
observe that
it is only the matrix element $T_F^{(V)}$ that inherits derivative
terms, as a result of the collinear expansion involving soft gluon poles.
Soft fermion poles, of the sort discussed in Refs.\ \cite{AN:ET2,AN:QS}
have no such derivatives at leading order.  
Soft-fermion poles also do not correspond
to the valence quark approximation identified above, since they
require one of the quark fields to carry zero momentum fraction.
These features of the calculation follow exactly the same pattern
as for direct photon production, as treated in Ref.\ \cite{AN:QS},
and we shall not repeat them here.  It is only necessary to 
emphasize that the $T_F^{(V)}$ contributions from Fig.\ \ref{fig7}
are the complete set of derivative contributions at twist three and
leading order, for the first (chiral even) term in Eq.\ (\ref{s3e11}).

\subsection{Spin-Averaged Cross Sections for Hadronic Pion Production} 

In order to evaluate the asymmetries, defined in Eq.~(\ref{s2e5}), 
we need to compute the leading-order spin-averaged cross section.
QCD perturbation theory has been generally successful with experimental
data on spin-averaged cross sections for inclusive single-pion
production at large transverse momentum \cite{JFO:Rev}.  
At leading order in $\alpha_s$, 
only $2\rightarrow 2$ Feynman diagrams, shown in Fig.~\ref{fig4}, 
contribute to $E_{c}d^3\hat{\sigma}_{a+b\rightarrow c}/d^3p_c$.
In terms of scattering amplitudes, the leading order 
$E_{c}d^3\hat{\sigma}_{a+b\rightarrow c}/d^3p_c$ can be 
expressed as \cite{JFO:Rev}  
\begin{equation}
E_c \frac{d\hat{\sigma}_{a+b\rightarrow c}}{d^3p_c} =
\frac{1}{16\pi^2\hat{s}} 
\left| \overline{M}_{a+b\rightarrow c} \right|^2 \,
\delta\left(\hat{s}+\hat{t}+\hat{u}\right) \ ,
\label{s3e6}
\end{equation}
where $\overline{M}$ is the spin-averaged amplitude.  
In Eq.~(\ref{s3e6}), invariants at the parton level are 
given by
\begin{eqnarray}
\hat{s} &=& (xP+x'P')^2 = x\, x'\, S \ , \nonumber \\
\hat{t} &=& (xP-p_c)^2 = x\, T\, /z \ , 
\label{s3e7} \\
\hat{u} &=& (x'P'-p_c)^2 = x'\, U\, /z \ , \nonumber
\end{eqnarray}
where $S,T$ and $U$ are defined in Eq.~(\ref{s3e2}).

In the valence quark approximation, using the $\delta$-function 
in Eq.~(\ref{s3e6}) to fix the $x'$-integration in Eq.~(\ref{s3e5}), 
we find the spin-averaged cross section for pion production at leading 
order in $\alpha_s$,
\begin{eqnarray}
E_\ell\frac{d^3\sigma}{d^3\ell} 
&=& \frac{\alpha_s^2}{S}\, \sum_{a,c} 
    \int_{z_{\rm min}}^1 \frac{dz}{z^2}\, D_{c\rightarrow\pi}(z)
    \int_{x_{\rm min}}^1 \frac{dx}{x}\, \frac{1}{xS + U/z} 
    \int \frac{dx'}{x'}\, \delta\left(x'-\frac{-xT/z}{xS+U/z}\right)
\nonumber \\
&\times& q_a(x)\, \left[
    G(x')\, \hat{\sigma}_{ag\rightarrow c}\, 
  + \sum_q q(x')\, \hat{\sigma}_{aq\rightarrow c} \right] \ ,
\label{s3e8}
\end{eqnarray}
where $\sum_a$ runs over up and down valence quarks, and 
$\sum_q$ over quarks and antiquarks. 
In Eq.~(\ref{s3e8}), the integration limits $z_{\rm min}$ and 
$x_{\rm min}$, and variable $x$ are given by
\begin{eqnarray}
z_{\rm min} &=& \frac{-(T+U)}{S} = \sqrt{x_F^2 + x_T^2} \ ,
   \nonumber \\
x_{\rm min} &=& \frac{-U/z}{S+T/z} \ , 
\label{s3e9} \\
x' &=& \frac{-x\, T/z}{x\, S + U/z}\ , \nonumber 
\end{eqnarray}
where $S,T,U$ are defined in Eq.~(\ref{s3e2}), and $x_F$ and $x_T$ 
in Eq.~(\ref{s3e3}).  
The short-distance partonic parts,
$\hat{\sigma}_{ag\rightarrow c}$ and 
$\hat{\sigma}_{aq\rightarrow c}$, in Eq.~(\ref{s3e8}), are given by
\cite{JFO:Rev}
\begin{mathletters}
\begin{eqnarray}
\hat{\sigma}_{ag\rightarrow c} &=& \delta_{ac} \left[
     2 \left(1-\frac{\hat{s}\hat{u}}{\hat{t}^2}\right)
   + \frac{4}{9} \left( \frac{-\hat{u}}{\hat{s}} + 
                        \frac{\hat{s}}{-\hat{u}} \right)
   +             \left( \frac{\hat{s}}{\hat{t}} + 
                         \frac{\hat{u}}{\hat{t}} \right) \right]\, ;
\label{s3e10a} \\
\hat{\sigma}_{aq\rightarrow c} &=& 
    \delta_{ac}\, \frac{4}{9} 
      \left(\frac{\hat{s}^2+\hat{u}^2}{\hat{t}^2} \right)
  + \delta_{qc}\, \frac{4}{9} 
      \left(\frac{\hat{s}^2+\hat{t}^2}{\hat{u}^2} \right) 
        \nonumber \\
 && \quad  
  +  \delta_{aq}\delta_{qc}\, \frac{-8}{27} 
      \left(\frac{\hat{s}^2}{\hat{u}\hat{t}} \right) 
  +  \delta_{a\bar{q}}\, \frac{4}{9} 
      \left(\frac{\hat{t}^2+\hat{u}^2}{\hat{s}^2} \right) \, ,
\label{s3e10b}
\end{eqnarray}
\label{s3e10}
\end{mathletters}

\noindent
where $\hat{s}, \hat{t}, \hat{u}$ are defined in Eq.~(\ref{s3e7}).

Since we are interested in the large $x_F$ region, we have
$\ell^{+} \gg \ell^{-}$ and $T\ll U \leq S$.  Therefore, leading
contributions to the cross section given in Eq.~(\ref{s3e8}) come from
the $t$-channel diagrams (the first diagrams in Fig.~\ref{fig4}a 
and Fig.~\ref{fig4}b), or equivalently, the $1/\hat{t}^2$ term (i.e., 
first term) in Eq.~(\ref{s3e10}a) and Eq.~(\ref{s3e10}b).  
Consequently, for leading contributions in the forward region, 
incoming parton $a$ has the same flavor as fragmenting parton $c$.
Therefore, in the valence quark approximation,
we keep only $D_{u\rightarrow\pi^+}$  
for $\pi^+$ production; $D_{d\rightarrow\pi^-}$ for 
$\pi^-$ production, although we keep both $D_{u\rightarrow\pi^0}$ and 
$D_{d\rightarrow\pi^0}$ for $\pi^0$ production.



\section{Calculation of the Asymmetry}
\label{sec:3}

In this section, we present our calculation of the single transverse-spin
asymmetries in pion production in the valence quark-soft gluon approximation
described in the previous section.  
We derive analytic expressions for the
spin-dependent cross section, $\Delta\sigma(\ell,s_T)$, which is needed 
to evaluate the asymmetries.

\subsection{Quark-Gluon Subprocesses with Initial-State 
Interactions}

\begin{figure}
\epsfysize=4.2in
\epsfbox{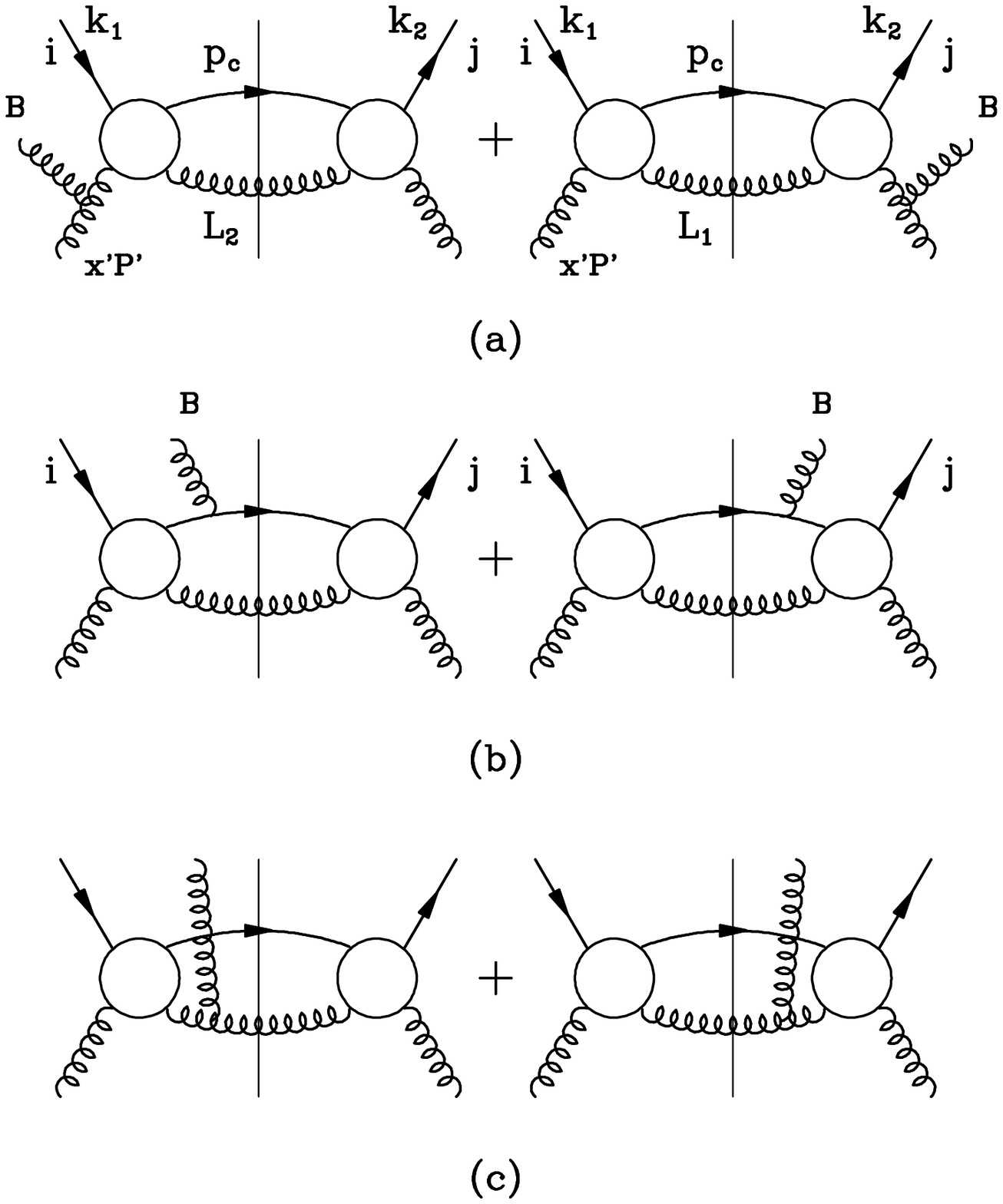}
\caption{Three classes of quark-gluon diagrams contributing to the
spin-dependent cross section $\Delta\sigma(\protect\vec{s}_T)$:
(a) diagrams with an initial-state 
pole, (b) and (c) diagrams with a final-state pole. 
Symbols $B$ and $ij$ are color indices for the gluon and quarks.}
\label{fig7}
\end{figure}

\begin{figure}
\epsfysize=2.2in
\epsfbox{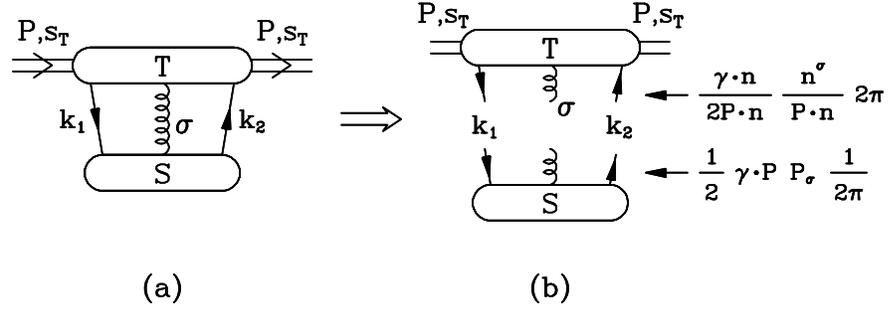}
\caption{General diagram that gives a leading contribution to 
$\Delta\sigma(\vec{s}_T)$: (a) before separation of spinor trace and 
Lorentz indices, (b) leading contribution after separation of spinor
trace and Lorentz indices.}
\label{fig8}
\end{figure}

\begin{figure}
\epsfysize=3.3in
\epsfbox{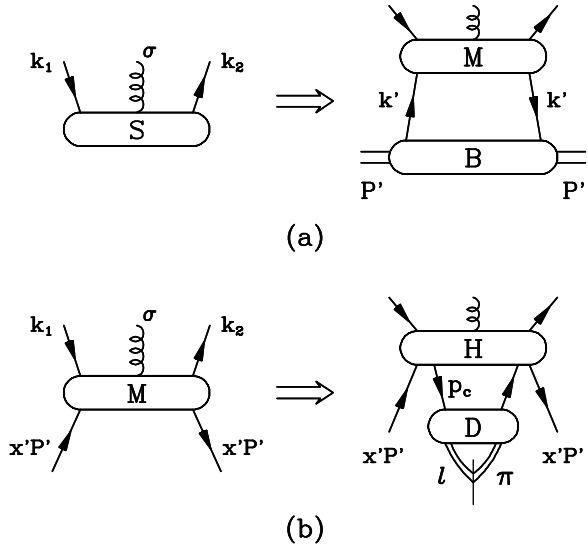}
\caption{Factorization of a general diagram contributing to 
$S_a(k_1,k_2)$ of Eq.~(\protect\ref{s3e16}): 
(a) separation of target hadron,
(b) separation of final-state pion.}
\label{fig9}
\end{figure}

\begin{figure}
\epsfysize=2.4in
\epsfbox{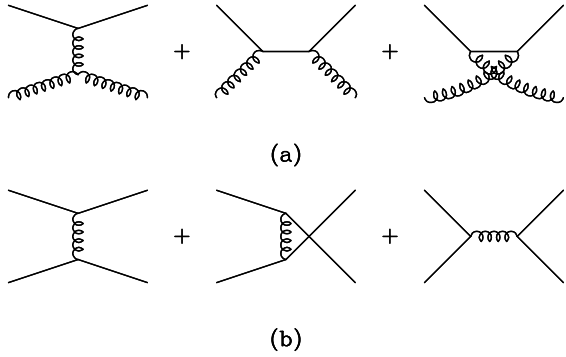}
\caption{Sample leading order Feynman diagrams contributing to the 
cross section of hadronic single pion production. }
\label{fig4}
\end{figure}

Consider the two diagrams with poles from initial-state interactions, 
as shown in Fig.~\ref{fig7}a.  We parameterize the parton momenta $k_i$ as
\begin{equation}
k_1 = x_1P + k_{1_T},  \quad\quad
\mbox{and} \quad\quad
k_2 = x_2P + k_{2_T}\, ,
\label{s3e23}
\end{equation}
with the $k_{i_T}$ two-dimensional transverse momenta.  The remaining
momentum components do not enter at twist three.
The pole in the diagram at the left of Fig.\ \ref{fig7}a is given 
in these terms by
\begin{equation}
\frac{1}{(x'P'+k_2-k_1)^2+i\epsilon}
\approx \frac{1}{(x_2-x_1)x'S+(k_{2_T}-k_{1_T})^2+i\epsilon} \ .
\label{s3e24}
\end{equation}
The derivative of this pole with respect to $k_{2_T}$ (or $k_{1_T}$) 
vanishes as $k_{i_T}\rightarrow 0$.  The diagram on the 
right has the same feature.  Therefore, 
following the arguments of Sec.\ \ref{sec:2d} above, the
leading contribution to 
$\Delta\sigma(\vec{s}_T)$ in the diagrams in 
Fig.~\ref{fig7}a is from the 
derivative of the phase space $\delta$-function only.

\begin{figure}
\epsfysize=1.5in
\epsfbox{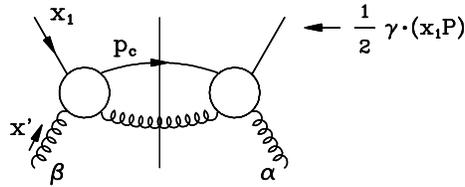}
\caption{Two-parton forward scattering amplitude contributing to 
the partonic hard part $H_{2\rightarrow 2}(x,x',p_c)$ in 
Eq.~(\protect\ref{s3e28n}).}
\label{fig10}
\end{figure}

Let $L_1$ and $L_2$ be  the momenta of the unobserved partons in the diagrams 
to the left and right, respectively in Fig.\ \ref{fig7}a.  We have 
\begin{equation}
L_1 \equiv x'P'+x_1P+k_{1_T}-p_c, \quad\quad
\quad\quad\quad\quad
L_2 \equiv x'P'+x_2P+k_{2_T}-p_c\ .
\label{s3e25}
\end{equation}
Taking the derivative with respect to $k_{1}^{\rho}$ and 
$k_{2}^{\rho}$, we obtain
\begin{mathletters}
\label{s3e26}
\begin{eqnarray}
\frac{\partial}{\partial k_1^{\rho}}\, \delta(L_1^2) 
&=& (-2p_{c_{\rho}}) \delta'(L_1^2) \ ,
\label{s3e26a} \\
\frac{\partial}{\partial k_2^{\rho}}\, \delta(L_1^2) 
&=& 0 \ ,
\label{s3e26b} \\
\frac{\partial}{\partial k_1^{\rho}}\, \delta(L_2^2) 
&=& 0 \ ,
\label{s3e26c} \\
\frac{\partial}{\partial k_2^{\rho}}\, \delta(L_2^2) 
&=& (-2p_{c_{\rho}}) \delta'(L_2^2) \ .
\label{s3e26d}
\end{eqnarray}
\end{mathletters}
In deriving these relations, we have used that 
$\rho$ is a transverse index.  After taking the derivative with respect to
the $k_i$ on the $\delta$-functions, we can set $k_{i_T}$ to
zero in the remainder of each diagram.  For the diagrams in Fig.~\ref{fig7}a, 
the poles giving the leading contributions are from 
\begin{mathletters}
\label{s3e26n}
\begin{eqnarray}
L(x_1,x_2)&\equiv &
g_s\,\left[(2x'P'\cdot P)\, g_{\rho\beta}
          -(x'P')_{\rho}P_{\beta}
          -(x'P')_{\beta}P_{\rho}\right]\,
     \frac{-i}{[x'P'+(x_2-x_1)P]^2+i\epsilon} \nonumber \\
&=& g_s\, (-i)\, g_{\rho\beta}\, \left(
     \frac{-1}{x_1-x_2-i\epsilon}\right)\ ,
\label{s3e26na} \\
R(x_1,x_2)
&=& g_s\, (-i)\, g_{\rho\beta}\, \left(
     \frac{-1}{x_2-x_1+i\epsilon}\right)\ ,
\label{s3e26nb}
\end{eqnarray}
\end{mathletters}
where $L$ and $R$ represent the diagrams at left and at right, 
respectively.  In Eq.~(\ref{s3e26n}), 
$g_s=\sqrt{4\pi\alpha_s}$ is the strong coupling.  In 
the following discussion, we absorb the overall 
$(-i)$ in Eq.~(\ref{s3e26n}) into the color factor
for the subprocess.  Using the distribution identity
\begin{equation}
\frac{1}{x_2-x_1\pm i\epsilon} = 
P\left[\frac{1}{x_2-x_1}\right] \mp i\pi\delta(x_2-x_1)\ ,
\label{s3e27}
\end{equation}
for the poles in Eq.~(\ref{s3e26n}) 
and keeping the imaginary contribution of 
the pole, we can express the contributions of the diagrams in Fig.~\ref{fig7}a
as
\begin{mathletters}
\label{s3e28}
\begin{eqnarray}
\frac{\partial}{\partial k_2^{\rho}} &&\left(
      H_{a_L}(x_1,x_2,x',p_c) + H_{a_R}(x_1,x_2,x',p_c) \right)
 \nonumber \\
&&= \frac{g_s}{2\pi x_2} H_{2\rightarrow 2}(x_2,x',p_c) 
    \left[i\pi\delta(x_1-x_2)(2p_{c_{\rho}})\right]
    \delta'(L_2^2) \ ,
\label{s3e28a} \\
-\frac{\partial}{\partial k_1^{\rho}} && \left(
      H_{a_L}(x_1,x_2,x',p_c) + H_{a_R}(x_1,x_2,x',p_c) \right)
 \nonumber \\
&&= \frac{g_s}{2\pi x_1} H_{2\rightarrow 2}(x_1,x',p_c) 
    \left[i\pi\delta(x_2-x_1)(2p_{c_{\rho}})\right]
    \delta'(L_1^2) \ ,
\label{s3e28b}
\end{eqnarray}
\end{mathletters}
where subscripts $a_L$ and $a_R$ represent the left and right diagrams 
of Fig.~\ref{fig7}a.  In Eq.~(\ref{s3e28}), 
$H_{2\rightarrow 2}(x_i,x',p_c)$ with $i=1,2$ 
is proportional to the imaginary part of the
$2\rightarrow 2$ partonic forward scattering amplitude 
shown in Fig.~\ref{fig10},
\begin{equation}
H_{2\rightarrow 2}(x_i,x',p_c) = \frac{1}{16\pi^2}\,
\left|\overline{M}^I_{a+g\rightarrow c}\right|^2\, C^I_g\ ,
\label{s3e28n}
\end{equation}
where the matrix element squared, 
$\left| \overline{M}^I_{a+g\rightarrow c} \right|^2$, 
is the same as that in Eq.~(\ref{s3e6}), except for the color 
factor, $C^I_g$, due to the extra initial-state interaction.  Combining
Eqs.~(\ref{s3e12n}) and (\ref{s3e26n}), the factor $C^I_g$ is given by 
the color structure of the partonic diagrams shown in Fig.~\ref{fig7}a,
contracted with a common factor $[(-i)2/(N^2-1)]\,(t^B)_{ij}$, where
$B$ and $ij$ are color indices for the gluon and quarks 
from the polarized hadron.  
The factor $1/2\pi$ in Eq.~(\ref{s3e28}) was explained in the
text following Eq.~(\ref{s3e12}), and the factor $1/x_1$ is due to 
the definition of $H_{2\rightarrow 2}(x_1,x',p_c)$, where incoming 
quark lines are contracted with $(1/2)\gamma\cdot (x_1P)$.  
Eq.~(\ref{s3e28}) shows that Eq.~(\ref{s3e22}) is satisfied 
when $k_{i_T} = 0$.

Substituting Eq.~(\ref{s3e28}a) into Eq.~(\ref{s3e21}), we have a complete 
factorized form for the spin-dependent cross section from the diagrams shown 
in Fig.~\ref{fig7}a,
\begin{equation}
E_\ell \frac{d\Delta\sigma^I_{g}(\vec{s}_T)}{d^3\ell} 
= \sum_{a,c} 
\int \frac{dz}{z^2}\, D_{c\rightarrow\pi}(z)
\int dx'\, G(x')
\int dx\, T_{F_a}^{(V)}(x,x) 
\left(E_c\frac{d\Delta\hat{\sigma}^I_{a+g\rightarrow c}(\vec{s}_T)}
{d^3p_c}\right)\ ,
\label{s3e30}
\end{equation}
where the factor $1/z^2$ is due to the phase space difference 
between $d^3\ell/(2\pi)^3 2E_{\ell}$ and $d^3p_c/(2\pi)^3 2E_{c}$, 
and the partonic hard part, 
$E_c d\Delta\hat{\sigma}^I_{a+g\rightarrow c}/d^3p_c$, is given by
\begin{equation}
E_c\frac{d\Delta\hat{\sigma}^I_{a+g\rightarrow c}(\vec{s}_T)}{d^3p_c}
= g_s\, \epsilon^{s_T p_c n \bar{n}}\, C^I_g\, \left[
   \frac{1}{16\pi^2\hat{s}} 
   \left| \overline{M}^I_{a+g\rightarrow c} \right|^2
   \delta'(\hat{s}+\hat{t}+\hat{u}) \right] \ .
\label{s3e31}
\end{equation}
In Eqs.~(\ref{s3e30}) and (\ref{s3e31}), superscript $I$  
indicates the contribution from a partonic subprocess with an initial-state 
pole,
and subscript $g$ represents the quark-gluon subprocess.  
In deriving Eq.~(\ref{s3e30}), we renamed the integration variable 
$x_1$ in Eq.~(\ref{s3e28}) as $x$.  The factorized 
spin-dependent cross section given in Eq.~(\ref{s3e30})
is very similar to the factorized form for the spin-averaged cross section 
in Eq.~(\ref{s3e5}), with the unpolarized parton distribution $\phi_{a/A}(x)$
replaced by the twist-three correlation function $T^{(V)}_F(x,x)$.  The partonic
hard part in Eq.~(\ref{s3e31}) is also very similar to that in 
Eq.~(\ref{s3e6}).  For the spin-dependent case, the derivative of
the $\delta$-function is just the derivative 
with respect to the  parton momentum $k_i$
in Eq.\ (\ref{s3e13}),
which comes from the collinear expansion. The factor
$\epsilon^{s_T p_c n \bar{n}}$ in Eq.~(\ref{s3e31})
is necessary for a nonvanishing asymmetry.  

After partial integration over $x$, we can reexpress the derivative of the
$\delta$-function as
\begin{equation}
\int dx\, \delta'(\hat{s}+\hat{t}+\hat{u}) F(x)
= \int \frac{dx}{x'S+T/z} \delta(\hat{s}+\hat{t}+\hat{u}) 
  \left[ -\frac{\partial}{\partial x}F(x) \right] 
\label{s3e32}
\end{equation}
for any smooth function $F(x)$.  Using Eq.~(\ref{s3e31}), we thus rewrite
$E_\ell d\Delta\sigma^I_{g}(s_T)/d^3\ell$ as
\begin{eqnarray}
E_\ell \frac{d\Delta\sigma^I_{g}(\vec{s}_T)}{d^3\ell} 
&=& \frac{\alpha_s^2}{S} \sum_{a,c} 
\int_{z_{\rm min}}^1 \frac{dz}{z^3}\, D_{c\rightarrow\pi}(z)
\int_{x_{\rm min}}^1 \frac{dx}{x}\, \frac{1}{xS+U/z}
\int \frac{dx'}{x'}\, \delta\left(x'-\frac{-xT/z}{xS+U/z}\right) 
\nonumber\\
& \times & g_s\, 
   \epsilon^{s_T \ell n \bar{n}}\, \left(\frac{1}{x'S+T/z}\right)\,
   G(x')\, \left[-x\frac{\partial}{\partial x}
                \left(\frac{T^{(V)}_{F_a}(x,x)}{x}\,
                      H^I_{ag\rightarrow c}(\hat{s},\hat{t},\hat{u}) 
                \right) \right]\, ,
\label{s3e33}
\end{eqnarray} 
where $z_{\rm min}$ and $x_{\rm min}$ 
are given in Eq.~(\ref{s3e9}),
and $S,T$ and $U$ are defined in Eq.~(\ref{s3e2}).
In Eq.~(\ref{s3e33}), the spin-dependent cross section 
$E_\ell d\Delta\sigma^I_{g}(s_T)/d^3\ell$ has almost the same factorized 
form as the spin-averaged cross section shown in Eq.~(\ref{s3e8}).  The
extra factor of $1/z$ is due to the replacement of $p_c$ by $\ell$ in 
the $\epsilon$-tensor of Eq.~(\ref{s3e31}).  
The dimension of $1/(x'S+T/z)$ due to 
the derivative of the $\delta$-function is balanced by the dimension of
$\ell$ in the $\epsilon$-tensor and the dimension of the twist-three 
correlation function $T^{(V)}_{F_a}(x,x)$.  In our definition, the
twist-three correlation function has the dimensions of energy.  
The partonic hard part,
$H^I_{ag\rightarrow c}(\hat{s},\hat{t},\hat{u})$, 
in Eq.~(\ref{s3e33}) plays the role 
of $\hat{\sigma}_{ag\rightarrow c}$ in Eq.~(\ref{s3e8}). It
is given by $C^I_g\, |\overline{M}^I_{a+g\rightarrow c}|^2$ in 
Eq.~(\ref{s3e31}), 
which represents the $2\rightarrow 2$ matrix element squared  
in Eq.~(\ref{s3e28n}), but with a different color factor due to the extra 
initial-state interaction.

\subsection{Quark-Gluon Subprocesses with Final-State 
Interactions}

\begin{figure}
\epsfysize=1.6in
\epsfbox{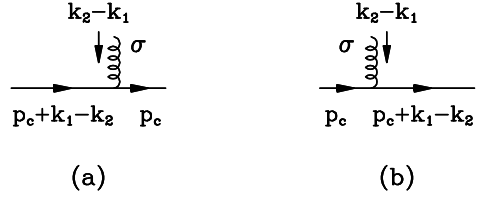}
\caption{Sketch for the effective diagrams giving the leading poles 
in Eq.~(\protect\ref{s3e33n}): (a) pole to the left of the cut;
(b) pole to the right of the cut.}
\label{fig11}
\end{figure}

The diagrams shown in Fig.~\ref{fig7}b represent
final-state interactions of the fragmenting parton.  
As with the contributions from initial-state interactions, these diagrams
also have a derivative
with respect to $k_{1}^{\rho}$ and $k_{2}^{\rho}$
of the phase space $\delta$-function associated with 
the unobserved final-state parton, of momentum $L_1$ or $L_2$.  
Similarly to Eq.~(\ref{s3e26n}), the final-state poles giving leading
contributions are given by, as sketched in Fig.~\ref{fig11}, 
\begin{mathletters}
\label{s3e33n}
\begin{eqnarray}
L(x_1,x_2)
&\equiv & g_s\,(\gamma\cdot p_c)\, 
         \frac{\gamma\cdot P\,\gamma\cdot(p_c+(x_1-x_2)P)}
        {(p_c+(x_1-x_2)P)^2+i\epsilon} \nonumber \\
&\approx& g_s\,(\gamma\cdot p_c)\, 
         \left(\frac{1}{x_1-x_2+i\epsilon}\right)\ ,
\label{s3e33na} \\
R(x_1,x_2)
&\approx& g_s\,(\gamma\cdot p_c)\, 
         \left(\frac{1}{x_2-x_1-i\epsilon}\right)\ ,
\label{s3e33nb}
\end{eqnarray}
\end{mathletters}
where the factor $(\gamma\cdot p_c)$ will be absorbed into the $2\rightarrow 2$
hard-scattering function.  Similarly to Eq.~(\ref{s3e33}), 
we obtain the contribution from the derivative of the $\delta$-function
for a final-state interaction,
\begin{eqnarray}
E_\ell \frac{d\Delta\sigma^F_{g}(\vec{s}_T)}{d^3\ell} 
&=& \frac{\alpha_s^2}{S} \sum_{a,c} 
\int_{z_{\rm min}}^1 \frac{dz}{z^3}\, D_{c\rightarrow\pi}(z)
\int_{x_{\rm min}}^1 \frac{dx}{x}\, \frac{1}{xS+U/z}
\int \frac{dx'}{x'}\, \delta\left(x'-\frac{-xT/z}{xS+U/z}\right) 
\nonumber\\
& \times & g_s\,
   \epsilon^{s_T \ell n \bar{n}}\, \left(\frac{1}{x'S+T/z}\right)\,
   G(x')\, \left[-x\frac{\partial}{\partial x}
                \left(\frac{T^{(V)}_{F_a}(x,x)}{x}\,
                      H^F_{ag\rightarrow c}(\hat{s},\hat{t},\hat{u}) 
                \right) \right]\, ,
\label{s3e34}
\end{eqnarray} 
where superscript $F$ denotes the final-state interaction.
The only difference between 
$E_\ell d\Delta\sigma^F_{g}(s_T)/d^3\ell$ in Eq.~(\ref{s3e34}) and
$E_\ell d\Delta\sigma^I_{g}(s_T)/d^3\ell$ in Eq.~(\ref{s3e33}) 
is the color factors in the partonic hard parts.  The hard
part $H^F_{ag\rightarrow c}(\hat{s},\hat{t},\hat{u})$ 
in Eq.~(\ref{s3e34}) is given by 
$C^F_g\, |\overline{M}^I_{a+g\rightarrow c}|^2$, which
has the same kinematic dependence as 
$H^I_{ag\rightarrow c}(\hat{s},\hat{t},\hat{u})$ in Eq.~(\ref{s3e33}), 
but a different color factor, $C^F_g$, due to different color structures
in final-state compared to initial-state interactions.  Similarly to
$C^I_g$, $C^F_g$ is computed by contracting the matrix 
$[2/(N^2-1)]\,(t^B)_{ij}$, Eq.~(\ref{s3e12n}), into the diagrams.

\begin{figure}
\epsfysize=1.6in
\epsfbox{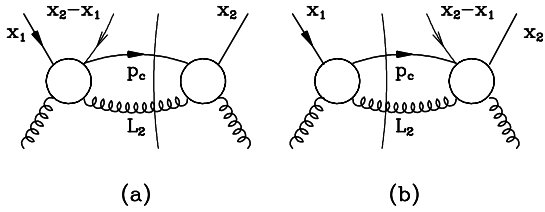}
\caption{Effective quark-gluon $2\rightarrow 2$ diagrams with the 
thin line of momentum $(x_2-x_1)P$ representing momentum flow 
that is a result of the extra final-state interaction.}
\label{fig12}
\end{figure}

In addition to the contribution from the derivative of the $\delta$-function, 
the diagrams shown in Fig.~\ref{fig7}b also give leading contributions, 
proportional to $(\partial/\partial x)T^{(V)}_F(x,x)$, from 
the double pole which results when the derivative $(\partial/\partial k_i)$ acts 
on a propagator that goes on-shell at $x_1=x_2$.
Consider the final-state interaction in the diagram at the left 
in Fig.~\ref{fig11}a. The pole giving the leading contribution is from the 
factor
\begin{eqnarray}
L(k_{1_T},k_{2_T}) &\equiv & g_s\,
\gamma\cdot p_c\, \left[
  \frac{\gamma\cdot P\, \gamma\cdot(p_c+k_1-k_2)}
       {(p_c+k_1-k_2)^2+i\epsilon} \right]
\nonumber \\
&=& g_s\, \gamma\cdot p_c\, \left[
  \frac{1}{x_1-x_2+x_0(k_{1_T},k_{2_T})+i\epsilon} \right.
\nonumber \\
&& {\hskip 0.4in} \left.
- \frac{\gamma\cdot(k_{1_T}-k_{2_T})\, \gamma\cdot P}
       {2P\cdot p_c\,[x_1-x_2+x_0(k_{1_T},k_{2_T})+i\epsilon]}
  \right]\ ,
\label{s3e35}
\end{eqnarray}
where $x_0$ is defined as
\begin{eqnarray}
x_0(k_{1_T},k_{2_T}) &\equiv &
\frac{2(k_{1_T}-k_{2_T})\cdot p_c + (k_{1_T}-k_{2_T})^2}{2P\cdot p_c}
\nonumber \\
& \rightarrow & 0 \quad 
\mbox{as $k_{1_T}$ and $k_{2_T} \rightarrow 0$}\ .
\label{s3e36}
\end{eqnarray}
In deriving Eq.~(\ref{s3e35}), we used the parameterization of 
Eq.~(\ref{s3e23}),  
and the relations $p_c^2=0$, $P^2\approx 0$, and $2P\cdot p_c > 0$.
Applying $(\partial/\partial k_{i}^{\rho})$ to
$L(k_{1_T},k_{2_T})$, and letting $k_{i_T}$ ($i=1,2$) go to zero, 
the first term in Eq.~(\ref{s3e35}) develops a double pole, 
while the second term remains a single pole,
\begin{eqnarray}
\frac{\partial}{\partial k_{1}^{\rho}} L(k_{1_T}=0,k_{2_T}=0) 
&=& \gamma\cdot p_c \left(\frac{g_s}{2P\cdot p_c}\right) \left[
    (-2p_{c_\rho})\,\frac{1}{(x_1-x_2+i\epsilon)^2} \right.
\nonumber \\
&& {\hskip 1.1in} \left.
   -(\gamma_{\rho}\,\gamma\cdot P)\,\frac{1}{(x_1-x_2+i\epsilon)}
  \right]
\nonumber \\
&=& -\frac{\partial}{\partial k_{2}^{\rho}} L(k_{1_T}=0,k_{2_T}=0)\ .
\label{s3e37}
\end{eqnarray}
Since we keep only contributions proportional to 
$(\partial/\partial x)T^{(V)}_F(x,x)$, we neglect
the single-pole term in Eq.~(\ref{s3e37}) in the following discussion, and
use
\begin{eqnarray}
\frac{\partial}{\partial k_{2}^{\rho}} L(k_{1_T}=0,k_{2_T}=0) 
&=& - \frac{\partial}{\partial k_{1}^{\rho}} L(k_{1_T}=0,k_{2_T}=0) 
\nonumber \\
&\approx & \gamma\cdot p_c\, \left(\frac{g_s}{2P\cdot p_c}\right) 
\left[ (2p_{c_\rho})\, \frac{1}{(x_1-x_2+i\epsilon)^2} \right]\, .
\label{s3e38}
\end{eqnarray}
Similarly, for the diagram at the right in Fig.~\ref{fig11}b, we have
\begin{eqnarray}
R(k_{1_T},k_{2_T}) &\equiv & g_s\, \left[
  \frac{\gamma\cdot(p_c+k_2-k_1)\, \gamma\cdot P} 
       {(p_c+k_2-k_1)^2-i\epsilon} \right]\, \gamma\cdot p_c 
\nonumber \\
&=& g_s\, \left[
  \frac{1}{x_2-x_1+x_0(k_{2_T},k_{1_T})-i\epsilon} \right.
\nonumber \\
&& \quad \left.
- \frac{\gamma\cdot P\, \gamma\cdot(k_{2_T}-k_{1_T})} 
       {2P\cdot p_c\,[x_2-x_1+x_0(k_{2_T},k_{1_T})-i\epsilon]}
  \right]\, \gamma\cdot p_c\, ,
\label{s3e39}
\end{eqnarray}
where $x_0$ is defined in Eq.~(\ref{s3e36}).  Taking the derivative 
with respect to $k_{i}^{\rho}$, we have
\begin{eqnarray}
\frac{\partial}{\partial k_{2}^{\rho}} R(k_{1_T}=0,k_{2_T}=0) 
&=& - \frac{\partial}{\partial k_{1}^{\rho}} R(k_{1_T}=0,k_{2_T}=0) 
\nonumber \\
&\approx& \gamma\cdot p_c\, \left(\frac{g_s}{2P\cdot p_c}\right) 
\left[(-2p_{c_\rho})\,\frac{1}{(x_2-x_1-i\epsilon)^2}\right] \ .
\label{s3e40}
\end{eqnarray}
Eqs.~(\ref{s3e38}) and (\ref{s3e40}) show that 
the double-pole contributions from 
diagrams in Fig.~\ref{fig7}b satisfy Eq.~(\ref{s3e22}).  
Keeping only these double-pole terms, as in 
Eq.~(\ref{s3e28a}), we now have
\begin{mathletters}
\label{s3e41} 
\begin{eqnarray}
\frac{\partial H_{D_L}}{\partial k_2^{\rho}}(x_1,x_2,x',p_c)
&\approx & \frac{g_s}{2\pi}\,H^L_{2\rightarrow 2}(x_1,x_2,x',p_c)
\left[(2p_{c_\rho})\,\frac{1}{(x_1-x_2+i\epsilon)^2}\right]
\left(\frac{1}{2P\cdot p_c}\right)\ ,
\label{s3e41a} \\
\frac{\partial H_{D_R}}{\partial k_2^{\rho}}(x_1,x_2,x',p_c)
&\approx & \frac{g_s}{2\pi}\,H^R_{2\rightarrow 2}(x_1,x_2,x',p_c)
\left[(-2p_{c_\rho})\frac{1}{(x_2-x_1-i\epsilon)^2}\right]
\left(\frac{1}{2P\cdot p_c}\right)\ ,
\label{s3e41b}
\end{eqnarray}
\end{mathletters}

\noindent
where $D_L$ and $D_R$ denote the double-pole contributions from 
the left and right diagram in Fig.~\ref{fig7}b, respectively.  In
Eq.~(\ref{s3e41}), $H^L_{2\rightarrow 2}(x_1,x_2,x',p_c)$ and 
$H^R_{2\rightarrow 2}(x_1,x_2,x',p_c)$ are $2\rightarrow 2$ partonic
parts corresponding to the left and right diagrams shown 
in Fig.~\ref{fig12}.  They have the limits
\begin{mathletters}
\label{s3e42}
\begin{eqnarray}
H^L_{2\rightarrow 2}(x_1,x_2,x',p_c)_{x_1\rightarrow x_2} 
& = & \frac{1}{x_2}\, H_{2\rightarrow 2}(x_2,x',p_c) \ ,
\label{s3e42a} \\
H^R_{2\rightarrow 2}(x_1,x_2,x',p_c)_{x_2\rightarrow x_1} 
& = & \frac{1}{x_1}\, H_{2\rightarrow 2}(x_1,x',p_c) \ ,
\label{s3e42b}
\end{eqnarray}
\end{mathletters}

\noindent
where $H_{2\rightarrow 2}(x_i,x',p_c)$ with $i=1,2$ are the same as 
in Eq.~(\ref{s3e28}).

Recalling that $T_F$ is real, it is evident from Eq.~(\ref{s3e21})
that we need the imaginary part of 
$(\partial/\partial k_2^{\rho}) H_{D_L}$ and
$(\partial/\partial k_2^{\rho}) H_{D_R}$ in order to get
a real contribution to the spin-dependent cross section.  
For double pole terms like those in Eq.\ (\ref{s3e41a}) and (\ref{s3e41b}),
the imaginary part is given by
\begin{equation}
\int dx_1\, \frac{1}{(x_1-x_2+i\epsilon)^2}\, F(x_1,x_2)
=\int dx_1\, \left[-i\pi\,\delta(x_1-x_2)\right]\,
\left[ \frac{\partial}{\partial x_1}\, F(x_1,x_2) \right]
\label{s3e43}
\end{equation}
for any smooth function $F(x_1,x_2)$.  Using Eq.~(\ref{s3e42}), 
we have following relation,
\begin{eqnarray}
\int dx_1\, dx_2\, && T^{(V)}_{F_a}(x_1,x_2)\left[
i\epsilon^{\rho s_T n \bar{n}}\, 
\frac{\partial}{\partial k_2^{\rho}}\left( 
H_{D_L}(x_1,x_2,x',p_c) + H_{D_R}(x_1,x_2,x',p_c) \right) \right]
\nonumber \\
&&= g_s\, \frac{\epsilon^{s_T p_c n\bar{n}}}{2P\cdot p_c}\left\{
    \int dx_2\,\left[-\frac{\partial}{\partial x_1}\left(
      H^L_{2\rightarrow 2}(x_1,x_2,x',p_c)\, T^{(V)}_{F_a}(x_1,x_2)
      \right) \right]_{x_1=x_2} \right.
\nonumber \\
&& {\hskip 0.8in} \left.   
   +\int dx_1\,\left[-\frac{\partial}{\partial x_2}\left(
      H^R_{2\rightarrow 2}(x_1,x_2,x',p_c)\, T^{(V)}_{F_a}(x_1,x_2)
      \right) \right]_{x_2=x_1} \right\}
\nonumber \\
&&\approx g_s\, \frac{\epsilon^{s_T\ell n\bar{n}}}{2P\cdot\ell}\,
    \int \frac{dx}{x}\, H_{2\rightarrow 2}(x,x',p_c)\,
    \left[ -\frac{\partial}{\partial x}\left(T^{(V)}_{F_a}(x,x)\right)
    \right]\ .
\label{s3e44}
\end{eqnarray}
In deriving Eq.~(\ref{s3e44}), we have used the symmetry property
$T^{(V)}_{F}(x_1,x_2) = T^{(V)}_{F}(x_2,x_1)$, Eq.\ (\ref{TFiseven}), 
\cite{AN:QS}
\begin{mathletters}
\label{s3e45}
\begin{equation}
\frac{\partial}{\partial x}\left(T^{(V)}_{F_a}(x,x)\right)
=2\,
 \left[\frac{\partial}{\partial x_1}\left(T^{(V)}_{F_a}(x_1,x)\right)
 \right]_{x_1=x}
=2\,
 \left[\frac{\partial}{\partial x_2}\left(T^{(V)}_{F_a}(x,x_2)\right)
 \right]_{x_2=x}\ ,
\label{s3e45b}
\end{equation}
\end{mathletters}
and Eq.~(\ref{s3e42}).  In addition, we have used the approximation
\begin{eqnarray}
&& \left[ -\frac{\partial}{\partial x_1} \left(
    H^L_{2\rightarrow 2}(x_1,x_2,x',p_c)\, T^{(V)}_{F_a}(x_1,x_2)
    \right) \right]_{x_1=x_2} \nonumber\\
&&\quad\quad \approx \frac{1}{x_2}\,H_{2\rightarrow 2}(x_2,x',p_c)
 \left[-\frac{\partial}{\partial x_1}
        \left(T^{(V)}_{F_a}(x_1,x_2)\right)
 \right]_{x_1=x_2}\ ,
\label{s3e46}
\end{eqnarray}
demanding as usual a derivative of $T_F^{(V)}$.

Substituting Eq.~(\ref{s3e44})
into the cross section Eq.~(\ref{s3e21}), we obtain the leading double-pole 
contributions from the diagrams shown in Fig.~\ref{fig7}b
\begin{eqnarray}
E_\ell \frac{d\Delta\sigma^D_{g}(\vec{s}_T)}{d^3\ell} 
&=& \frac{\alpha_s^2}{S} \sum_{a,c} 
\int_{z_{\rm min}}^1 \frac{dz}{z^2}\, D_{c\rightarrow\pi}(z)
\int_{x_{\rm min}}^1 \frac{dx}{x}\, \frac{1}{xS+U/z}
\int \frac{dx'}{x'}\, \delta\left(x'-\frac{-xT/z}{xS+U/z}\right) 
\nonumber\\
& \times & g_s\,
   \epsilon^{s_T \ell n \bar{n}}\, \left(\frac{1}{-T}\right)\,
   G(x')\, \left[-\frac{\partial}{\partial x}
                  \left(T^{(V)}_{F_a}(x,x)\right)\right]\,
                   H^D_{ag\rightarrow c}(\hat{s},\hat{t},\hat{u})\, ,
\label{s3e47}
\end{eqnarray} 
where $T=-2P\cdot\ell$\ is defined in Eq.~(\ref{s3e2}), 
and the partonic hard part 
$H^D_{ag\rightarrow c}(\hat{s},\hat{t},\hat{u})$ is normalized to have
\begin{equation}
H^D_{ag\rightarrow c}(\hat{s},\hat{t},\hat{u}) 
= H^F_{ag\rightarrow c}(\hat{s},\hat{t},\hat{u}) \, ,
\label{s3e48}
\end{equation}
with $H^F_{ag\rightarrow c}(\hat{s},\hat{t},\hat{u})$ 
 the same partonic hard part
derived from the contribution of the derivative of the
$\delta$-function, and given in Eq.~(\ref{s3e34}).

In addition to the diagrams in Fig.~\ref{fig7}b, there is
another type of diagram with final-state interactions, as shown in 
Fig.~\ref{fig7}c.  In this case the final-state interactions taken place on 
an unobserved final-state parton.  The diagrams on the left and 
right are the same, except for the final-state propagator and the 
argument of the phase space $\delta$-function.
The total partonic contribution from these two diagrams 
can be expressed as
\begin{eqnarray}
&& H_{c_L}(k_1,k_2,x',p_c) + H_{c_R}(k_1,k_2,x',p_c)
\nonumber\\
&& \quad 
= \left[\frac{1}{L_1^2+i\epsilon}\, \delta(L_2^2)
       +\frac{1}{L_2^2-i\epsilon}\, \delta(L_1^2)\right]\,
  F(k_1,k_2,x',p_c)\, ,
\label{s3e49}
\end{eqnarray}
where the momenta $L_1$ and $L_2$ are defined in Eq.~(\ref{s3e25}).
The function $F(k_1,k_2,x',p_c)$ represents the common factor 
of two diagrams in Fig.~\ref{fig7}c; it has the symmetry property
\begin{equation}
F(k_1,k_2,x',p_c) = F(k_2,k_1,x',p_c)\, .
\label{s3e50}
\end{equation}
>From Eq.~(\ref{s3e49}), combining the symmetry properties of 
Eqs.~(\ref{TFiseven}) and (\ref{s3e50}), we 
readily show that the leading contribution of 
the diagrams in Fig.~\ref{fig7}c to the spin-dependent cross section 
(or Eq.~(\ref{s3e21})) vanishes.

\subsection{Quark-Quark and Quark-Antiquark Subprocesses}

In this subsection, we present
the leading contributions to the spin-dependent cross section 
from quark-quark and quark-antiquark subprocesses. 

\begin{figure}
\epsfysize=3.8in
\epsfbox{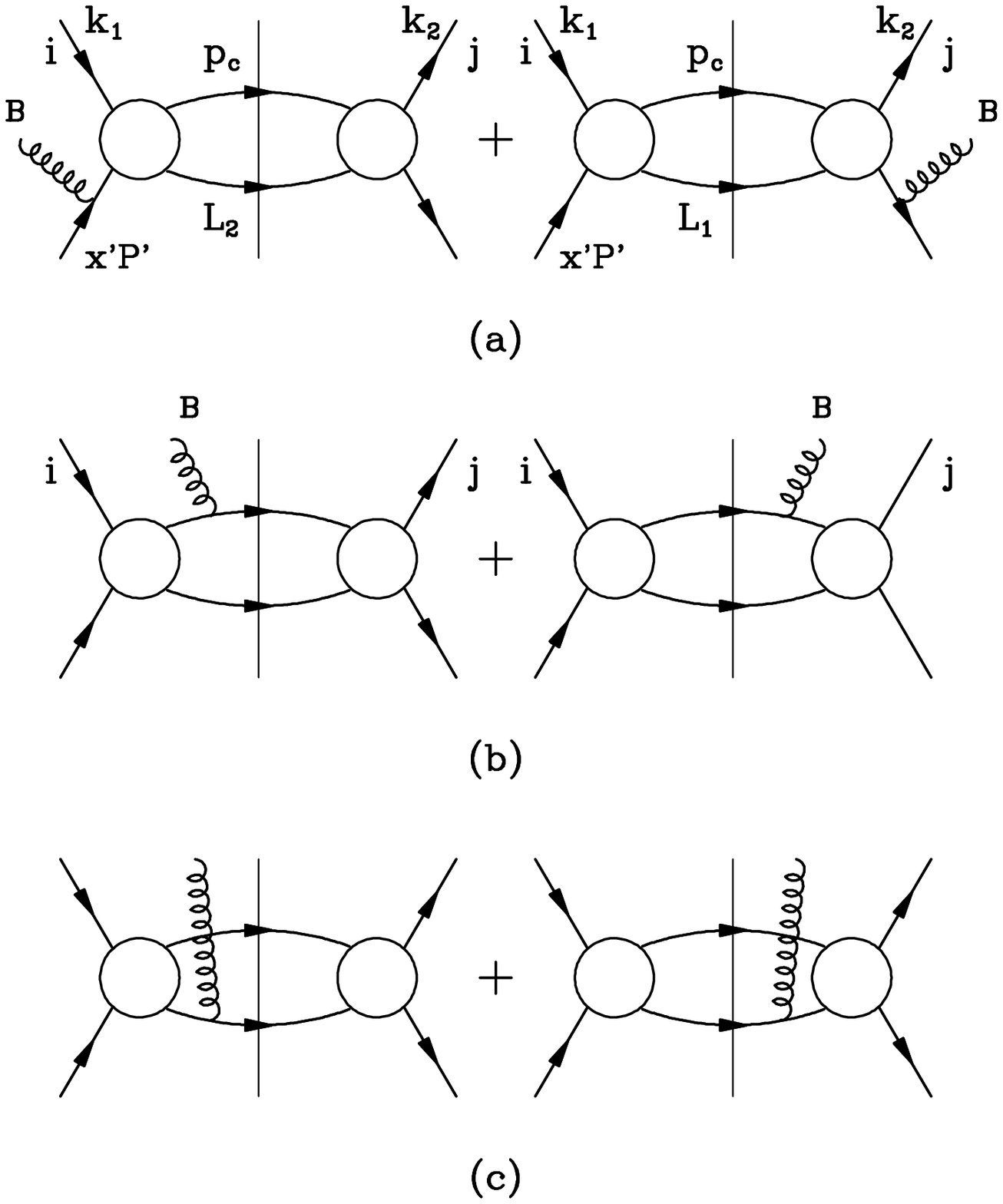}
\caption{Three classes of quark-quark (or antiquark) diagrams 
contributing to the spin-dependent cross section 
$\Delta\sigma(\protect\vec{s}_T)$:
(a) diagrams with an initial-state 
pole, (b) and (c) diagrams with a final-state pole. 
Symbols $B$ and $ij$ are color indices for the gluon and quarks.}
\label{fig13}
\end{figure}

\begin{figure}
\epsfysize=1.4in
\epsfbox{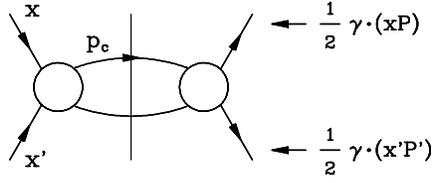}
\caption{Effective quark-quark (and antiquark) $2\rightarrow 2$ 
diagrams contributing to the partonic hard parts, 
$H_{aq\rightarrow c}$.}
\label{fig14}
\end{figure}

Based on the same arguments following Eq.~(\ref{s3e24}), the leading
contributions from diagrams with initial-state interactions, 
shown in Fig.~\ref{fig13}a, come only from the derivative of 
the phase space $\delta$-function.  By analogy to Eq.~(\ref{s3e33}), we obtain
\begin{eqnarray}
E_\ell \frac{d\Delta\sigma^I_{q}(\vec{s}_T)}{d^3\ell} 
&=& \frac{\alpha_s^2}{S} \sum_{a,c} 
\int_{z_{\rm min}}^1 \frac{dz}{z^3}\, D_{c\rightarrow\pi}(z)
\int_{x_{\rm min}}^1 \frac{dx}{x}\, \frac{1}{xS+U/z}
\int \frac{dx'}{x'}\, \delta\left(x'-\frac{-xT/z}{xS+U/z}\right) 
\nonumber\\
& \times & g_s\,
   \epsilon^{s_T \ell n \bar{n}}\, \left(\frac{1}{x'S+T/z}\right)\,
   \sum_q q(x')\, \left[-x\frac{\partial}{\partial x}
                \left(\frac{T^{(V)}_{F_a}(x,x)}{x}\,
                     H^I_{aq\rightarrow c}(\hat{s},\hat{t},\hat{u}) 
                \right) \right]\, ,
\label{s3e51}
\end{eqnarray} 
where the partonic hard part, 
$H^I_{aq\rightarrow c}(\hat{s},\hat{t},\hat{u})$ is given by the 
$2\rightarrow 2$ quark-quark (quark-antiquark) diagrams shown in 
Fig.~\ref{fig14}.  Compared to the spin-averaged case,
$H^I_{aq\rightarrow c}(\hat{s},\hat{t},\hat{u})$ plays the same 
role as $\hat{\sigma}_{aq\rightarrow c}$ in Eq.~(\ref{s3e8}).  
In fact, $H^I_{aq\rightarrow c}(\hat{s},\hat{t},\hat{u})$ 
is given by the same Feynman diagrams needed to calculate 
$\hat{\sigma}_{aq\rightarrow c}$, but, with different color factors,
$C^I_q$, due to the extra initial-state interactions.  Similarly 
to $C^I_g$, $C^I_q$ is given by the color structures of the diagrams
shown in Fig.~\ref{fig13}a, contracted with $[2/(N^2-1)]
(t^B)_{ij}$, where $B$ and $ij$ are color indices for the gluon 
and quarks from the polarized hadron, respectively.

Contributions from the derivatives of 
the phase space $\delta$-functions of 
the diagrams with final-state interactions
shown in Fig.~\ref{fig13}b are given
by
\begin{eqnarray}
E_\ell \frac{d\Delta\sigma^F_{q}(\vec{s}_T)}{d^3\ell} 
&=& \frac{\alpha_s^2}{S} \sum_{a,c} 
\int_{z_{\rm min}}^1 \frac{dz}{z^3}\, D_{c\rightarrow\pi}(z)
\int_{x_{\rm min}}^1 \frac{dx}{x}\, \frac{1}{xS+U/z}
\int \frac{dx'}{x'}\, \delta\left(x'-\frac{-xT/z}{xS+U/z}\right) 
\nonumber\\
& \times & g_s\,
   \epsilon^{s_T \ell n \bar{n}}\, \left(\frac{1}{x'S+T/z}\right)\,
   \sum_q q(x')\, \left[-x\frac{\partial}{\partial x}
                \left(\frac{T^{(V)}_{F_a}(x,x)}{x}\,
                     H^F_{aq\rightarrow c}(\hat{s},\hat{t},\hat{u}) 
                \right) \right]\, ,
\label{s3e52}
\end{eqnarray} 
where superscript $F$ represents the final-state interactions. 
The partonic hard-scattering function, $H^F_{aq\rightarrow 
c}(\hat{s},\hat{t},\hat{u})$,
has the same functional form as 
$H^I_{aq\rightarrow c}(\hat{s},\hat{t},\hat{u})$ in
Eq.~(\ref{s3e51}), with a different color factor $C^F_q$, 
because of the final-state interactions.  As with $C^F_g$, 
$C^F_q$ is given by the color structure of the diagrams
shown in Fig.~\ref{fig13}b, contracted with $[2/(N^2-1)]
(t^B)_{ij}$.

In addition to the contributions given in Eq.~(\ref{s3e52}) from 
the derivative of the $\delta$-function, the diagrams in Fig.~\ref{fig13}b
also have leading contributions from double-pole terms. 
Just as for the contributions from the quark-gluon subprocesses, 
given in Eq.~(\ref{s3e47}), the quark-quark and 
quark-antiquark double-pole contributions take the form
\begin{eqnarray}
E_\ell \frac{d\Delta\sigma^D_{q}(\vec{s}_T)}{d^3\ell} 
&=& \frac{\alpha_s^2}{S} \sum_{a,c} 
\int_{z_{\rm min}}^1 \frac{dz}{z^2}\, D_{c\rightarrow\pi}(z)
\int_{x_{\rm min}}^1 \frac{dx}{x}\, \frac{1}{xS+U/z}
\int \frac{dx'}{x'}\, \delta\left(x'-\frac{-xT/z}{xS+U/z}\right) 
\nonumber\\
& \times & g_s\,
   \epsilon^{s_T \ell n \bar{n}}\, \left(\frac{1}{-T}\right)\,
   \sum_q q(x')\, \left[-\frac{\partial}{\partial x}
                  \left(T^{(V)}_{F_a}(x,x)\right)\right]\,
                   H^D_{aq\rightarrow c}(\hat{s},\hat{t},\hat{u})\, ,
\label{s3e53}
\end{eqnarray} 
where the hard-scattering function found from the double pole, 
$H^D_{aq\rightarrow c}(\hat{s},\hat{t},\hat{u})$ is equal to 
$H^F_{aq\rightarrow c}(\hat{s},\hat{t},\hat{u})$ in
Eq.~(\ref{s3e52}).

\subsection{Calculation of the Partonic Hard Scattering Functions}

In Eqs.~(\ref{s3e33}), (\ref{s3e34}), (\ref{s3e47}), (\ref{s3e51}), 
(\ref{s3e52}) and (\ref{s3e53}), we have presented 
factorized expressions for leading contributions to the spin-dependent
cross section, $E_\ell d\Delta\sigma(\vec{s}_T)/d^3\ell$, for
quark-gluon, quark-quark and quark-antiquark subprocesses.  To
complete our derivation of the spin-dependent cross section, 
in this subsection we outline the
calculation of the partonic hard scattering functions
$H^I_{ag\rightarrow c}$, $H^F_{ag\rightarrow c}$, 
$H^I_{aq\rightarrow c}$ and $H^F_{aq\rightarrow c}$.
We recall that the subscripts $I$ and  $F$ refer to initial-
and final-state interactions, respectively.  The other two
hard scattering functions, associated with derivatives on final-state
propagators only, $H^D_{ag\rightarrow c}$ and $H^D_{aq\rightarrow c}$, 
are equal to $H^F_{ag\rightarrow c}$ and $H^F_{aq\rightarrow c}$,
respectively.

\begin{figure}
\epsfysize=4.0in
\epsfbox{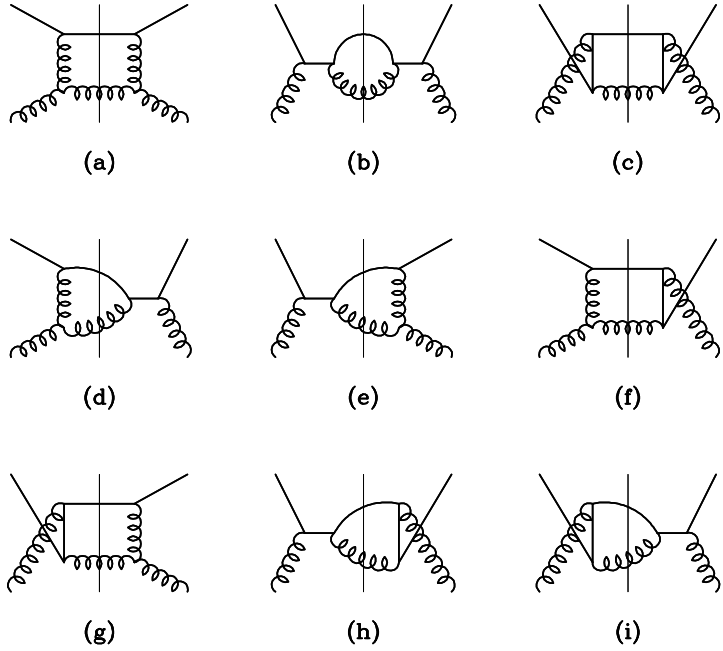}
\caption{All $2\rightarrow 2$ quark-gluon diagrams contributing to 
partonic hard parts, $H_{ag\rightarrow c}$.}
\label{fig15}
\end{figure}

For the quark-gluon subprocesses, the partonic hard scattering functions 
$H^I_{ag\rightarrow c}$ and $H^F_{ag\rightarrow c}$ are 
given by the same quark-gluon $2\rightarrow 2$ Feynman diagrams 
as shown in Fig.~\ref{fig15}, which are actually the same diagrams
contributing to the spin-averaged partonic part, 
$\hat{\sigma}_{ag\rightarrow c}$, in Eq.~(\ref{s3e10a}).
Incoming quark lines are contracted by $(1/2)\gamma\cdot(xP)$, 
and incoming gluon lines are contracted by $(1/2)(-g_{\alpha\beta})$.

Let $C_g$, $C^I_g$, and $C^F_g$ be the color factors for 
processes that are spin-averaged, 
spin-dependent with an initial-state interaction, and spin-dependent
with a final-state interaction, respectively.  The factor
$C_g$ for each diagram shown in Fig.~\ref{fig15} is simply 
the standard color factor for that diagram, with an average over initial-state
quark and gluon color.  

\begin{figure}
\epsfysize=1.6in
\epsfbox{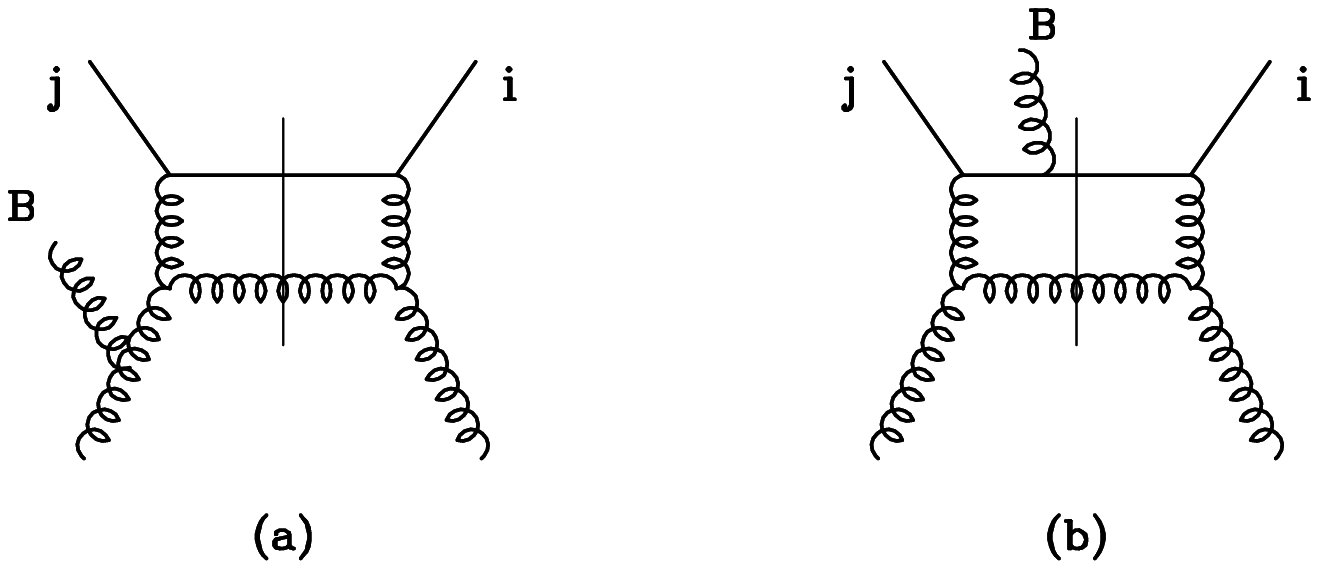}
\caption{Sample diagrams with initial-state and final-state interactions,
used to calculate the color factors, $C^I_g$ and $C^F_g$ in 
Table~\protect\ref{table1}.}
\label{fig16}
\end{figure}

Each $C^I_g$ is given by the color factor
of the diagram with one extra initial-state three-gluon vertex.
An example is shown in Fig.~\ref{fig16}a.  The color of the
incoming gluon from the unpolarized hadron is averaged, and the colors of 
the incoming quarks and the extra gluon from the polarized hadron are 
contracted with $[-2i/(N^2-1)](t^B)_{ij}$, as explained in the text 
following Eq.~(\ref{s3e28n}).  

Finally, the $C^F_g$ are the color 
factors of the same $2\rightarrow 2$ diagrams with an extra 
final-state quark-gluon interaction, illustrated by the diagram shown 
in Fig.~\ref{fig16}b.  Similarly to $C^I_g$, the color of the
incoming gluon from the unpolarized hadron is averaged, and the colors of 
the incoming quarks and the extra gluon from the polarized hadron are 
contracted with $[2/(N^2-1)](t^B)_{ij}$, as mentioned in the text 
after Eq.~(\ref{s3e34}).  Our results for all these color factors
are collected in Table~\ref{table1}.  

\begin{figure}
\epsfysize=4.8in
\epsfbox{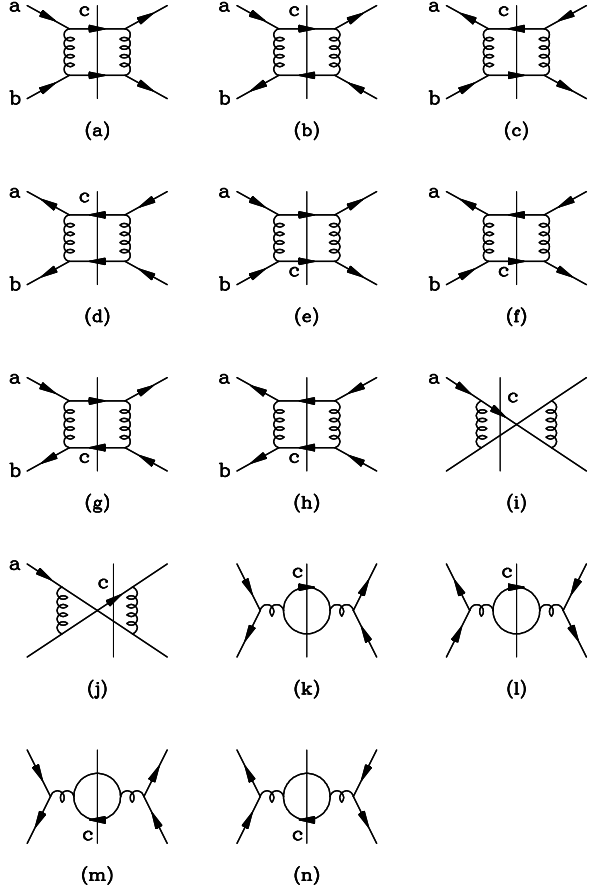}
\caption{All $2\rightarrow 2$ quark-quark (and antiquark) diagrams 
contributing to partonic hard parts, $H_{aq\rightarrow c}$.}
\label{fig17}
\end{figure}

For quark-quark (or quark-antiquark) subprocesses, 
the partonic hard scattering functions, 
$H^I_{aq\rightarrow c}$ and $H^F_{aq\rightarrow c}$ are 
given by the same quark-quark (or quark-antiquark) 
$2\rightarrow 2$ Feynman diagrams 
as shown in Fig.~\ref{fig17}, which are the same diagrams
contributing to the spin-averaged partonic cross section, 
$\hat{\sigma}_{aq\rightarrow c}$ in Eq.~(\ref{s3e10b}).
Incoming quark lines from the polarized hadron are contracted by 
$(1/2)\gamma\cdot(xP)$, and incoming quark (or antiquark) lines 
from the unpolarized hadron are contracted by $(1/2)\gamma\cdot(x'P')$.

\begin{figure}
\epsfysize=1.2in
\epsfbox{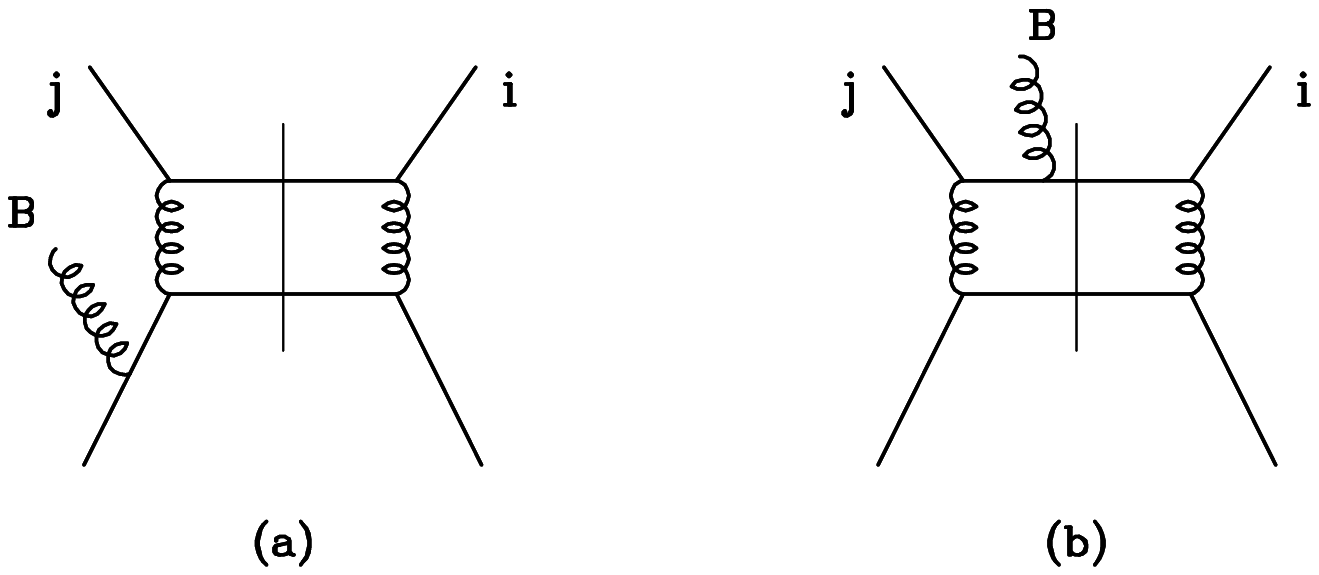}
\caption{Sample diagrams with initial-state and final-state interactions,
used to calculate the color factors, $C^I_q$ and $C^F_q$ in 
Table~\protect\ref{table2}.}
\label{fig18}
\end{figure}

As with the quark-gluon subprocesses, $C_q$, $C^I_q$, 
and $C^F_q$ are respectively the color factors for 
subprocesses that are spin-averaged, spin-dependent 
with an initial-state interaction, and spin-dependent
with a final-state interaction.  
The $C_q$ for the individual diagrams shown in Fig.~\ref{fig17} are 
the color factors for each diagram, with a standard average over 
initial-state
quark (or antiquark) color.  The $C^I_q$'s are found by including 
an extra initial-state three-gluon interaction in the 
$2\rightarrow 2$ process,
(for example, Fig.~\ref{fig18}a) averaging  the color of 
the quark (or antiquark) from the unpolarized hadron, and
contracting the colors 
of the incoming quarks and the extra gluon from the polarized hadron 
with $[2/(N^2-1)](t^B)_{ij}$, as mentioned 
following Eq.~(\ref{s3e51}).  The $C^F_q$ are 
found from the same $2\rightarrow 2$ diagrams, now with one extra 
final-state quark-gluon interaction (illustrated by the diagram shown 
in Fig.~\ref{fig18}b).  In exactly the same fashion as for $C^I_q$, 
the colors from the unpolarized hadron are averaged, and 
the colors from the polarized hadron 
are contracted with $[2/(N^2-1)](t^B)_{ij}$
(as mentioned in connection with Eq.~(\ref{s3e52})).  

Our results for the
quark-quark and quark-antiquark color factors
are summarized in Table~\ref{table2}.  Notice the sign difference
for the coefficient of $4N$ in the color factor $(N^2\pm 4N-4)/(32N)$,
between graphs related by reversing the arrow of a quark or antiquark line.
These will give slight differences to the asymmetries in 
proton($\uparrow$)-proton compared with antiproton($\uparrow$)-proton 
collisions.

>From Table~\ref{table1} and Table~\ref{table2}, we can construct 
all the necessary partonic hard scattering functions.  For the 
spin-averaged cross section, 
the hard-scattering function for the quark-gluon subprocess, 
$\hat{\sigma}_{ag\rightarrow c}$ in Eq.~(\ref{s3e10a}), 
is found by combining the entries in the columns of 
{\it Partonic Parts} and $C_g$
in Table~\ref{table1}. For the quark-quark (or antiquark) 
subprocesses, $\hat{\sigma}_{aq\rightarrow c}$ in Eq.~(\ref{s3e10b}) 
is found by combining entries from the columns of 
{\it Partonic Parts} with $C_q$
in Table~\ref{table2}.  For the spin-dependent cross section,
the twist-three partonic hard scattering function 
$H^I_{ag\rightarrow c}$ is found by 
combining entries in the columns of {\it Partonic Parts} and $C^I_g$ in 
Table~\ref{table1}.  In the same way, one can read off other partonic hard 
scattering functions, $H^F_{ag\rightarrow c}$, $H^I_{aq\rightarrow c}$ and
$H^F_{aq\rightarrow c}$ from Table~\ref{table1} and Table~\ref{table2}.


\section{Numerical Results for Single Transverse-Spin Asymmetries}
\label{sec:4}

Having derived expressions for the single 
transverse-spin asymmetries in previous section, we 
are now ready to develop numerical 
estimates of $A_N$ for inclusive single 
pion production.

\subsection{Model for the Twist-3 Distribution: $T^{(V)}_F(x,x)$}

The application of perturbative QCD to observables involving 
hadrons in the initial state relies on factorization theorems \cite{CSS:FAC} 
and on the universality of the nonperturbative, long-distance distributions.  
For 
the single transverse-spin asymmetries discussed in this paper, a test 
of the perturbative formalism requires in principle an independent 
extraction of the 
spin-dependent twist-three distributions, $\phi^{(3)}_{a/A}(x_1,x_2)$ 
introduced in Eq.~(\ref{s3e11}).  As we have observed, there are 
a variety of twist-three distributions, dependent in general on 
a  pair of momentum fractions.  It would requires extensive measurements
to pin down all of these functions.  However, for  
single-spin asymmetries in the forward region, we have argued
above, and in Ref.\ \cite{AN:QS}, 
that the dominant contribution may depend primarily on only a single
twist-three distribution, $T^{(V)}_{F_a}(x,x)$, at equal values of its
two arguments.  Assuming this to be the case, it could be possible 
to infer the form of $T^{(V)}_{F_a}(x,x)$ from single transverse-spin 
asymmetries in $\pi^+$ and/or $\pi^-$ production, and then use it to 
predict asymmetries in the production of $\pi^0$, direct photon or
other particles, at least approximately.

In order to compare our calculated asymmetries to the
existing data, we need to assume an initial functional form for the 
twist-3 distribution, $T^{(V)}_{F_a}(x,x)$.  To help motivate our
model, we compare the operator definition
of $T^{(V)}_{F_a}(x,x)$ with that of a twist-2 quark distribution
$q_a(x)$ of flavor $a$.  From Eq.~(\ref{s3e18}), we have
\begin{eqnarray}
T_{F_a}^{(V)}(x,x) &=&
\int \frac{dy^-}{4\pi}\, \mbox{e}^{ixP^+y^-}\,
     \langle P,\vec{s}_T |\bar{\psi}_a(0)\gamma^+
\nonumber \\
&& \quad \times
\left[\int dy_2^-\, \epsilon^{s_T\sigma n \bar{n}}\, 
      F_{\sigma +}(y_2^-) \right]
 \psi_a(y^-) |P,\vec{s}_T\rangle \ ,
\label{s4e1}
\end{eqnarray}
where subscript $a$ is quark flavor.  Correspondingly, from 
Ref.\cite{CS:PDF} we have for the quark distribution
\begin{equation}
q_{a}(x) = \int \frac{dy^-}{4\pi}\, \mbox{e}^{ixP^+y^-}\,
     \langle P |\bar{\psi}_a(0)\gamma_+
                \psi_a(y^-) |P\rangle \ .
\label{s4e2}
\end{equation}
As above, we suppress ordered exponentials of the gauge field.
Comparing Eqs.~(\ref{s4e1}) with (\ref{s4e2}), 
the operator defining $T^{(V)}_{F_a}(x,x)$ is the same as 
for the spin-averaged quark distribution,
except for the term in the 
square brackets.  This factor, however, does not introduce 
explicit $x$-dependence (or $y$-dependence in coordinate space).
Based on this
similarity of the operators, we model the twist-3 distribution 
with the following functional form, inspired by the quark
distributions themselves,
\begin{equation}
T^{(V)}_{F_a}(x,x) \equiv \kappa_a\, \lambda\, q_a(x)\ ,
\label{s4e3}
\end{equation}
where $\lambda$ (with dimensions of energy) is a normalization 
constant, which will be fixed by the data; and where 
$\kappa_a =\pm 1, 0$, depending 
on flavor $a$.  Note that we propose the relation Eq.\ (\ref{s4e3}) 
only for relatively large
$x$, where the correlations of quarks with the gluon field may
be simplified.  This restriction limits somewhat the utility
of low moments of $T_F$ in estimates of its magnitude \cite{TF:Schafer}.

For the parameters $\kappa_a$ in Eq.~(\ref{s4e3}), we shall see that
the data suggest the choices
\begin{eqnarray}
\kappa_u &=& +1 \quad \mbox{and} \quad \frac{\kappa_u}{\kappa_d} = -1
\quad \mbox{(proton)}\ ,
\nonumber\\
\kappa_{\bar u} &=& -1 \quad \mbox{and} \quad \frac{\kappa_{\bar 
u}}{\kappa_{\bar d}} = -1
\quad \mbox{(antiproton)}\ ,
\label{s4e4}
\end{eqnarray}
where the second line follows from the first by using
charge conjugation invariance in $T_F$.
In the valence quark approximation, discussed in the previous sections, we
further assume that $\kappa_s=0$.
Of course, Eq.~(\ref{s4e3}) is simply a model, and the true functional
form of the twist-three distribution $T^{(V)}_{F_a}(x,x)$ should be 
determined by detailed comparison with experiment.  The purpose of our model
is to have a functional form that we can use to begin such
a comparison with the important, but still limited, data that are available.

\subsection{Single Transverse-spin Asymmetries in Pion Production}

Single transverse-spin asymmetries for pions were been measured
at Fermilab by the E704 Collaboration with 200 GeV polarized proton and 
antiproton beams on an unpolarized proton target \cite{EXP:Pion}.  In this
subsection, we use the Fermilab data to estimate the value of $\lambda$,
in Eq.~(\ref{s4e3}), and check the consistency of our model.

\subsubsection{Absolute Sign of the Single Transverse-Spin Asymmetry}

In order to compare the experimental data on the asymmetries, $A_N$, 
with our calculations in Sec.~\ref{sec:3}, we need to fix the 
absolute sign of $A_N$.  

\begin{figure}
\epsfysize=1.2in
\epsfbox{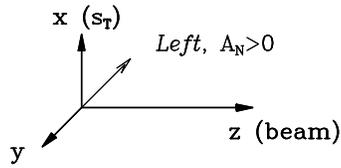}
\caption{Sketch for the coordinate system: the polarized beam is along
the $z$-axis and the beam particle spin along the $x$-axis. Positive $A_N$ 
corresponds to an excess of events in the $-y$-direction.}
\label{fig19}
\end{figure}

According to Ref.\cite{EXP:Pion}, positive values of $A_N$ correspond 
to larger cross sections for production of $\pi^0$ to the beam's 
{\it left} when the beam particle spin is vertically {\it upward}, 
as sketched in Fig.~\ref{fig19}.  We choose our coordinate system
such that the beam direction is along the $z$-axis, and the direction of 
the beam spin is along the $x$-axis, as shown in Fig.~\ref{fig19}.
Consequently, the experimental beam's {\it left} corresponds to 
the $-y$-direction in our coordinate system, and 
\begin{equation}
\left(A_N\right)_{\rm exp} > 0 
\quad \Longleftrightarrow \quad
\epsilon^{\ell_T s_T n \bar{n}} > 0\ .
\label{s4e5}
\end{equation}
Eq.~(\ref{s4e5}) fixes the absolute sign of $A_N$ presented in 
Sec.~\ref{sec:3}, and dictates our choice $\kappa_u=+1$ in Eq.\ (\ref{s4e4}).

\subsubsection{Leading Single Transverse-Spin Asymmetry 
[$(\partial/\partial x)T^{(V)}_{F_a}(x,x)$ only]}

As explained in Sec.~\ref{sec:3}, we are interested in 
$A_N$ in the forward region, where
it is largest experimentally.  In deriving Eqs.~(\ref{s3e33}), 
(\ref{s3e34}), (\ref{s3e47}), (\ref{s3e51}), (\ref{s3e52}) and 
(\ref{s3e53}), we kept only contributions from the
terms discussed in items (1) and 
(2) of Sec.~\ref{sec:2d}, because 
those discussed in items (3) and (4)
lack a derivative on the twist-three
distribution.  To be consistent with our approximation, 
we rewrite the contributions in Eqs.~(\ref{s3e33}), 
(\ref{s3e34}), (\ref{s3e47}), (\ref{s3e51}), (\ref{s3e52}) and 
(\ref{s3e53}) in terms of an explicit factor of
$(\partial/\partial x)T^{(V)}_{F_a}(x,x)$, 
neglecting derivatives of other factors.  Combining all leading 
contributions to the spin-dependent cross section, 
in a manner similar to the spin-averaged cross section in Eq.~(\ref{s3e8}), 
we obtain
\begin{eqnarray}
E_\ell\frac{d^3\Delta\sigma(\vec{s}_T)}{d^3\ell} 
&=& \frac{\alpha_s^2}{S}\, \sum_{a,c} 
    \int_{z_{\rm min}}^1 \frac{dz}{z^2}\, D_{c\rightarrow\pi}(z)
    \int_{x_{\rm min}}^1 \frac{dx}{x}\, \frac{1}{xS + U/z} 
    \int \frac{dx'}{x'}\, \delta\left(x'-\frac{-xT/z}{xS+U/z}\right)
\label{s4e6} \\
&\times& \sqrt{4\pi\alpha_s}\, 
    \left(\frac{\epsilon^{\ell s_T n \bar{n}}}{z(-\hat{u})}\right)
    \left[-x\frac{\partial}{\partial x}T^{(V)}_{F_a}(x,x)\right]
    \left[G(x')\, \Delta\hat{\sigma}_{ag\rightarrow c}\, 
  + \sum_q q(x')\, \Delta\hat{\sigma}_{aq\rightarrow c} \right] \ ,
\nonumber
\end{eqnarray}
where $\sum_{a}$ runs over up and down valence quarks.  The integration 
limits in Eq.~(\ref{s4e6}) are the same as those defined in 
Eq.~(\ref{s3e8}).  The spin-dependent partonic cross sections,
$\Delta\hat{\sigma}_{ag\rightarrow c}$ and 
$\Delta\hat{\sigma}_{aq\rightarrow c}$ are given 
by
\begin{mathletters}
\label{s4e7}
\begin{eqnarray}
\Delta\hat{\sigma}_{ag\rightarrow c} &=&
-\left[ H^I_{ag\rightarrow c}(\hat{s},\hat{t},\hat{u}) +
        H^F_{ag\rightarrow c}(\hat{s},\hat{t},\hat{u}) +
        \left(\frac{\hat{u}}{\hat{t}}\right)
        H^D_{ag\rightarrow c}(\hat{s},\hat{t},\hat{u}) \right] \ ,
\label{s4e7a} \\
\Delta\hat{\sigma}_{aq\rightarrow c} &=&
-\left[ H^I_{aq\rightarrow c}(\hat{s},\hat{t},\hat{u}) +
        H^F_{aq\rightarrow c}(\hat{s},\hat{t},\hat{u}) +
        \left(\frac{\hat{u}}{\hat{t}}\right)
        H^D_{aq\rightarrow c}(\hat{s},\hat{t},\hat{u}) \right] \ ,
\label{s4e7b}
\end{eqnarray}
\end{mathletters}

\noindent
where the minus sign is from $\epsilon^{s_T\ell n\bar{n}} = -\, 
\epsilon^{\ell s_T n\bar{n}}$, and where all the partonic 
hard-scattering functions have been given in 
Sec.~\ref{sec:3}.  In deriving Eq.~(\ref{s4e7}), 
$(x'S+T/z)/(-T/z)=\hat{u}/\hat{t}$ was used.  From the information
given in Table~\ref{table1} and Table~\ref{table2}, we find 
the following explicit
expression for $\Delta\hat{\sigma}_{ag\rightarrow c}$ taking $N=3$,
\begin{mathletters}
\label{s4e8}
\begin{eqnarray}
\Delta\hat{\sigma}_{ag\rightarrow c} &=&
\delta_{ac}\left\{
 2\left(1-\frac{\hat{s}\hat{u}}{\hat{t}^2}\right)
  \left[\frac{9}{16} + \frac{1}{8} 
        \left(1+\frac{\hat{u}}{\hat{t}}\right)\right]
\right. 
\nonumber \\ 
&& \quad +
 \frac{4}{9} \left( \frac{-\hat{u}}{\hat{s}} + 
                        \frac{\hat{s}}{-\hat{u}} \right)
  \left[\frac{63}{128} -\frac{1}{64}
        \left(1+\frac{\hat{u}}{\hat{t}}\right)\right] 
\nonumber \\
&& \quad +
             \left( \frac{\hat{s}}{\hat{t}} + 
                         \frac{\hat{u}}{\hat{t}} \right)
  \left[\frac{9}{16} + \frac{1}{8} 
        \left(1+\frac{\hat{u}}{\hat{t}}\right)\right] 
\nonumber \\
&& \quad + \left.
 \left[\frac{9}{32}\left( \frac{-\hat{u}}{\hat{s}} - 
                        \frac{\hat{s}}{-\hat{u}} \right)\right]
+\left[\frac{9}{16}\left( \frac{\hat{s}}{\hat{t}} - 
                         \frac{\hat{u}}{\hat{t}} \right)\right] 
\right\}\ .
\label{s4e8a}
\end{eqnarray}
For quark-quark (or antiquark) scattering, the color factors of 
individual subprocess depend on quark or antiquark, as shown
in Table~\ref{table2}.  For parton $a$ a quark  
(corresponding to a polarized proton beam), we have
\begin{eqnarray}
\Delta\hat{\sigma}_{qq'\rightarrow q} &=&
    \frac{4}{9} 
      \left(\frac{\hat{s}^2+\hat{u}^2}{\hat{t}^2} \right)
  \left[\frac{21}{64} + \frac{1}{8} 
        \left(1+\frac{\hat{u}}{\hat{t}}\right)\right] 
\nonumber \\
\Delta\hat{\sigma}_{q\bar{q}'\rightarrow q}
&=& \frac{4}{9} 
      \left(\frac{\hat{s}^2+\hat{u}^2}{\hat{t}^2} \right)
  \left[ \frac{51}{64} + \frac{1}{8} 
        \left(1+\frac{\hat{u}}{\hat{t}}\right)\right] 
\nonumber \\
\Delta\hat{\sigma}_{qq'\rightarrow q'}
&=& \frac{4}{9} 
      \left(\frac{\hat{s}^2+\hat{t}^2}{\hat{u}^2} \right) 
  \left[\frac{21}{64} - \frac{51}{64} 
        \left(1+\frac{\hat{u}}{\hat{t}}\right)\right]
\nonumber \\
\Delta\hat{\sigma}_{q\bar q'\rightarrow \bar q}
&=&  \frac{4}{9} 
      \left(\frac{\hat{s}^2+\hat{t}^2}{\hat{u}^2} \right) 
  \left[\frac{51}{64} - \frac{21}{64} 
        \left(1+\frac{\hat{u}}{\hat{t}}\right)\right]  
\nonumber \\
\Delta\hat{\sigma}_{qq\rightarrow q}
&=&   \frac{-8}{27} 
      \left(\frac{\hat{s}^2}{\hat{u}\hat{t}} \right) 
  \left[\frac{10}{8} + \frac{1}{8} 
        \left(1+\frac{\hat{u}}{\hat{t}}\right)\right] 
\nonumber \\
\Delta\hat{\sigma}_{q\bar{q}\rightarrow q'}
&=&  \frac{4}{9} 
      \left(\frac{\hat{t}^2+\hat{u}^2}{\hat{s}^2} \right)
  \left[-\frac{1}{8} - \frac{51}{64} 
        \left(1+\frac{\hat{u}}{\hat{t}}\right)\right] 
\nonumber \\
\Delta\hat{\sigma}_{q\bar{q}\rightarrow \bar{q}'}
&=&  \frac{4}{9} 
      \left(\frac{\hat{t}^2+\hat{u}^2}{\hat{s}^2} \right)
  \left[-\frac{1}{8} - \frac{21}{64} 
        \left(1+\frac{\hat{u}}{\hat{t}}\right)\right] \ .
\label{s4e8b}
\end{eqnarray}
\end{mathletters}

\noindent
For a polarized 
antiproton beam, similar formulas for 
$\Delta\hat{\sigma}_{\bar{q}b\rightarrow c}$ can be derived from 
Table~\ref{table2}.  From Eqs.~(\ref{s3e8})
and (\ref{s4e8}), we see that the underlying partonic 
cross sections for spin-dependent and 
spin-averaged cases are similar, other than the
factors in square brackets. 

\subsubsection{Comparison with the Fermilab Data}

Because of limited phase space, 
most of the Fermilab data in Ref.\ \cite{EXP:Pion} were collected  
at relatively small values of transverse momenta, ranging up 
to 4~GeV for $\pi^0$ in the central
region (where $A_N$ is small), and up to only 1.5 GeV
for $\pi^\pm,\pi^0$ in the forward region, where
$A_N$ is large.  In general, a transverse
momentum of even 2~GeV is considered too small 
to apply perturbative QCD reliably to single-particle
inclusive cross sections, because of
their steep dependence on  $\ell_T$.  This strong dependence makes the
cross sections sensitive to 
higher-twist effects not associated directly with
spin, such as intrinsic transverse momentum,
hadronic scales, and, of course, yet higher powers in $1/\ell_T$.
One consequence of these effects is to regularize the cross section at  
$\ell_T=0$. 
For the asymmetry, however, the strongest power dependence on
$1/\ell_T$ cancels in the ratio of the spin-dependent and spin-averaged 
cross sections, leaving at most $\lambda/\ell_T$ in $A_N$.  
In fact, as we will show below, $A_N$ does not behave numerically even as 
steeply as $1/\ell_T$ in 
most of the range where the data were collected.  This suggests that our
calculation for $A_N$ is perturbatively 
stable and may be meaningfully compared with the data.

  For simplicity in our numerical estimates, we 
employed the following 
simple parametrizations, without scaling violation,
for twist-two parton distributions \cite{AN:QS},
\begin{mathletters}
\label{s4e9}
\begin{eqnarray}
xu_v(x) &=& \frac{2}{B(0.5,4)}\, x^{0.5}\,(1-x)^3\ ,
\\
xd_v(x) &=& \frac{1}{B(0.5,4.5)}\, x^{0.5}\,(1-x)^{3.5}\ ,
\\
xS(x) &=& 8\left[\frac{1}{2} - 2\frac{B(1.5,4)}{B(0.5,4)}
                 -\frac{B(1.5,4.5)}{B(0.5,4.5)}\right]\,
           (1-x)^7\ ,
\\
xG(x) &=& 3\,(1-x)^5\ .
\end{eqnarray}
\end{mathletters}

\noindent
Here, $B(x,y)$ is the beta function.
For pion fragmentation
functions, we rely on Ref.~\cite{PI:frag}.
Using the simplified parton distributions 
of Eqs.\ (\ref{s4e9}) in the spin-averaged cross section,
and in the model for the twist-three distribution given by
Eq.~(\ref{s4e3}), we evaluated $A_N$
as the ratio of the spin-dependent cross section in Eq.~(\ref{s4e6}) to the
spin-averaged cross section, Eq.~(\ref{s3e8}).  

\begin{figure}
\epsfysize=3.6in
\epsfbox{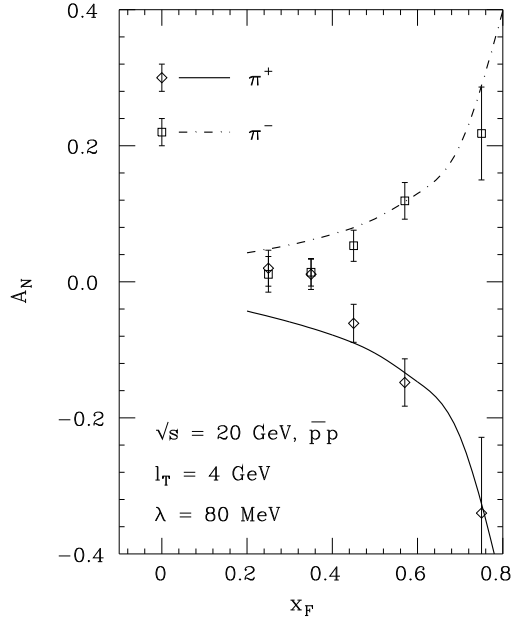}
\caption{Single transverse-spin asymmetry as a function of $x_F$
for $\pi^+$ and $\pi^-$ production with a polarized {\it antiproton} 
beam.  Here and in the following five
figures, data are from Ref.~\protect\cite{EXP:Pion}
at $\protect\sqrt{S}=20$ GeV and $l_T$ up
to 1.5 GeV. Theory curves 
are evaluated at transverse momentum $l_T=4$~GeV and
$\lambda=0.080$~GeV at the same center-of-mass energy.}
\label{fig20}
\end{figure}

\begin{figure}
\epsfysize=3.6in
\epsfbox{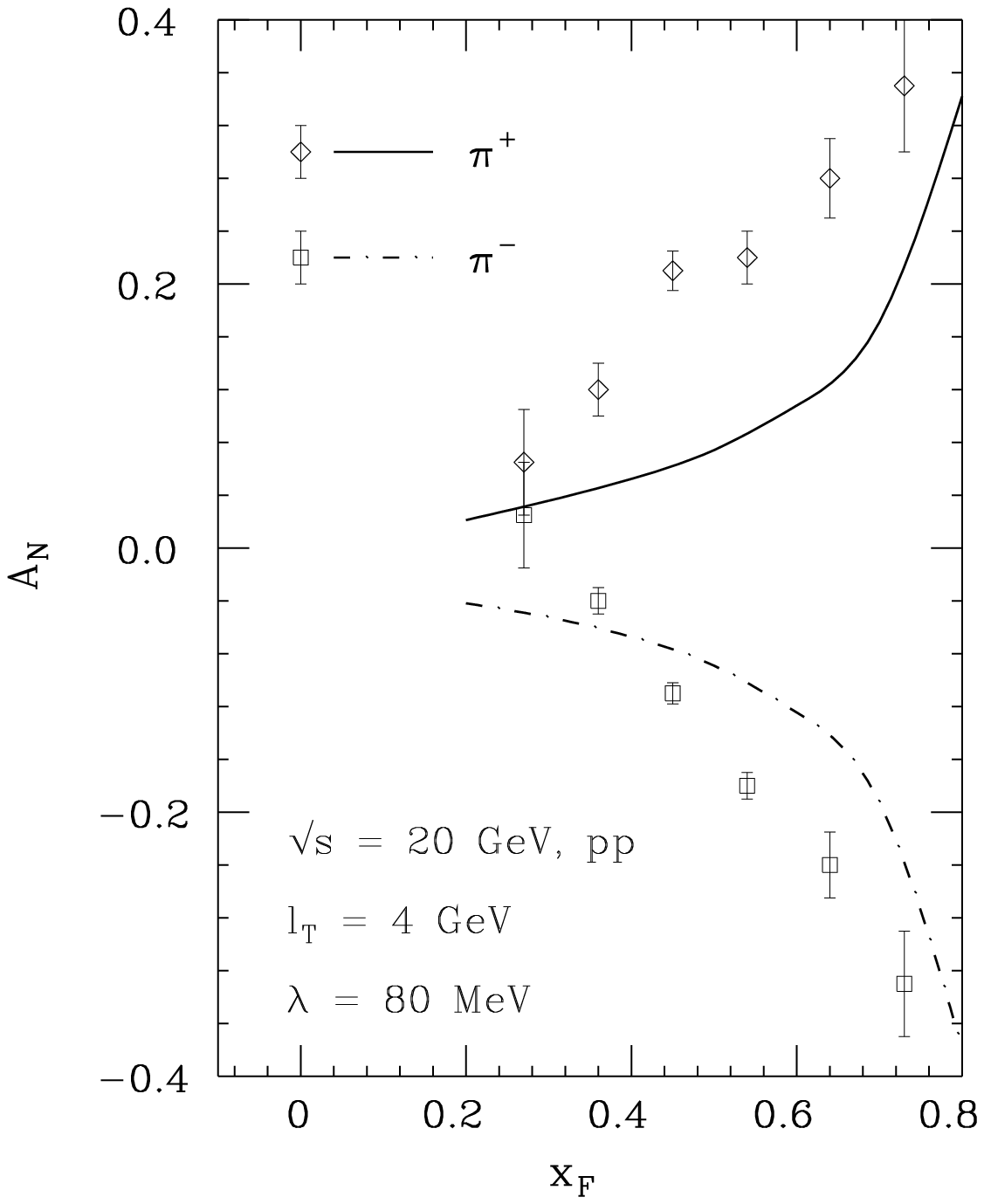}
\caption{Single transverse-spin asymmetry as a function of $x_F$
for $\pi^+$ and $\pi^-$ production with a polarized {\it proton} 
beam.  Data are from Ref.~\protect\cite{EXP:Pion}. Theory curves 
are evaluated at transverse momentum $l_T=4$~GeV and with $\lambda=0.0.80$~GeV.}
\label{fig21}
\end{figure}

\begin{figure}
\epsfysize=3.6in
\epsfbox{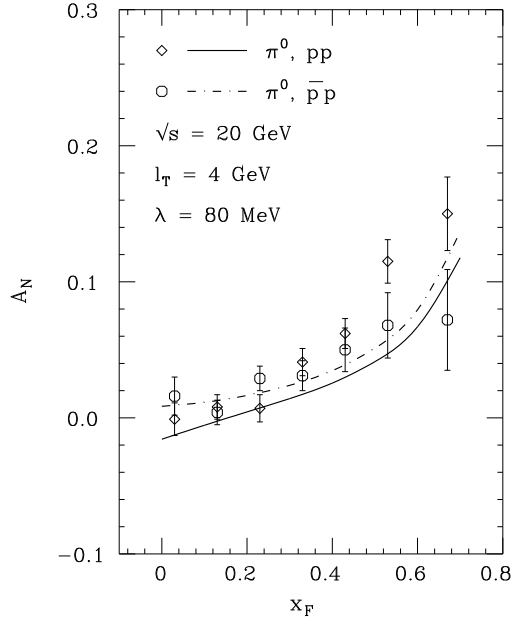}
\caption{Single transverse-spin asymmetry as a function of $x_F$
for $\pi^0$ production with a polarized antiproton 
and proton beams.  Data are from Ref.~\protect\cite{EXP:Pion}. Theory curves 
are evaluated at transverse momentum $l_T=4$~GeV and with $\lambda=0.080$~GeV.}
\label{fig22}
\end{figure}

In 
Fig.~\ref{fig20}, along with experimental data from 
Ref.\cite{EXP:Pion}, we have plotted our calculated $A_N$ for $\pi^+$ and 
$\pi^-$ production in the scattering of a polarized antiproton beam 
on an unpolarized proton target.  Similarly, in Fig.~\ref{fig21}, 
we plot the asymmetries with a polarized proton beam.  In 
Fig.~\ref{fig22}, we compare theory and experiment in the asymmetries for 
$\pi^0$ 
production with a polarized antiproton beam and 
a polarized proton beam.  The data presented in Figs.~\ref{fig20}, \ref{fig21} 
and
\ref{fig22} are averaged over the range of transverse 
momenta, up to 1.5 GeV.  
All of the calculations in these figures, however, were carried out 
at $\ell_T\sim  4$ GeV, with a normalization
constant $\lambda=0.080$~GeV, adjusted to give a rough match
to the data. \footnote{For the purpose of this comparison, we
neglect correlations between $x_F$ and $\ell_T$ in the data.}
We will come back to the choice of $\ell_T$ in a  moment.  
This limitation notwithstanding, 
fixing the single overall normalization 
constant, $\lambda$, is enough to give theoretical predictions that are 
consistent
with the shapes and relative signs and
normalizations of all the experimental data.

\begin{figure}
\epsfysize=3.6in
\epsfbox{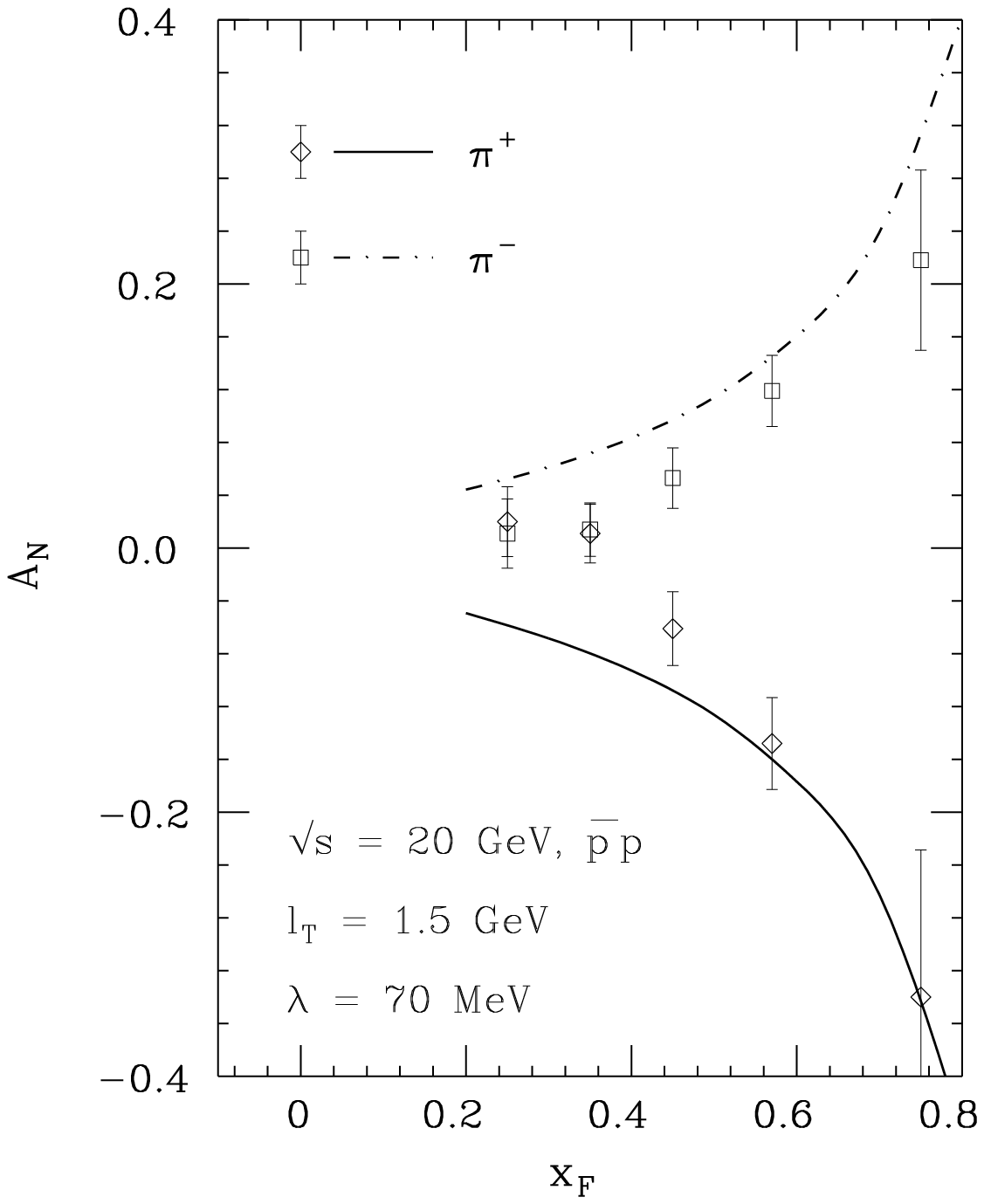}
\caption{Single transverse-spin asymmetry as a function of $x_F$
for $\pi^+$ and $\pi^-$ production with a polarized {\it antiproton} 
beam.  Data are from Ref.~\protect\cite{EXP:Pion}. Theory curves 
are evaluated at $l_T=1.5$~GeV and $\lambda=0.070$~GeV.}
\label{fig23}
\end{figure}

\begin{figure}
\epsfysize=3.6in
\epsfbox{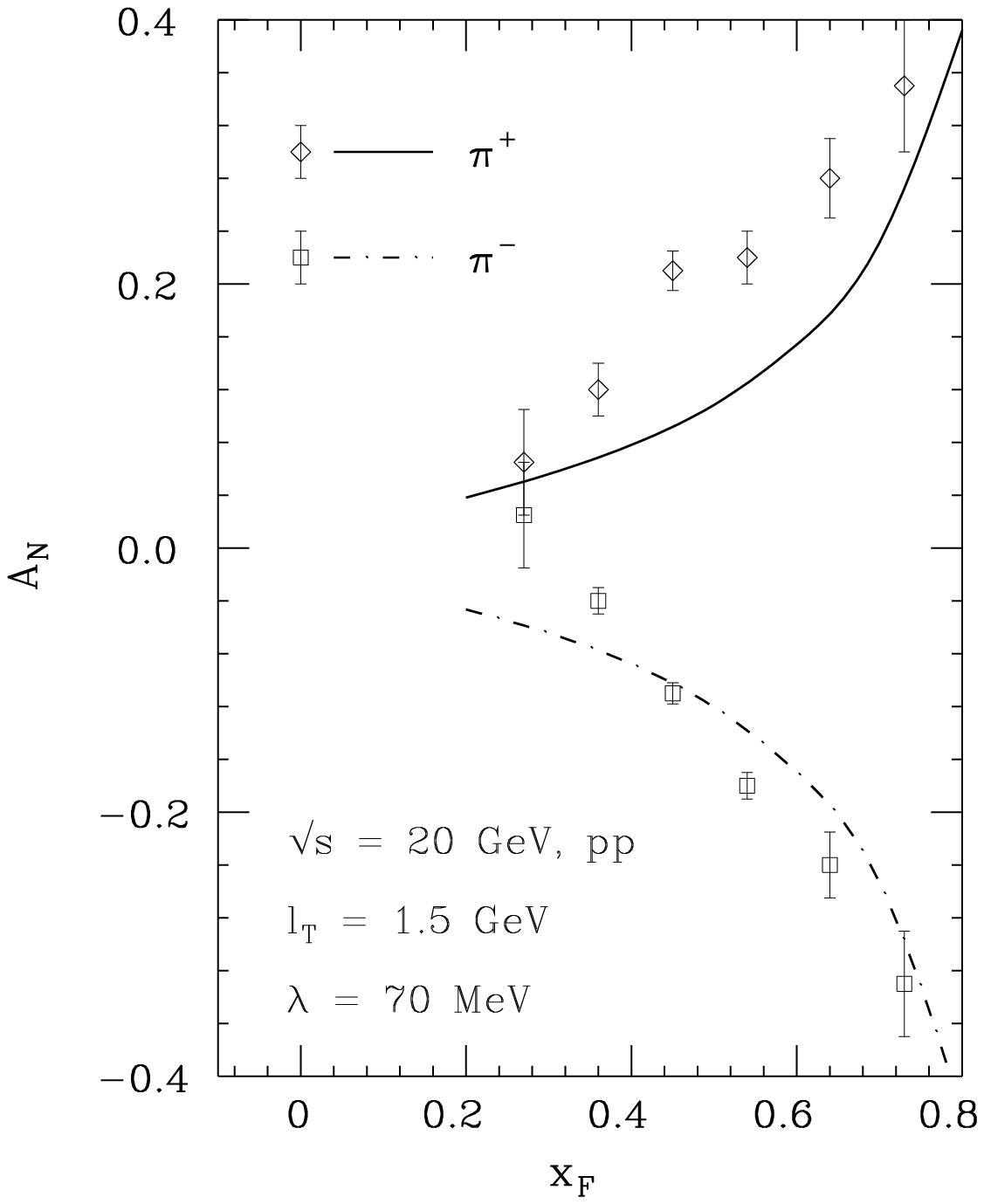}
\caption{Single transverse-spin asymmetry as a function of $x_F$
for $\pi^+$ and $\pi^-$ production with a polarized {\it proton} 
beam.  Data are from Ref.~\protect\cite{EXP:Pion}. Theory curves 
are evaluated at $p_T=1.5$~GeV and $\lambda=0.070$~GeV.}
\label{fig24}
\end{figure}

\begin{figure}
\epsfysize=3.6in
\epsfbox{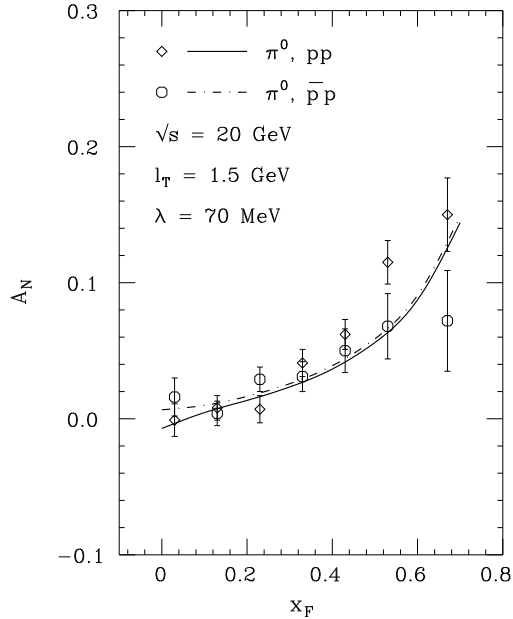}
\caption{Single transverse-spin asymmetry as a function of $x_F$
for $\pi^0$ production with a polarized antiproton, along  
with the same asymmetry obtained with a polarized proton
beam.  Data are from Ref.~\protect\cite{EXP:Pion}. Theory curves 
are evaluated at $l_T=1.5$~GeV and $\lambda=0.070$~GeV.}
\label{fig25}
\end{figure}

Now let us consider to the question of how to best to choose $\ell_T$ for
the comparison with the data.
Given the naive expectation that $A_N\sim  1/\ell_T$,
the extracted value of $\lambda$ might be expected to depend strongly on the 
value of 
$\ell_T$ at which we evaluate the asymmetries.  
Surprisingly, however, the perturbative 
prediction for the asymmetries in this momentum region is  
not very sensitive the 
precise value of $\ell_T$.  Thus, in Figs.~\ref{fig23},
\ref{fig24}, and \ref{fig25}, we present the same asymmetries 
as in the foregoing three figures, now
evaluated at $\ell_T=1.5$~GeV.  
For this value, we find a good match to
the data by choosing 
$\lambda=0.070$~GeV, not too different from the
value found at $\ell_T=4$~GeV.  
Clearly, the normalizations and shapes of the 
asymmetries at $\ell_T=4$~GeV and $\ell_T=1.5$~GeV are very similar, 
with an only slightly different normalization factor.
We consider
this stability very encouraging.
 Such consistency is strong evidence that 
the twist-three formalism of perturbative QCD can be applied to single 
transverse-spin asymmetries at moderate transverse
momenta.  We will give a further discussion of this
point in the next section.  

We close this section with a 
few comments on the consequences of our model of $T_F$ 
(Eqs.~(\ref{s4e3}) and (\ref{s4e4}), with $\lambda\sim 0.080$~GeV)
for single-spin asymmetries in
direct photon production.  Compared to the ansatz for $T_F$ proposed
in Ref.\ \cite{AN:QS}, the two main differences are, first, the relative
minus sign between the down and up quark matrix elements, and, second,
a decrease in the overall normalization $\lambda$, below
100 MeV.  Both of these features are suggested by comparison to the data
for pion production, which is only now possible.  The effects of
the both changes would be to reduce the cross
section estimates given in Ref.\ \cite{AN:QS}, although the second
is more important than the first, because the
down quark's charge is small.  In any case, the data \cite{EXP:Photon} 
which limit
the direct photon asymmetry is at low $x_F$, where either
model predicts a small effect.


\section{Summary and Discussion}
\label{sec:5}

In this section, we summarize and interpret the main features of our results and 
provide a  few thoughts on future development on this subject.

We have presented a calculation of single transverse-spin asymmetries, $A_N$, 
for
hadronic pion production at large $x_F$.   
This calculation was based on a ``valence quark-soft gluon" 
approximation.  In this 
approximation, we kept only those contributions to $A_N$ 
proportional to the derivative of the
twist-3 quark gluon correlation function,
$(\partial/\partial x)T^{(V)}_{Fa}(x,x)$,  
where  $a$ denotes a valence quark flavor, and 
where the equal arguments in $T_F$ imply zero gluon momentum fraction. Our  
results for spin-dependent single-spin cross sections are given in 
Eq.~(\ref{s4e6}).  The ratio of the spin-dependent cross section in 
Eq.~(\ref{s4e6}) and the spin-averaged cross section in 
Eq.~(\ref{s3e8}) defines $A_N$ for
hadronic pion production.  
The spin-dependent cross section, Eq.~(\ref{s4e6}) has two types 
of contributions: quark-gluon and quark-quark (or antiquark), which 
are given by $\Delta\hat{\sigma}_{ag\rightarrow c}$ in 
Eq.~(\ref{s4e8a}) and $\Delta\hat{\sigma}_{ab\rightarrow c}$
in Eq.~(\ref{s4e8b}), respectively.  All of these
calculations are strictly leading order; we anticipate
that a large part of higher order corrections
will cancel in the asymmetry.  
Our model for the 
twist-3 matrix element $T_F$ is given in Eqs.\ (\ref{s4e3})
and (\ref{s4e4}).  We have not investigated the
evolution properties of these matrix elements here.
We expect this to be an interesting subject, but we
do not anticipate that evolution will 
require qualitative changes in our
conclusions.

Single transverse-spin asymmetries are a twist-three effect in QCD 
perturbation theory.   After taking the  ratio of Eqs.~(\ref{s4e6}) and 
(\ref{s3e8}), the asymmetry has following schematic dependence on kinematic
variables
in the large $x_F$ region,
\begin{equation}
A_N \sim \lambda \, \frac{\ell_T}{(-U)}\,
\left[ 1 + O\left(\frac{U}{T}\right) \right]\, 
\frac{1}{1-x_F} \ ,
\label{s5e1}
\end{equation}
where the invariants, $U$ and $T$ are defined in Eq.~(\ref{s3e2}).
In Eq.~(\ref{s5e1}), the prefactor $\ell_T/(-U)$ comes directly from the factor 
$\epsilon^{\ell s_T n \bar{n}}/(-\hat{u})$ in the spin-dependent 
cross section in Eq.~(\ref{s4e6}).  The combination $[1+O(U/T)]$ is  
left over from the partonic cross sections in Eq.~(\ref{s4e8}),
after the cancelation of the dominant $1/\hat{t}^2$ dependence
in the ratio.
The normalization parameter $\lambda$ comes from our model of 
the twist-three correlation functions, $T_F^{(V)}(x,x)$ in 
Eq.~(\ref{s4e3}).   Finally, the factor $1/(1-x_F)$ for
$x_F$ large is associated with $(\partial/\partial x)T^{(V)}_F(x,x)$ 
in Eq.~(\ref{s4e6}).  The approximate $1/(1-x_F)$ behavior in the ratio
of the derivative 
of the twist-three correlation function to the corresponding
twist-two parton distribution is the dominant feature of the 
twist-three asymmetry, and is responsible for the observed 
growth of $A_N$ in the large-$x_F$ region.  

The factors $\ell_T/(-U)$ and $\ell_T/(-T)$ in Eq.~(\ref{s5e1}) 
reflect the twist-three nature of the asymmetry, $A_N$.  
Combining Eqs.~(\ref{s3e2}) and (\ref{s3e3}), we express the 
invariants, $U$ and $T$, in terms of $x_F$ and $\ell_T$,
\begin{mathletters}
\label{s5e2}
\begin{eqnarray}
U&=& -\frac{S}{2}\, \left[\sqrt{x_F^2 + x_T^2} + x_F \right] 
\label{s5e2a} \\
T&=& -\frac{S}{2}\, \left[\sqrt{x_F^2 + x_T^2} - x_F \right]\, .
\label{s5e2b}
\end{eqnarray}
\end{mathletters}

\noindent
When $x_F=0$, both $U$ and $T$ are equal to $\ell_T\,\sqrt{S}$.
>From Eq.~(\ref{s5e1}), we conclude that the asymmetry at $x_F=0$
should have a very mild, 
probably linear dependence on the pion's transverse 
momentum ($A_N$ must vanish at $\ell_T=0$).  Our analytical results in 
Eq.~(\ref{s4e6}), however, are 
not accurate for the asymmetry near $x_F=0$, because of the large 
$x_F$ approximation used in our calculations.  But, from the general 
structure of the asymmetry, we believe that weak transverse 
momentum dependence at $x_F=0$ for $A_N$ should be a more general
conclusion.
 
\begin{figure}
\epsfysize=3.6in
\epsfbox{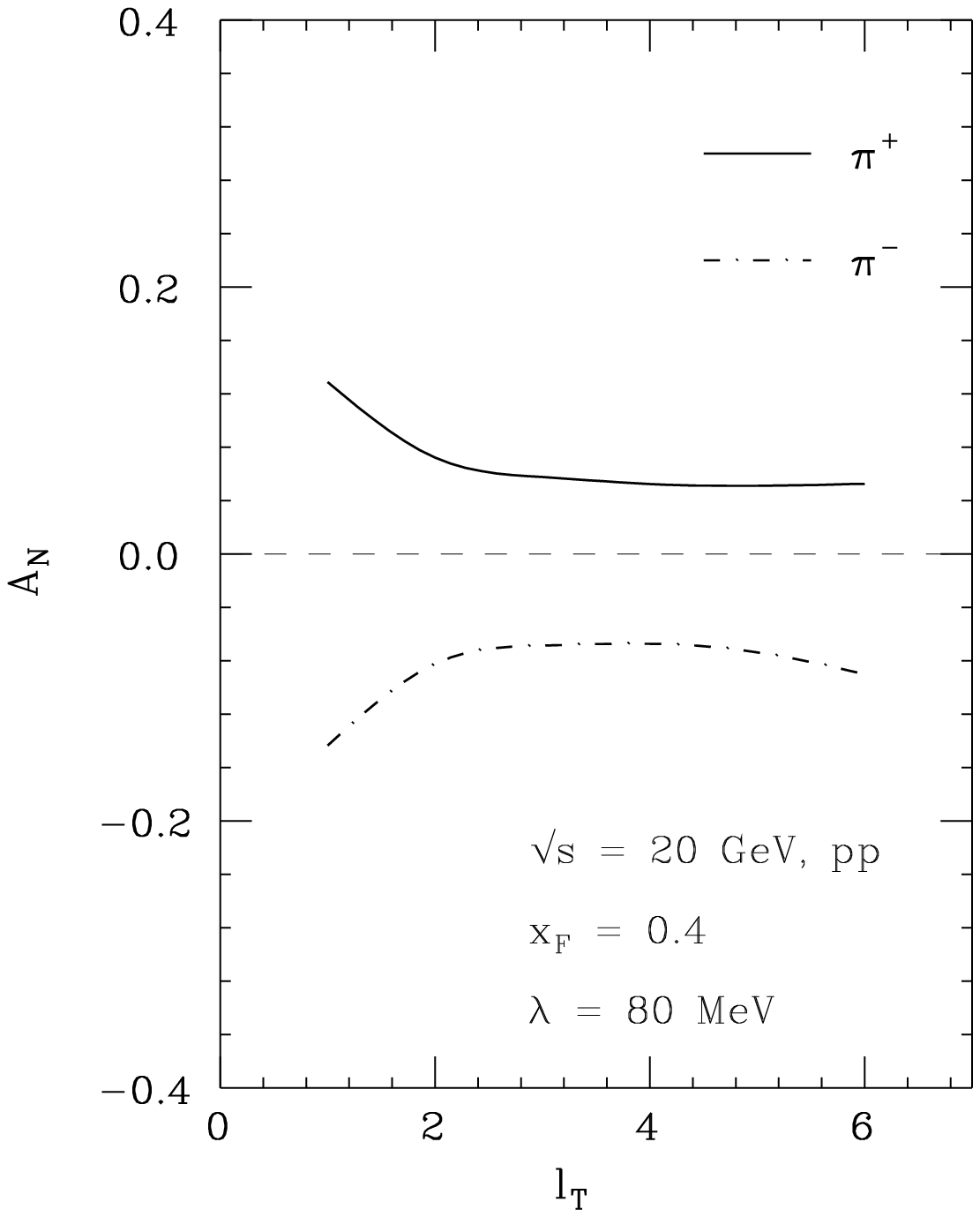}
\caption{Single transverse-spin asymmetry for $\pi^+$ and $\pi^-$ 
production with a polarized {\it proton} beam as a function of pion's 
transverse momentum $l_T$.  Theory curves are evaluated at $x_F=0.4$,
$\protect\sqrt{S}=20$ GeV 
and $\lambda=0.080$~GeV.}
\label{fig26}
\end{figure}

If $x_F\gg x_T$, the invariants $U$ and $T$ in Eq.~(\ref{s5e2})
have the following approximate dependence on $\ell_T$ and $x_F$,
\begin{mathletters}
\label{s5e3}
\begin{eqnarray}
U &\rightarrow &  - x_F\, S \ ,
\label{s5e3a} \\
T &\rightarrow &  - \frac{\ell_T^2}{x_F} \ .
\label{s5e3b}
\end{eqnarray}
\end{mathletters}
Consequently, in the large $x_F$ region, the asymmetry, $A_N$, 
will have two typical contributions, $\lambda/\ell_T$ and 
$\lambda\ell_T/S$, respectively.  If the $\lambda/\ell_T$ contribution
dominates, perturbative QCD calculations of the asymmetry may be 
relatively sensitive to  nonperturbative 
effects, because of its singular behavior at $\ell_T=0$.  On
the other hand, QCD perturbation theory may provide a reliable 
calculation of the asymmetries when the $\lambda\ell_T/S$ term is relatively 
important.  
In Fig.~\ref{fig26}, we plot the transverse momentum dependence 
of the asymmetry at $x_F=0.4$, where most data were collected.  
The asymmetries for both $\pi^+$ and $\pi^-$ have a quite weak
dependence on pion's transverse momentum for $\ell_T > 2$~GeV.  This 
suggests that perturbative calculations 
for the asymmetries are reliable for a wide range of 
the experimental kinematics.
 
\begin{figure}
\epsfysize=3.6in
\epsfbox{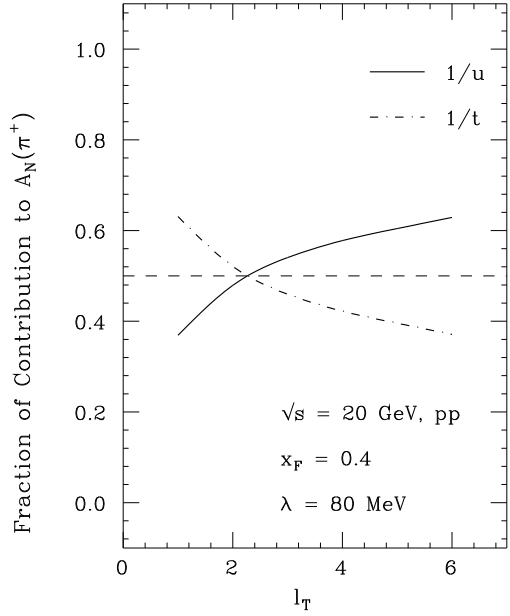}
\caption{Fractional contribution from $1/(-U)$ and $1/(-T)$ terms to
the single transverse-spin asymmetry of $\pi^+$ production as a 
function of pion transverse momentum.  Theory curves are evaluated 
at $x_F=0.4$, $\protect\sqrt{S}=20$ GeV  and $\lambda=0.080$~GeV.}
\label{fig27}
\end{figure}

The remarkable feature of mild transverse momentum
dependence,  shown in Fig.~\ref{fig26}, can be easily 
traced to Eq.~(\ref{s4e8}).  For the quark-gluon 
subprocess, once the dominant $1/\hat{t}^2$ dependence has
canceled in the asymmetry, the
coefficient of $\hat{u}/\hat{t}$ is much smaller than
the corresponding constant term.  Similarly, for quark-quark and
antiquark subprocess, the coefficient of $\hat{u}/\hat{t}$ is also
much smaller than the constant term, except for terms 
proportional to $1/\hat{u}^2$ and $1/\hat{s}^2$.  The latter, however,   
are suppressed by $\hat{t}^2/\hat{s}^2$
relative to the leading terms in the forward region.  In 
summary, the small coefficients for $\hat{u}/\hat{t}$ terms
assure that the $\ell_T/(-T)$ dependence in Eq.~(\ref{s5e1}) does not dominate
the $\ell_T/(-U)$ dependence.  We verify this conclusion by plotting
the fractional contributions to the $\pi^+$ asymmetry from
$1/(-U)$ term and $1/(-T)$ term, respectively, 
as a function of pion's transverse momentum in 
Fig.~\ref{fig27}.  It is evident that contribution from $1/(-U)$ term 
is comparable with $1/(-T)$ term for the region of our interest.

\begin{figure}
\epsfysize=3.6in
\epsfbox{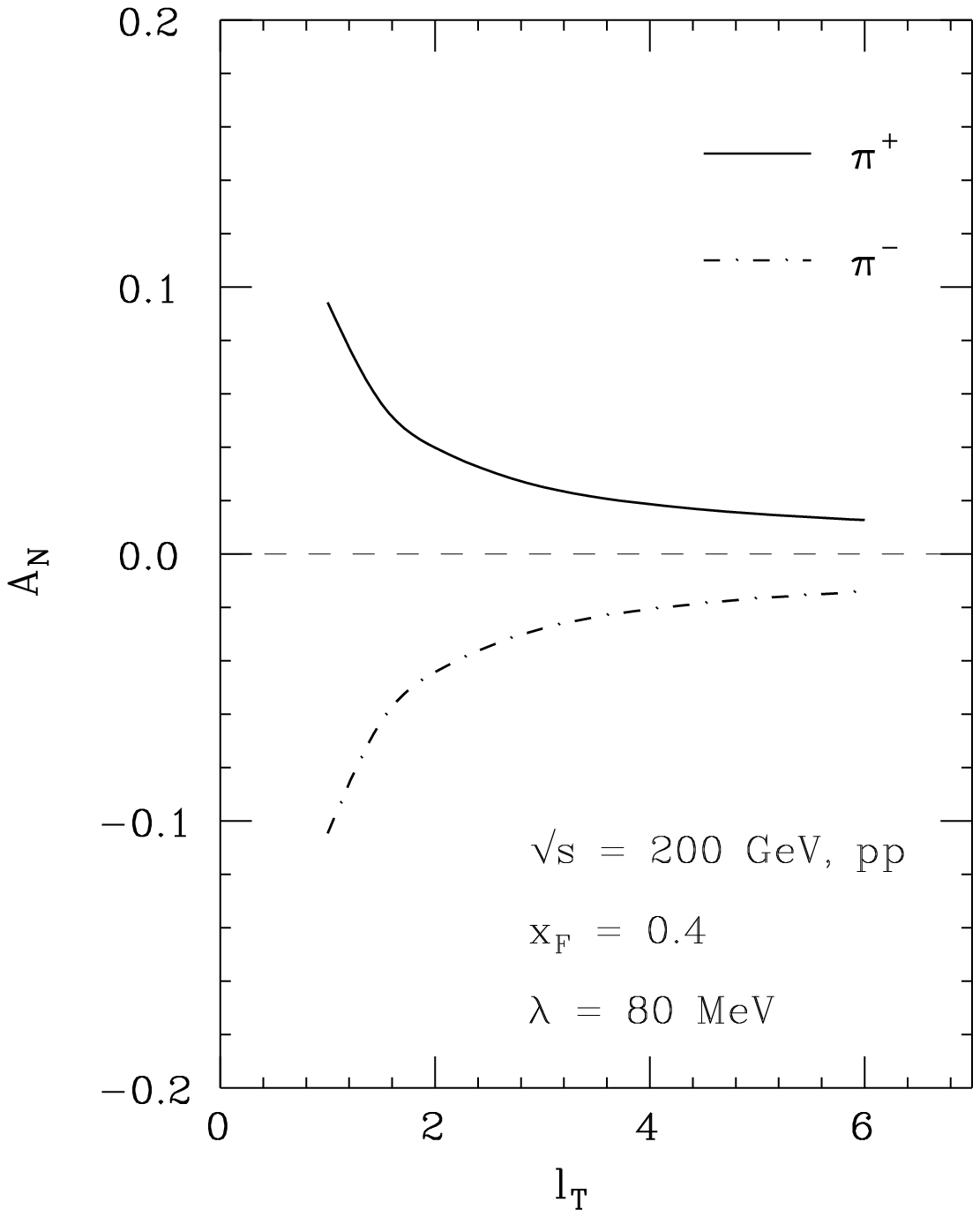}
\caption{Single transverse-spin asymmetry for $\pi^+$, $\pi^-$ and $\pi^0$ 
production with a polarized proton beam, as a function of pion
transverse momentum $l_T$.  Theory curves are evaluated at $x_F=0.4$,
$\protect\sqrt{S}=200$ GeV 
and $\lambda=0.0.80$~GeV.}
\label{fig28}
\end{figure}

\begin{figure}
\epsfysize=3.6in
\epsfbox{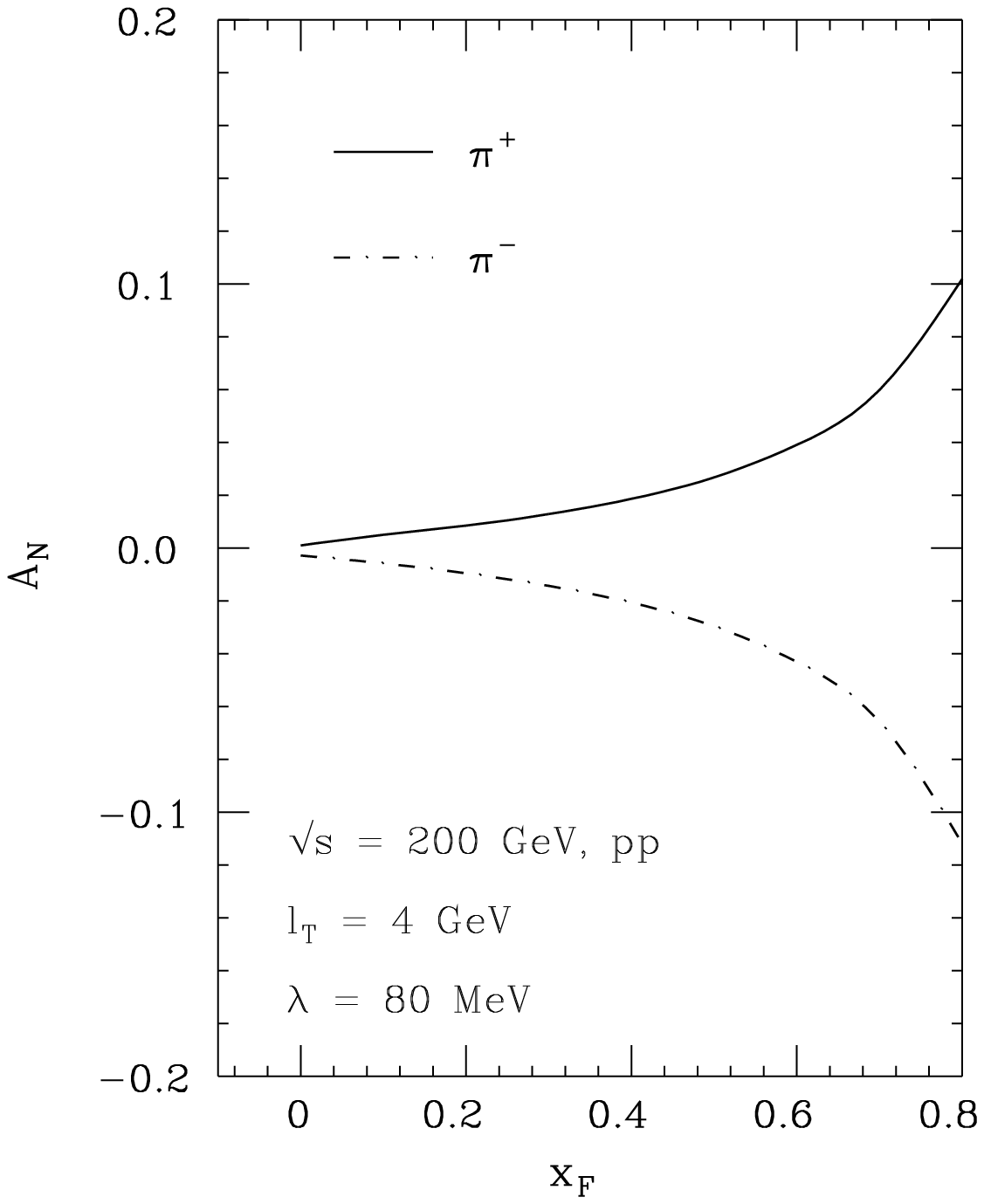}
\caption{Single transverse-spin asymmetry for $\pi^+$, $\pi^-$ and $\pi^0$
production with a polarized proton beam, as a function of pion
transverse momentum $l_T$.  Theory curves are evaluated at $x_F=0.4$,
$\protect\sqrt{S}=200$ GeV
and $\lambda=0.080$~GeV.}
\label{fig29}
\end{figure}

If $x_F \rightarrow 1$, or $U/T\sim x_F^2\,S/\ell_T^2 \gg 1$, the 
asymmetry will be eventually dominated by the $\lambda/\ell_T$ terms.  
Therefore, the asymmetry will scale with $1/\ell_T$ in this 
region.  For the kinematics of the Fermilab data, this
scaling region is not yet reached.  In Fig.~\ref{fig26}, the steep
increase of the asymmetries for $\ell_T < 2$~GeV indicates the 
dominance of the $\lambda/\ell_T$ contribution, and probably signals that 
the perturbative calculations are relatively less reliable if $\ell_T$ is much 
less than 2~GeV.  The slight increase when $\ell_T \rightarrow 
6$~GeV signals an effect of the edge of phase space.  
Clearly, the high energies of the polarized RHIC proton
beam would make it possible to check these predictions.
In Fig.\ \ref{fig28}, we show the $\ell_T$-dependence of
$A_N$ for $x_F=0.4$ at $\sqrt{S}=200$ GeV.  Compared to
Fig.\ \ref{fig26} at Fermilab energies, the $\ell_T/U$ term 
is relatively suppressed, and the model predicts a
steeper $\ell_T$-dependence and, in general, a smaller,
but still substantial, asymmetry.  Fig.\ \ref{fig29} shows the
asymmetry as a function of $x_F$ at $\ell_T=4$ GeV.
These are examples only; the model can be used to
predict $A_N$ over any kinematic range that is experimentally
convenient, so long as it is in the forward region.

In summary, we have calculated the single 
transverse-spin asymmetry for hadronic pion production
in perturbative QCD.  With only
one normalization parameter $\lambda$ and a relative sign of 
polarized twist-3 valence quark distributions, our numerical results
are consistent with Fermilab data on the asymmetry for both
the sign and shape, as well as relative normalizations.  
In addition, we demonstrated that perturbative calculation of the 
asymmetries is applicable even for pion momenta as small as 
a few GeV.  This conclusion is very encouraging for future 
applications of perturbative QCD beyond the leading twist.  Our method can be 
easily generalized 
to calculate the single transverse-spin asymmetries for inclusive 
production of other particles.  
The planned polarized beam at RHIC
affords an exciting opportunity to test these, and related
ideas on the spin structure of the nucleon.


\section*{Acknowledgments}

We thank Aki Yokosawa for helpful communications regarding 
the Fermilab data, and R.L.\ Jaffe and P.\ Ratcliffe 
for useful conversations.  This work was supported in part by 
the U.S. Department of Energy under Grant Nos. DE-FG02-87ER40731
and DE-AC02-98CH10886 and by the National Science Foundation,
under grant PHY9722101.


\section*{Appendix}

In this appendix, we identify twist-3 distributions and fragmentation 
functions that can
contribute to the sums in the collinear expansion, Eq.\ (\ref{s3e11}).
The factorization in (\ref{s3e11}) enables us to 
apply parity and time-reversal (PT) invariance
to hadron-hadron scattering in a manner similar to their
classic application to inclusive DIS, reviewed in Sec.\ \ref{sec:2a}.
Thus, it will be natural to
study the symmetry properties of possible matrix elements.

We will identify terms of the type 
discussed in connection with Eq.\ (\ref{s3e21}), that is, with
integrals over two quark momentum fractions, $x_1$ and $x_2$.
Furthermore, we require that $x_1$ be set equal to $x_2$
by a ``gluonic" pole at $x_1=x_2$ in the hard  
scattering (see Eq.\ (\ref{s3e33})), in accordance with
our valence quark-soft gluon approximation.
Let us concentrate first on parton distributions,
 and return at the end to
fragmentation functions.

\subsection*{Twist-3 Distributions}

As mentioned in Sec.\ \ref{sec:2c}, the derivation of terms
in Eq.\ (\ref{s3e11}) involving quarks requires a Fierz 
projection of the Dirac indices linking the distribution
or fragmentation function and the hard scattering.  
A schematic illustration was given in Fig.\ \ref{fig8}.
The collinear expansion
then isolates twist-three 
fermion matrix elements with two quark fields 
and either a  covariant derivative or a field strength.
It will be convenient to start by discussing the
expectation values of combinations of these fields in position space.
We thus introduce
\begin{eqnarray}
D_\Gamma^i(y_1,y_2,s)
&=&
\langle P,s|\; \bar{\psi}(0)\; \Gamma\; D^i(y_2)\; \psi(y_1)\; |P,s\rangle
\label{Dgamdef}\\
F_\Gamma^i(y_1,y_2,s)
&=&
\langle P,s|\; \bar{\psi}(0)\; \Gamma\; 
n_\mu F^{i\mu}(y_2)\; \psi(y_1)\; |P,s\rangle\, ,
\label{Egamdef}
\end{eqnarray}
with $\Gamma$ a Dirac matrix.  We define $D^i\equiv i\partial^i-gA^i$,
and we adopt the kinematics and notation of Sec.\ \ref{sec:2a};
in particular, $n^\mu$ is defined in Eq.\ (\ref{s3e1}).  
In these matrix elements, the index $i$ is assumed
to be transverse.  This alone is enough to make the matrix element 
twist-3; the Dirac projection must not 
raise the twist further.  The relevant terms in the 
Fierz projection between the distribution for a hadron of momentum
$P^\mu=\bar{n}^\mu \sqrt{S/2}$ and the hard
scattering are then given by
\begin{eqnarray}
\delta_{aa'}\delta_{bb'}
&=&
{1\over 4}
(\gamma\cdot n)_{ab} (\gamma\cdot\bar{n})_{b'a'} +
{1\over 4}(\gamma\cdot n \gamma_5)_{ab} 
          (\gamma_5 \gamma\cdot\bar{n})_{b'a'} 
\nonumber\\
&\ & \quad +
{1\over 4}\sum_\beta 
  \left( (n\sigma)_\beta\right)_{ab}
  \left((\bar{n}\sigma)^\beta\right)_{b'a'}+\dots \, ,
\label{Fierz}
\end{eqnarray}
where omitted terms raise the twist, and where we define
\begin{equation}
(n\sigma)^\beta
\equiv
n_\mu\sigma^{\mu\beta}\, .
\label{nsigdef}
\end{equation}
For an opposite-moving hadron, with momentum 
$P'^{\mu}=n^\mu \sqrt{S/2}$,
we exchange the roles of $n^\mu$ and $\bar{n}^\mu$. The
matrices above have the properties
\begin{eqnarray}
\Gamma
&=&
\gamma^0\; \Gamma^\dagger\; \gamma^0\, ,
\label{Epsdef}\\
\Gamma
&=&
\delta_\Gamma\; \left({\cal T}\Gamma^*{\cal T}\right)^\dagger\, ,
\label{Deltadef}
\end{eqnarray}
with $\delta_\Gamma=\pm1$,
where  ${\cal T}\equiv i\gamma^1\gamma^3={\cal T}^{-1}$ is a time-reversal 
matrix that
acts as
\begin{equation}
{\cal T}\left(\gamma^\mu\right)^*{\cal T}=\gamma_\mu\, .
\label{calTact}
\end{equation}
Specifically, for the vector, axial-vector and tensor cases we have
\begin{eqnarray}
&n\cdot\gamma:  \quad \delta_{n\cdot \gamma}=1
\label{gam}
\\
&n\cdot\gamma\gamma_5: \quad \delta_{n\cdot \gamma\gamma_5}=-1
\label{gamgam5}
\\
&(n\sigma)^{\nu}:  \quad
\delta_{(n\sigma)^\nu}=-1\, .
\label{nsiggam5}
\end{eqnarray}
>From the expectation values $F^i_\Gamma$ and $D^i_\Gamma$ we define
parton distributions by Fourier transforms with respect to 
light-cone momenta, and if desired transverse momenta as well,
\begin{eqnarray}
t_\Gamma^{(D)i}(x_1,{\bf k}_1,x_2,{\bf k}_2,s)
&=&
\int dy_1dy_2\; {\rm e}^{ik_1\cdot y_1+i(k_2-k_1)\cdot y_2}
D_\Gamma^i(y_1,y_2,s)\, ,
\label{MfromD}\\
t_\Gamma^{(F)i}(x_1,{\bf k}_1,x_2,{\bf k}_2,s)
&=&
\int dy_1dy_2\; {\rm e}^{ik_1\cdot y_1+i(k_2-k_1)\cdot y_2}
F_\Gamma^i(y_1,y_2,s)\, ,
\label{NfromE}
\end{eqnarray}
where we define $dy_i\equiv dy^-d^2 {\bf y}$, with $\bf y$ a two-dimensional
transverse vector,  and $k_i\cdot y\equiv x_ipy^--{\bf k}\cdot {\bf y}$.
In the following, we study constraints on spin-dependence that
follow from the reality and symmetry properties of these matrix
elements in QCD.  This will enable us to identify the relevant
contributions to the sums in Eq.\ (\ref{s3e11}).

\subsection*{Reality and Symmetry}

The reality properties of the matrix elements (\ref{Dgamdef}) and
(\ref{Egamdef}) are conveniently expressed as
\begin{eqnarray}
\left[\; D_\Gamma^i(-y_1,-y_2,s)\; \right ]^*
&=&
D_\Gamma^i(y_1,y_1-y_2,s)\, ,
\label{RealforD}\\
\left[\; F_\Gamma^i(-y_1,-y_2,s)\; \right]^*
&=&
F_\Gamma^i(y_1,y_1-y_2,s)\, ,
\end{eqnarray}
which relate, of course, expectation values with the same spins.
 Invariance under time reversal  and parity, on
the other hand imply that
\begin{eqnarray}
D_\Gamma^i(y_1,y_2,s)
&=&
\delta_\Gamma\; D_\Gamma^i(y_1,y_1-y_2,-s)\, ,
\label{PTforD}\\
F_\Gamma^i(y_1,y_2,s)
&=&
-\delta_\Gamma F_\Gamma^i(y_1,y_1-y_2,-s)\, ,
\end{eqnarray}
in which spins are reversed.

Relations for parton distributions $t_\Gamma^{(D)i}$ and $t_\Gamma^{(F)i}$ are 
easy to derive by
inserting the reality and symmetry relations into the Fourier transforms
of Eqs.\ (\ref{MfromD}) and (\ref{NfromE}), and changing integration
variables.  Because in this paper we are concentrating on the collinear
expansion, with convolutions in light-cone momenta only, we shall
suppress transverse momenta in the arguments of the distributions, and
exhibit only the momentum fraction variables $x_i$ in the following
formulas.  Relations for transverse-momentum distributions are
found by simply reinserting the ${\bf k}_i$ arguments, alongside
the corresponding momentum fractions.
With this understood, the reality conditions give
\begin{eqnarray}
\left[\; t_\Gamma^{(D)i}(x_1,x_2,s)\; \right]^*
&=&
t_\Gamma^{(D)i}(x_2,x_1,s)\, ,
\label{RealforM}\\
\left[\; t_\Gamma^{(F)i}(x_1,x_2,s)\; \right]^*
&=&
t_\Gamma^{(F)i}(x_2,x_1,s)\, ,
\label{RealforN}
\end{eqnarray}
in which we note that  the momentum arguments are exchanged.
Referring to Eqs.\ (\ref{gam}-\ref{nsiggam5}) above, we
see that the even parts of the twist-3 distributions 
are real, the odd parts imaginary.  

Similarly, from PT invariance, we find
\begin{eqnarray}
t_\Gamma^{(D)i}(x_1,x_2,s)
&=&
\delta_\Gamma\; t_\Gamma^{(D)i}(x_2,x_1,-s)\, ,
\label{PTforM}\\
t_\Gamma^{(F)i}(x_1,x_2,s)
&=&
-\delta_\Gamma\; 
t_\Gamma^{(F)i}(x_2,x_1,-s)\, ,
\label{PTforN}
\end{eqnarray}
with $\delta_\Gamma$ defined in Eq.\ (\ref{Deltadef}).
Note the extra minus sign in the second case, which reflects
the PT properties of the field strength tensor.   

>From Eqs.\ (\ref{RealforM}-\ref{PTforN}), we can derive
the constraints on the spin-averaged,
\begin{equation}
\langle t_\Gamma^{(O)i}\rangle(x_1,x_2)
\equiv
{1\over 2}\; \left[ t_\Gamma^{(O)i}(x_1,x_2,s) + 
t_\Gamma^{(O)i}(x_1,x_2,-s)\right]
\label{spinavg}
\end{equation}
and spin-dependent
\begin{equation}
\Delta t_\Gamma^{(O)i}(x_1,x_2)
\equiv
{1\over 2}\; \left[ t_\Gamma^{(O)i}(x_1,x_2,s) - 
t_\Gamma^{(O)i}(x_1,x_2,-s)\right]
\label{spindpt}
\end{equation}
distributions for each choice of operator $O=D,F$ and Dirac structure 
$\Gamma$.  Specifically, the spin-dependent distributions
$\Delta t_{n\cdot\gamma}^{(D)i}$, $\Delta t_{n\cdot\gamma\gamma_5}^{(F)i}$
and $\Delta t_{(n\sigma)^j}^{(F)i}$ 
and the spin-averaged distributions $\langle 
t_{n\cdot\gamma\gamma_5}^{(D)i}\rangle$,
$\langle t_{(n\sigma)^j}^{(D)i}\rangle$ and $\langle 
t_{n\cdot\gamma}^{(F)i}\rangle$
are imaginary and vanish
at $x_1=x_2$.  They therefore cannot be associated with gluon poles
in Eq.\ (\ref{s3e11}), and are nonleading in the
valence quark-soft gluon approximation introduced in Sec.\ \ref{sec:2b}.

\subsection*{Leading Terms at Twist-3}

The remaining distributions are real and nonzero at $x_1=x_2$ in general.
For the first sum in Eq.\ (\ref{s3e11}), we need a real, chiral-even, 
spin-dependent 
parton distribution. 
The only one is $\Delta t_{n\cdot\gamma}^{(F)i}(x_1,x_2)$,
which is equal, up to  a constant, to $T_F^{(V)}$, Eq.\ (\ref{s3e18}),
\begin{equation}
\Delta t_{n\cdot\gamma}^{(F)i}(x_1,x_2)
=
-4\pi\epsilon^{n\bar{n}is}\; T_F^{(V)}(x_1,x_2)\, ,
\label{normtngam}
\end{equation}
where the tensor structure follows from parity
invariance applied to the matrix element.

For the second sum in Eq.\ (\ref{s3e11}),
we need a chiral-odd {\it spin-averaged} distribution,
to give a nonzero trace in the hard-scattering amplitude
when paired with the transversity distribution \cite{Jaffe:Ji},
\begin{equation}
\delta q(x)=\int{dy^-\over 2\pi}\; {\rm e}^{ixP^+y^-}\; 
\langle P,s|\bar \psi (0)\; {i\over 2}(n\sigma)^is_i\gamma_5\; 
\psi(y^-n)\; |P,s\rangle\, .
\label{transversity}
\end{equation}  
Here again there is only a single contribution, $\langle t_{(\bar 
n\sigma)^j}^{(F)i}\rangle(x_1,x_2)$.
Parity invariance implies that
$\langle t_{(\bar n\sigma)^j}^{(F)i}\rangle(x_1,x_2)$ is of the form
\begin{equation}
\langle t_{(\bar n\sigma)^j}^{(F)i}\rangle(x_1,x_2) 
=
4\pi{\delta_{ij}\over 2}\; T_F^{(\sigma)}(x_1,x_2)\, , 
\end{equation}
where the scalar distribution $T_F^{(\sigma)}$ is defined
by analogy to  $T_F^{(V)}$, Eq.\ (\ref{s3e18}), 
\begin{eqnarray}
T_F^{(\sigma)}(x_1,x_2)
&=&
\int {dy_1^+dy_2^+\over 4\pi}\; 
{\rm e}^{ix_1 P'{}^-y_1^++i(x_2-x_1)P'{}^-y_2^+}
\nonumber\\
&\ & \quad \times
{1\over 2}\sum_{s'}\langle P',s'|\; \bar \psi(0)\; (\bar n \sigma)_{j}\; \bar 
n_\rho F^{j\rho}(y_2^+)\;
\psi(y_1^+)\; |P',s'\rangle\, .
\label{Tsigdef}
\end{eqnarray}
In these expressions we have taken $P'{}^\mu$ in the minus-$z$ direction, in 
accordance
with the kinematics of the unpolarized hadron in Sec.\ \ref{sec:2a}.
In Ref.\ \cite{AN:QS} the possibility of such a term was noted.

\subsection*{Fragmentation at Twist-3}

Turning to the third term in Eq.\ (\ref{s3e11}), we must
deal with twist-3 chiral-even fragmentation functions, which are transforms of
matrix elements of the general form
\begin{eqnarray}
\bar d^{(\sigma)}(y_1,y_2,\ell)
&=&
 \sum_X\, {\rm Tr}\; \left[\; (n_\ell\sigma)_i\;
\langle 0|\; \bar{\psi}(0)\; |\ell,X\rangle \langle \ell,X|\; D^i(y_2)\; 
\psi(y_1)\; |0\rangle\; 
\right]
\nonumber\\
\label{bardsigdef}\\
\bar f^{(\sigma)}(y_1,y_2,\ell)
&=&
 \sum_X\, {\rm Tr}\; \left[\; (n_\ell\sigma)_i\;
\langle 0|\; \bar{\psi}(0)\; |\ell,X\rangle \langle \ell,X|\; n_\ell{}_\mu 
F^{\mu i}(y_2)\; \psi(y_1)\; |0\rangle\;
\right]\, ,
\nonumber\\
\label{barfsigdef}
\end{eqnarray}
with the sum over inclusive final (out) states $|X,\ell\rangle$, where $\ell$ 
is the momentum of the observed
particle. The vector $n_\ell^\mu$ is defined by analogy to $n^\mu$ in Eq.\ 
(\ref{s3e1}),
as a lightlike velocity vector in the direction opposite to $\bar n_\ell\equiv 
\ell^\mu/\ell_0$.
The trace is over Dirac indices.
There is no analog of the spin variable in this case,
although extensions to production of polarized particles \cite{EXP:Lambda} 
should 
be straightforward. 
We have used the constraints of parity in forming scalar fragmentation 
functions,
depending on two momentum fractions.
In momentum space they are
\begin{eqnarray}
 d^{(\sigma)}(z_1,z_2)
&=&
\int {dy_1dy_2 \over 4\pi} {\rm e}^{-i\ell\cdot ny_1/z_1-i\ell\cdot 
ny_2(1/z_2-1/z_1)}\; 
\bar d^{(\sigma)}(y_1,y_2,\ell)
\label{dGammadef}
\\
 f^{(\sigma)}(z_1,z_2)
&=&
\int {dy_1dy_2 \over 4\pi} {\rm e}^{-i\ell\cdot ny_1/z_1-i\ell\cdot 
ny_2(1/z_2-1/z_1)}\; 
\bar f^{(\sigma)}(y_1,y_2,\ell)\, .
\label{fGammadef}
\end{eqnarray}

The
constraints of reality are different for these fragmentation functions
than for the distributions,  because the sums over states in Eqs\ 
(\ref{bardsigdef})
and (\ref{barfsigdef}) are incomplete.  We find that
\begin{eqnarray}
\bar d^{(\sigma)}{}^*(-y_1,-y_2,\ell)
&=&
  \sum_X\, {\rm Tr}\; \big[\; (\bar n_\ell\sigma)_i\;
\langle 0|\; \bar{\psi}(0)\; D^i(y_1-y_2)\; |\ell,X\rangle 
 \langle \ell,X|\; \psi(y_1)\; |0\rangle\; 
\big]
\nonumber\\
\label{dsigdef}\\
\bar f^{(\sigma)}{}^*(-y_1,-y_2,\ell)
&=&
 \sum_X\, {\rm Tr}\; \big[\; (\bar n_\ell\sigma)_i\;
\langle 0|\; \bar{\psi}(0)\; \bar n_\mu F^{\mu i}(y_1-y_2)\;  |\ell,X\rangle 
\langle \ell,X|\;  \psi(y_1)\; |0\rangle\;
\big]\, .
\nonumber\\
\label{fsigdef}
\end{eqnarray}
As Collins has emphasized \cite{FRAG:Collins},  time-reversal 
does not constrain fragmentation functions in the same
manner as distributions, because T reverses
the roles of in and out states.   To the extent
that a sum over in and out states is the same
in these functions,   symmetry under PT would
imply that $\bar d^{(\sigma)}$ is purely imaginary, while
 $\bar f^{(\sigma)}$ is real.  
These properties can, however, be modified
by phases associated with final state interactions.  Indeed, this is the 
mechanism
by which Artru {\it et al.} \cite{AN:Artru} derive single-spin asymmetries 
starting from 
a model for fragmentation
functions with intrinsic transverse momenta.  Such functions 
can be thought of as extensions of 
$\bar d^{(\sigma)}$,  finite distances from the light cone.
Following the procedure of Sec.\ \ref{sec:3} above,
we can derive hard-scattering coefficients for either function.
We reserve this for future investigation.



\newpage

\begin{table}
\caption{Partonic hard parts and corresponding color factors for
quark-gluon and antiquark-gluon (indicated by
barred letters) subprocesses.  Feynman diagrams are shown in 
Fig.~\protect\ref{fig15}.  $C_g$, $C^I_g$, and $C^F_g$ are color factors
for spin-averaged, spin-dependent with initial-state interaction, and 
spin-dependent with final-state interaction subprocess, respectively.  
The explicit factors (-1) are due to the sign difference 
between
quark and antiquark propagators with
the same momentum.  Calculations were done in Feynman 
gauge.} 

\label{table1} 
\begin{tabular}{c|cc|cc|cc|cc}
Diagram & Partonic Parts && $C_g$ && $C^I_g$ && $C^F_g$ &\\ \hline
(a) & $4\left[1-\frac{\hat{s}\hat{u}}{\hat{t}^2}\right]$ &
    & $\frac{1}{2}$ &
    & $-\frac{N^2}{4(N^2-1)}$ &
    & $-\frac{1}{2(N^2-1)}$ &
    \\ \hline
(b) & $2\left[\frac{-\hat{u}}{\hat{s}}\right]$ & 
    & $\frac{N^2-1}{4N^2}$ &
    & $-\frac{1}{4}$ &
    & $\frac{1}{4N^2(N^2-1)}$ & 
    \\ \hline
(c) & $2\left[\frac{\hat{s}}{-\hat{u}}\right]$ &
    & $\frac{N^2-1}{4N^2}$ &
    & $\frac{1}{4(N^2-1)}$ &
    & $\frac{1}{4N^2(N^2-1)}$  & 
    \\ \hline
(d) & $(+i)2\left[\frac{\hat{s}}{\hat{t}}\right]$ & 
    & $(-i)\frac{1}{4}$ &
    & $(+i)\frac{N^2}{4(N^2-1)}$ &
    & $(+i)\frac{1}{4(N^2-1)}$  &
    \\ \hline
(e) & $(-i)2\left[\frac{\hat{s}}{\hat{t}}\right]$ & 
    & $(+i)\frac{1}{4}$ &
    & $(-i)\frac{N^2}{4(N^2-1)}$ &
    & $(-i)\frac{1}{4(N^2-1)}$ &
    \\ \hline
(f) & $(-i)2\left[\frac{\hat{u}}{\hat{t}}\right]$ & 
    & $(+i)\frac{1}{4}$ &
    & $ 0 $ &
    & $(-i)\frac{1}{4(N^2-1)}$ &
    \\ \hline
(g) & $(+i)2\left[\frac{\hat{u}}{\hat{t}}\right]$ & 
    & $(-i)\frac{1}{4}$ &
    & $ 0 $ &
    & $(+i)\frac{1}{4(N^2-1)}$ & 
    \\ \hline
(h) & $ 0 $ &
    & $ - $ &
    & $ - $ &
    & $ - $ &
    \\ \hline
(i) & $ 0 $ &
    & $ - $ &
    & $ - $ &
    & $ - $ &
    \\ \hline
($\bar{\rm a}$) 
  & $4\left[1-\frac{\hat{s}\hat{u}}{\hat{t}^2}\right]$ &
    & $\frac{1}{2}$ &
    & $\frac{N^2}{4(N^2-1)}$ &
    & $-\frac{1}{2(N^2-1)}\,(-1)$ &
    \\ \hline
($\bar{\rm b}$) 
& $2\left[\frac{-\hat{u}}{\hat{s}}\right]$ & 
    & $\frac{N^2-1}{4N^2}$ &
    & $\frac{1}{4}$ &
    & $\frac{1}{4N^2(N^2-1)}\,(-1)$ & 
    \\ \hline
($\bar{\rm c}$) 
    & $2\left[\frac{\hat{s}}{-\hat{u}}\right]$ &
    & $\frac{N^2-1}{4N^2}$ &
    & $-\frac{1}{4(N^2-1)}$ &
    & $\frac{1}{4N^2(N^2-1)}\,(-1)$  & 
    \\ \hline
($\bar{\rm d}$) 
    & $(-i)2\left[\frac{\hat{s}}{\hat{t}}\right]$ & 
    & $(+i)\frac{1}{4}$ &
    & $(+i)\frac{N^2}{4(N^2-1)}$ &
    & $(-i)\frac{1}{4(N^2-1)}\,(-1)$  &
    \\ \hline
($\bar{\rm e}$) 
    & $(+i)2\left[\frac{\hat{s}}{\hat{t}}\right]$ & 
    & $(-i)\frac{1}{4}$ &
    & $(-i)\frac{N^2}{4(N^2-1)}$ &
    & $(+i)\frac{1}{4(N^2-1)}\,(-1)$ &
    \\ \hline
($\bar{\rm f}$) 
    & $(+i)2\left[\frac{\hat{u}}{\hat{t}}\right]$ & 
    & $(-i)\frac{1}{4}$ &
    & $ 0 $ &
    & $(+i)\frac{1}{4(N^2-1)}\,(-1)$ &
    \\ \hline
($\bar{\rm g}$) 
    & $(-i)2\left[\frac{\hat{u}}{\hat{t}}\right]$ & 
    & $(+i)\frac{1}{4}$ &
    & $ 0 $ &
    & $(-i)\frac{1}{4(N^2-1)}\,(-1)$ & 
    \\ \hline
($\bar{\rm h}$) 
    & $ 0 $ &
    & $ - $ &
    & $ - $ &
    & $ - $ &
    \\ \hline
($\bar{\rm i}$) 
    & $ 0 $ &
    & $ - $ &
    & $ - $ &
    & $ - $ &
\end{tabular}
\end{table}

\newpage

\begin{table}
\caption{Partonic hard parts and corresponding color factors for
subprocesses involving quarks and/or antiquarks.  Feynman diagrams are shown in 
Fig.~\protect\ref{fig17}. 
In diagrams ($\bar{\rm i}$) and ($\bar{\rm j}$) both fermion arrows
have been reversed, relative to (i) and (j).
 $C_q$, $C^I_q$, and $C^F_q$ are color factors
for spin-averaged, spin-dependent with initial-state interaction, and 
spin-dependent with final-state interaction, respectively.  Flavor 
indices $a$ and $b$ correspond to the flavor of the quark (or antiquark)
from the polarized hadron and unpolarized hadron, respectively, and $c$
is the flavor of fragmenting quark.  
The explicit factors (-1) are due to the sign difference between
quark and antiquark propagators with
the same momentum.  Calculations were done in Feynman 
gauge.}

\label{table2} 
\begin{tabular}{c|cc|cc|cc|cc}
Diagrams & Partonic Parts && $C_q$ && $C^I_q$ && $C^F_q$ &\\ \hline
(a)   & $2\left[\frac{\hat{s}^2+\hat{u}^2}{\hat{t}^2}\right]
        \delta_{ac}$ &
      & $\frac{N^2-1}{4N^2}$ &
      & $\frac{N^2-4N-4}{32N}$ &
      & $-\frac{1}{4N^2}$ &
      \\ \hline
(b)   & $2\left[\frac{\hat{s}^2+\hat{u}^2}{\hat{t}^2}\right]
      \delta_{ac}$ &
      & $\frac{N^2-1}{4N^2}$ &
      & $\frac{N^2+4N-4}{32N}\,(-1)$ &
      & $-\frac{1}{4N^2}$ &
      \\ \hline
(c)   & $2\left[\frac{\hat{s}^2+\hat{u}^2}{\hat{t}^2}\right]
      \delta_{ac}$ &
      & $\frac{N^2-1}{4N^2}$ &
      & $\frac{N^2+4N-4}{32N}$ &
      & $-\frac{1}{4N^2}\,(-1)$ &
      \\ \hline
(d)   & $2\left[\frac{\hat{s}^2+\hat{u}^2}{\hat{t}^2}\right]
        \delta_{ac}$ &
      & $\frac{N^2-1}{4N^2}$ &
      & $\frac{N^2-4N-4}{32N}\,(-1)$ &
      & $-\frac{1}{4N^2}\,(-1)$ &
      \\ \hline
(e)   & $2\left[\frac{\hat{s}^2+\hat{t}^2}{\hat{u}^2}\right]
      \delta_{bc}$ &
      & $\frac{N^2-1}{4N^2}$ &
      & $\frac{N^2-4N-4}{32N}$ &
      & $\frac{N^2+4N-4}{32N}$ &
      \\ \hline
(f)   & $2\left[\frac{\hat{s}^2+\hat{t}^2}{\hat{u}^2}\right]
      \delta_{bc}$ &
      & $\frac{N^2-1}{4N^2}$ &
      & $\frac{N^2+4N-4}{32N}$ &
      & $\frac{N^2-4N-4}{32N}$ &
      \\ \hline
(g)   & $2\left[\frac{\hat{s}^2+\hat{t}^2}{\hat{u}^2}\right]
      \delta_{bc}$ &
      & $\frac{N^2-1}{4N^2}$ &
      & $\frac{N^2+4N-4}{32N}\,(-1)$ &
      & $\frac{N^2-4N-4}{32N}\,(-1)$ &
      \\ \hline
(h)   & $2\left[\frac{\hat{s}^2+\hat{t}^2}{\hat{u}^2}\right]
      \delta_{bc}$ &
      & $\frac{N^2-1}{4N^2}$ &
      & $\frac{N^2-4N-4}{32N}\,(-1)$ &
      & $\frac{N^2+4N-4}{32N}\,(-1)$ &
      \\ \hline
(i)   & $2\left[\frac{\hat{s}^2}{\hat{t}\,\hat{u}}\right]
      \delta_{ab}\delta_{ac}$ &
      & $-\frac{N^2-1}{4N^3}$ &
      & $\frac{N^2+1}{4N^3}$ &
      & $\frac{1}{4N^3}$ &
      \\ \hline
(j)   & $2\left[\frac{\hat{s}^2}{\hat{t}\,\hat{u}}\right]
      \delta_{ab}\delta_{bc}$ &
      & $-\frac{N^2-1}{4N^3}$ &
      & $\frac{N^2+1}{4N^3}$ &
      & $\frac{1}{4N^3}$ &
      \\ \hline
($\bar{\rm i}$)   
      & $2\left[\frac{\hat{s}^2}{\hat{t}\,\hat{u}}\right]
      \delta_{ab}\delta_{ac}$ &
      & $-\frac{N^2-1}{4N^3}$ &
      & $\frac{N^2+1}{4N^3}\,(-1)$ &
      & $\frac{1}{4N^3}\,(-1)$ &
      \\ \hline
($\bar{\rm j}$)   
      & $2\left[\frac{\hat{s}^2}{\hat{t}\,\hat{u}}\right]
      \delta_{ab}\delta_{bc}$ &
      & $-\frac{N^2-1}{4N^3}$ &
      & $\frac{N^2+1}{4N^3}\,(-1)$ &
      & $\frac{1}{4N^3}\,(-1)$ &
      \\ \hline
(k)   & $2\left[\frac{\hat{t}^2+\hat{u}^2}{\hat{s}^2}\right]
      \delta_{a\bar{b}}$ &
      & $\frac{N^2-1}{4N^2}$ &
      & $-\frac{1}{4N^2}\,(-1)$ &
      & $\frac{N^2+4N-4}{32N}$ &
      \\ \hline
(l)   & $2\left[\frac{\hat{t}^2+\hat{u}^2}{\hat{s}^2}\right]
      \delta_{a\bar{b}}$ &
      & $\frac{N^2-1}{4N^2}$ &
      & $-\frac{1}{4N^2}$ &
      & $\frac{N^2-4N-4}{32N}$ &
      \\ \hline
(m)   & $2\left[\frac{\hat{t}^2+\hat{u}^2}{\hat{s}^2}\right]
      \delta_{a\bar{b}}$ &
      & $\frac{N^2-1}{4N^2}$ &
      & $-\frac{1}{4N^2}\,(-1)$ &
      & $\frac{N^2-4N-4}{32N}\,(-1)$ &
      \\ \hline
(n)   & $2\left[\frac{\hat{t}^2+\hat{u}^2}{\hat{s}^2}\right]
      \delta_{a\bar{b}}$ &
      & $\frac{N^2-1}{4N^2}$ &
      & $-\frac{1}{4N^2}$ &
      & $\frac{N^2+4N-4}{32N}\,(-1)$ &
\end{tabular}
\end{table}


\begin{references}

\bibitem{EXP:EMC} J. Ashman et al., Phys. Lett. {\bf B206}, 364
(1988).

\bibitem{THY:EMC} For a recent review, see, for example, 
R.L. Jaffe, in the proceedings of {\it Ettore Majorana International 
School of Nucleon Structure: The Spin Structure of the
Nucleon}, Erice, Italy, 3-10 Aug 1995, hep-ph/9602236.

\bibitem{EXP:Lambda} 
G. Bunce et al., Phys. Rev. Lett. {\bf 36}, 1113 (1976); 
K. Heller et al., Phys. Lett. {\bf B68}, 480 (1977);
S.A. Gourlay et al., Phys. Rev. Lett. {\bf 56}, 2244 (1986).

\bibitem{EXP:Pion} 
D.L. Adams et al., Phys. Lett. {\bf B261}, 201 (1991); {\bf B264}, 462 (1991);
A. Bravar et al., Phys. Rev. Lett. {\bf 77}, 2626 (1996).

\bibitem{JFO:Rev} 
J.F. Owens, Rev. Mod. Phys. {\bf 59}, 465 (1987). 

\bibitem{AN:KPR} G.L. Kane, J. Pumplin and W. Repko, Phys. Rev. Lett.
                 {\bf 41}, 1689 (1978).
       
\bibitem{AN:ET1} A.V. Efremov and O.V. Teryaev, Sov. J. Nucl. Phys. {\bf 36}, 
140 (1982)
                 [Yad. Fiz. {\bf 36},
                 242, (1982) ].

\bibitem{AN:ET2} A.V. Efremov and O.V. Teryaev, Phys. Lett. {\bf 150B},
                 383 (1985);  Sov. J. Nucl. Phys. {\bf 36}, 557 (1982);
                 {\bf 39}, 962 (1984)[Yad. Fiz. {\bf 36}, 950 (1982); {\bf 39},
                 1517, (1984)].

\bibitem{AN:ET3} A.V.\ Efremov, V.\ Korotkiyan and O.\ Teryaev,
Phys.\ Lett.\ {\bf B348}, 577 (1995).

\bibitem{AN:QS} J.W. Qiu and G. Sterman, Phys. Rev. Lett. {\bf 67}, 
2264 (1991); Nucl. Phys. {\bf B378}, 52 (1992).

\bibitem{QS:FACs} J.W. Qiu and G. Sterman, in {\it Polarized 
Collider Workshop}, University Park, 
PA, 1990, AIP Conference Proceedings 233, ed.\ J.C. Collins, S.F. Heppelman, and 
R.W. Robinett (American Institute of Physics, New York, 1990).

\bibitem{AN:Ryskin} M.G.\ Ryskin, Sov.\ J.\ Nucl.\ Phys.\ {\bf 48}, 708 (1988);
D.I.\ D'yakanov and V.Yu.\ Petrov, Sov.\ Phys.\ JETP, {\bf 62}, 204 (1985).

\bibitem{TF:Schafer} A.\ Sch\"afer, L.\ Mankiewicz, P.\ Gornicki
and S.\ G\"ullenstern, Phys.\ Rev.\ {\bf D47}, 1 (1993); 
B.\ Ehrnsperger, A.\ Sch\"afer, W.\ Greiner and L.\  Mankiewicz, 
Phys.\ Lett.\ {\bf B321}, 121 (1994).

\bibitem{TFDY:Boer} D.\ Boer and P.J.\ Mulders, Phys.\ Rev.\ {\bf D57}, 3057 
(1998).

\bibitem{AN:Sivers} D.\ Sivers, Phys.\ Rev.\ {\bf D41}, 83 (1990); Phys.\ Rev., 
{\bf D43}, 261 (1991). 

\bibitem{AN:Anselmino} M.\ Anselmino, M.\ Boglione and F.\ Murgia,
Phys.\ Lett.\ {\bf B362}, 164 (1995).

\bibitem{FRAG:Collins} J.\ Collins, Nucl.\ Phys.\ {\bf B396}, 161 (1993);
J.C.\ Collins, S.F.\ Heppelmann and G.A.\ Ladinsky, Nucl.\ Phys.\ {\bf B420}, 
565
(1994).

\bibitem{FRAG:JJ} R.L.\ Jaffe and X.-d.\ Ji, Phys.\ Rev.\ Lett.\ {\bf 71}, 2547 
(1993). 

\bibitem{AN:Artru} X.\ Artru , J.\ Czyzewski and H.\ Yabuki, Z.\ Phys.\ {\bf 
C73}, 527 (1997);
X.\ Artru  and J.\ Czyzewski, hep-ph/9805463.

\bibitem{ktsmear} 
J.\ Huston {\it et al.}, Phys.\ Rev.\ {\bf D51}, 6139 (1995);
H.\ Baer and M.H.\ Reno, Phys.,\ Rev.\ {\bf D54}, 2017 (1996);
A.D.\ Martin, R.G.\ Roberts, W.J.\ Stirling and R.S.\ Thorne, hep-ph/9803445.

\bibitem{Ralston:Soper} J.P.\ Ralston and D.E.\ Soper, Nucl. Phys. 
{\bf B152}, 109 (1979).

\bibitem{Jaffe:Ji} R.L.\ Jaffe and X.-d.\ Ji, Phys. Rev. Lett. {\bf 67},
552 (1991); Nucl.\ Phys.\ {\bf B375}, 527 (1992); X.-d.\ Ji,
Phys. Lett. {\bf B284}, 137 (1992).

\bibitem{TT1} S.M.\ Troshin  and N.E.\ Tyurin, Phys.\ Rev.\ {\bf D52}, 3862 
(1995);
Phys.\ Rev., {\bf D54}, 838 (1996). 

\bibitem{Mengetal1} C.\ Boros, Z.-t.\ Liang, T.-c.\ Meng, Phys.\ Rev.\ {\bf 
D51}, 4867 (1995).

\bibitem{TT2} P.M.\ Nadolsky, S.M.\ Troshin and N.E.\ Tyurin,
Int.\ J.\ Mod.\ Phys.\ {\bf A9}, 2505 (1994).

\bibitem{Mengetal2} C.\ Boros, Z.-t.\ Liang, T.-c.\ Meng and R.\ Rittel, J.\ 
Phys.\ {\bf G24}, 75 
(1998).

\bibitem{Christ:Lee} N. Christ and T.D. Lee, Phys. Rev. {\bf 143}, 
1310 (1966).

\bibitem{EXP:Photon} 
D.L. Adams et al., Phys. Lett. {\bf B345}, 569 (1995).

\bibitem{Ratcliffe} P.G.\ Ratcliffe, Nucl.\ Phys.\ {\bf B264}, 493 (1986).

\bibitem{AN:JI} X.-d. Ji, Phys. Lett. {\bf B289}, 137 (1992).

\bibitem{CSS:FAC} J.C. Collins, D.E. Soper and G. Sterman, in 
{\it Perturbative  QCD}, ed.\ A.H. Mueller (World Scientific, Singapore, 1989).

\bibitem{QS:DYQ} J.W. Qiu and G. Sterman, Nucl. Phys. {\bf B353}, 
                 105 (1991). 

\bibitem{QS:FAC} J.W. Qiu and G. Sterman, Nucl. Phys. {\bf B353}, 
                 137 (1991).

\bibitem{HT:Ellis} R.K. Ellis, W. Furmanski and R. Petronzio,
Nucl. Phys. {\bf B207}, 1 (1982); Nucl. Phys. {\bf B212}, 29 (1983);
R.L.\ Jaffe, Nucl.\ Phys.\ {\bf B229}, 205 (1983).

\bibitem{HT:QIU} J.W. Qiu, Phys. Rev. {\bf D42}, 30 (1990).

\bibitem{IRRDIS1} Y.L.\ Dokshitzer, G.\ Marchesini and B.R.\ Webber,
Nucl.\ Phys.\ {\bf B469}, 93 (1996);
M.\ Dasgupta and B.R.\ Webber, Phys.\ Lett.\ {\bf B382}, 273 (1996);
Nucl.\ Phys.\ {\bf B484}, 247 (1997).

\bibitem{IRRDIS2} E.\ Stein, M.\ Meyer-Hermann, 
L.\ Mankiewicz and A.\ Sch\"afer, Phys.\ Lett.\ {\bf B376}, 177 (1996);
M.\ Meyer-Hermann {\it  et al.}, Phys.\ Lett.\ {\bf B383}, 463 (1996)
(E., {\bf B393}, 487 (1997));
E.\ Stein, M.\ Maul, L.\ Mankiewicz and A.\ Sch\"afer, hep-ph/9803342;

\bibitem{IRRFrag} M.\ Beneke,  V.M.\ Braun and L.\ Magnea, Nucl.\ Phys.\ {\bf 
B497}, 297 (1997).  
 
\bibitem{CS:PDF} 
J.C. Collins and D.E. Soper, Nucl. Phys. {\bf B194}, 445 (1982).

\bibitem{PI:frag} R. Baier, J. Engles and B. Petersson, Z. Phys.
{\bf C2}, 265 (1979).



\end{references}
\end{document}